\documentclass[twocolumn,prd,superscriptaddress,nofootinbib,preprintnumbers]{revtex4-1}

\usepackage{mathtools,slashed}
\usepackage{booktabs,multirow}
\usepackage{graphicx}
\usepackage{amsmath,bm,amssymb,amsfonts,dsfont,xspace}
\usepackage{textgreek}
\usepackage{derivative}
\usepackage[usenames,dvipsnames]{xcolor}
\usepackage[normalem]{ulem}
\usepackage{url}
\usepackage{footmisc}
\usepackage{multirow}
\usepackage[colorlinks  = true,
            linkcolor   = NavyBlue,
            urlcolor    = NavyBlue,
            citecolor   = NavyBlue,
            anchorcolor = NavyBlue, unicode]{hyperref}

\newcommand{\lsim}{\mathrel{\mathop{\kern 0pt \rlap
  {\raise.2ex\hbox{$<$}}}
  \lower.9ex\hbox{\kern-.190em $\approx$}}}
\newcommand{\gsim}{\mathrel{\mathop{\kern 0pt \rlap
  {\raise.2ex\hbox{$>$}}}
  \lower.9ex\hbox{\kern-.190em $\approx$}}}

\renewcommand{\vec}[1]{\boldsymbol{#1}}

\usepackage[normalem]{ulem} 

\def \prd {Phy. Rev. D}

\begin{document}

\preprint{APS/123-QED}

\title{A Comprehensive Study of WIMP Models Explaining the $\it Fermi$-LAT Galactic Center Excess}

\author{Chuiyang Kong}
\email{chuiyang\_kong@brown.edu}
\affiliation{Department of Physics, Brown University, 182 Hope Street, Providence, RI 02912, USA}
\author{Mattia Di Mauro}%
\email{dimauro.mattia@gmail.com}
\affiliation{%
 Istituto Nazionale di Fisica Nucleare, Sezione di Torino, Via P. Giuria 1, 10125 Torino, Italy
}%




\date{\today}

\begin{abstract}
The Galactic Center excess (GCE) of GeV $\gamma$ rays may hint at dark matter (DM), yet its origin remains debated. Motivated by this, we survey weakly interacting massive particle (WIMP) models that can fit the GCE while satisfying relic-density, direct-detection (DD), and indirect-detection (ID) bounds. We group candidates into \emph{hadronic} (Higgs portals; simplified scalar/vector mediators), \emph{leptonic} ($U(1)_{L_i-L_j}$), and \emph{mixed} ($U(1)_{B-L}$, $Z$-portal) classes. Across all cases, present DD and dwarf-spheroidal $\gamma$-ray limits exclude wide regions, leaving mainly narrow \emph{resonant funnels} with $m_{\rm DM}\!\simeq\! m_{\rm med}/2$ and portal couplings $\ll 1$. 
In hadronic setups, scalar and vector Higgs portals survive only in a thin strip near $m_h/2\simeq62.5$~GeV with portal couplings $\sim\!10^{-4}$, while the Dirac Higgs and $Z$ portals are essentially excluded. The UV-complete vector Higgs portal retains resonant bands whose viable portal strength depends on the mixing angle. Simplified scalars allow small windows for complex-scalar or vector DM; Dirac DM is strongly disfavored, whereas a pseudoscalar with Dirac DM remains viable over a broader parameter range. For a simplified $Z'$ mediator, a pure vector coupling leaves only a marginal region, while pure axial is excluded by DD/ID bounds. 
In leptonic scenarios, inverse-Compton emission is essential: $L_\mu\!-\!L_e$ (and, to a lesser extent, $B\!-\!L$) fits the GCE with near-thermal cross sections, while $L_\mu\!-\!L_\tau$ is disfavored. Overall, viable WIMP explanations are \emph{constrained to the finely tuned resonant regime}, with leptophilic vectors and pseudoscalar portals emerging as the most robust options.
\end{abstract}

\maketitle

\section{Introduction}
\label{sec:intro}

The nature of dark matter (DM) is one of the most puzzling problems in contemporary physics.
While the gravitational imprint of DM is firmly established across a wide range of scales—from galaxy rotation curves to the cosmic microwave background (CMB)—its particle nature remains elusive~\cite{Bertone:2010zza,Bertone:2016nfn,Cirelli:2024ssz}.
If DM consists of particles, new physics beyond the Standard Model (BSM) is required, as no known SM state can account for the observed abundance~\cite{Aghanim:2018eyx}.

A credible DM candidate must meet several broad criteria~\cite{Cirelli:2024ssz}: stability on cosmological timescales, electrical neutrality, nonrelativistic behavior by matter–radiation equality, and weak interactions with ordinary matter.
In addition, the underlying BSM framework must furnish a mechanism that yields the measured relic density, $\Omega_{\rm DM} h^2 \simeq 0.12$~\cite{Aghanim:2018eyx}.
A particularly compelling paradigm is that of Weakly Interacting Massive Particles (WIMPs), wherein the relic abundance arises from thermal freeze-out of pair annihilations in the early Universe.
In this picture, particles with masses in the range $\mathcal{O}(10$–$10^3)$~GeV and interaction strengths near the weak scale naturally reproduce the observed DM density—the so-called “WIMP miracle.”

Four complementary search strategies probe the WIMP hypothesis.
Cosmology determines the total DM abundance at the $\sim\!1\%$ level through precision CMB measurements~\cite{Aghanim:2018eyx}.
Direct detection (DD) experiments search for nuclear recoils from DM scattering in underground detectors~\cite{Schumann:2019eaa}.
Collider searches (at LEP and the LHC) look for events with missing transverse momentum consistent with DM production~\cite{Boveia:2018yeb}.
Finally, indirect detection (ID) seeks the stable final states of DM annihilation or decay—most notably $\gamma$ rays, antimatter, and neutrinos—above their astrophysical backgrounds~\cite{Gaskins:2016cha}.

Among ID channels, $\gamma$ rays are especially promising because they are not deflected by magnetic fields and retain the source’s spectral and morphological information.
Consequently, observations of regions with high expected DM density are particularly informative.
Among potential targets, the Galactic Center (GC) has the highest expected DM density~\cite{Pieri:2009je}.
Analyses of {\it Fermi}-LAT data have repeatedly identified an excess of GeV $\gamma$ rays—the Galactic Center Excess (GCE)~\cite{Goodenough:2009gk,Hooper:2010mq,Boyarsky:2010dr,Hooper:2011ti,Abazajian:2012pn,Gordon:2013vta,Abazajian:2014fta,Daylan:2014rsa,Calore:2014nla,Calore:2014xka,TheFermi-LAT:2015kwa,TheFermi-LAT:2017vmf,DiMauro:2019frs,DiMauro:2021raz,Cholis:2021rpp}.
Recent template-fitting studies~\cite{DiMauro:2021raz,Cholis:2021rpp} reveal an approximately spherical morphology, a spectrum peaking at a few GeV and extending to $\sim\!50$~GeV, and a radial profile consistent with a generalized NFW distribution with inner slope $\gamma\simeq1.2$–$1.3$.
Such characteristics are compatible with annihilating WIMPs of mass $m_{\rm DM}\!\sim\!30$–$60$~GeV into hadronic~\cite{DiMauro:2021qcf} or leptonic~\cite{Koechler:2025ryv} final states, with cross sections not far from the thermal benchmark.

Astrophysical explanations—most prominently unresolved millisecond pulsars—remain viable~\cite{Bartels:2015aea,Lee:2015fea,Macias:2016nev,Bartels:2017vsx,Manconi:2024tgh} even though recent work has emphasized that uncertainties in interstellar emission modeling can mimic or obscure population signatures~\cite{Leane:2019uhc,Chang:2019ars,Zhong:2019ycb,Calore:2021jvg,List:2025qbx}.
Consequently, a DM origin of the GCE remains a motivated possibility.

At the same time, present DD limits (e.g., LZ and XENONnT) constrain spin-independent (SI) scattering down to $\mathcal{O}(10^{-47}\text{--}10^{-48})\,\mathrm{cm}^2$ for weak-scale masses~\cite{LZ:2023,Aprile:2023XENONnT,LZ:2024zvo}, excluding broad regions of parameter space for models where the same portal controls both freeze-out and elastic scattering~\cite{Arcadi:2017kky,Arcadi:2019lka,DiMauro:2023tho,Arcadi:2024ukq,DiMauro:2025jia}.
Thermal annihilation through an $s$-channel resonance can relax these bounds, albeit with some mass tuning~\cite{DiMauro:2023tho,Koechler:2025ryv,DiMauro:2025jia}.
A more generic route is \emph{secluded} DM~\cite{Pospelov:2007mp,Pospelov:2008jd,DiMauro:2025jsb,DiMauro:2025uxt}, where $\Omega_{\rm DM}$ is set by annihilations into mediator pairs within a dark sector, while the SM–dark coupling (e.g.\ a Higgs mixing or a gauge kinetic mixing) is small.
In addition, the non-observation of $\gamma$ rays from dwarf spheroidal galaxies (dSphs) places stringent complementary limits on hadronic channels~\cite{DiMauro:2021qcf,McDaniel:2023bju}.
Given the tight constraints from DD and ID, and the plausibility of a millisecond-pulsar interpretation of the GCE, it is worthwhile to assess whether specific BSM models with DM can reproduce the excess while remaining compatible with all current bounds.

In this work we examine several WIMP frameworks capable of fitting the GCE spectrum while remaining consistent with relic abundance, DD, collider, and other ID constraints (including dSph limits)~\cite{McDaniel:2023bju}.
Our focus is on models where annihilation proceeds through an $s$-channel mediator into SM states.
In particular, we consider the following models:
\begin{itemize}
  \item \textbf{Canonical simplified models.}
  A single $s$-channel mediator—CP-even scalar $S$, CP-odd pseudoscalar $A$, or vector $Z'$—couples DM to SM fermions (Yukawa-like couplings for scalars, vector/axial for $Z'$)~\cite{ABDALLAH20158,Arina:2018zcq,Chang:2022jgo,Chang:2023cki,Arcadi:2024ukq,DiMauro:2025jia}.
  This setup captures broad collider and astroparticle phenomenology with four parameters (DM and mediator masses; mediator couplings to DM and to SM fermions) and serves as a benchmark for collider searches~\cite{ABDALLAH20158}.
  \item \textbf{Higgs portal.}
  A SM-singlet DM candidate interacts through the operator $H^\dagger H$, tying the dark sector to electroweak symmetry breaking~\cite{Cline:2013gha,Beniwal:2015sdl,Arcadi:2019lka,Arcadi:2024ukq,DiMauro:2023tho}.
  In its minimal form this model is highly predictive and tightly linked to Higgs measurements.
  \item \textbf{Vector Higgs portal (UV–complete).}
  A minimal, renormalizable realization with a dark gauge boson and a singlet scalar that mixes with the Higgs~\cite{Farzan:2012hh,Ko:2014gha,Baek:2014goa,Duch:2015jta,Arcadi:2020jqf}.
  Phenomenology is driven by Higgs–portal interactions: relic density from $s$–channel scalar exchange (often resonant) and coherent nuclear scattering via $t$–channel exchange, potentially softened by interference with the extra scalar.
  \item \textbf{$Z$ portal.}
  The SM $Z$ boson serves as the mediator between DM and SM fields~\cite{Arcadi:2014lta}, with strong constraints from LEP/LHC measurements and electroweak precision tests.
  \item \textbf{Anomaly-free \(U(1)\) extensions.} 
  
  \(\mathbf{U(1)_{L_i-L_j}}\): A new gauge symmetry acting on lepton-family differences introduces a light vector coupled to selected leptons \cite{Foot:1990mn,1991PhRvD..43...22H,Heeck:2011wj,Bauer:2018onh,Koechler:2025ryv}. It naturally yields leptophilic interactions and connects to neutrino and flavor sectors. This model features purely leptonic DM annihilation channels, in contrast to the previous cases that are mostly hadronic. Kinetic mixing provides the portal to nuclei in direct detection. 
  
  \(\mathbf{U(1)_{B-L}}\): A universal gauge symmetry for baryon minus lepton number extends the SM with a well-motivated vector boson and a simple charge assignment, enabling coherent tests from colliders, low-energy experiments, and cosmology \cite{Bauer:2018onh,Koechler:2025ryv}. This model mixes in a democratic manner DM annihilating into leptonic and hadronic channels.
\end{itemize}

In summary, we investigate whether WIMP models with $s$-channel mediators—spanning simplified scalar/vector mediators, Higgs and $Z$ portals, and anomaly-free $U(1)_{L_i-L_j}$ and $U(1)_{B-L}$ extensions—can simultaneously reproduce the GCE and satisfy current bounds on annihilation, scattering, and production.

The paper is organized as follows.
Section~\ref{sec:models_overview} presents the theoretical setup, including Lagrangians and dominant annihilation modes, and outlines the cosmological, DD, and ID constraints.
Results for each framework are given in Section~\ref{sec:results}.
We conclude in Section~\ref{sec:conclusions} with implications and prospects for future searches.

\section{Overview of the Models}
\label{sec:models_overview}

In this section we present the model ingredients used throughout the paper in a unified notation and with a common structure. 
For each setup we: (i) write the interaction Lagrangian and specify the particle content; 
(ii) summarize the dominant annihilation channels that control the relic abundance and set the main ID signals; 
(iii) provide the leading DD amplitudes. 
\textit{Notation is standardized across models.}
We adopt a minimal parameter set 
\(\{m_{\rm DM},\,m_{\rm med},\,g_{\rm DM},\,g_f\}\), 
where $m_{\rm DM}$ and $m_{\rm med}$ are the DM and mediator masses, while \(g_{\rm DM}\) and \(g_f\) denote the mediator couplings to the dark and visible sectors, respectively. 
When useful, we take $g_{\rm DM} \equiv g_f \equiv \lambda_{\rm portal}$ to reduce dimensionality.
When applicable we assume Minimal Flavour Violation (MFV), so fermionic couplings scale with the SM Yukawas, 
\(y_f=m_f/(\sqrt{2}\,v_h)\) with \(v_h=246.22~\mathrm{GeV}\).
All $s$-channel processes share the Breit--Wigner kernel
\begin{equation}
\label{eq:BW_den}
D_{\rm BW}(s;m_{\rm med}) \equiv \bigl(s-m_{\rm med}^{2}\bigr)^{2} + m_{\rm med}^{2}\Gamma_{\rm med}^{2},
\end{equation}
evaluated at \(s\simeq 4m_{\rm DM}^{2}\) for freeze-out and ID, implying a resonant enhancement near \(m_{\rm med}\simeq 2m_{\rm DM}\).
We use \(N_c^f=3\,(1)\) for quarks (leptons), \(\mu_N=m_{\rm DM}m_N/(m_{\rm DM}+m_N)\) for the DM--nucleon reduced mass, 
and the standard scalar nucleon factor (including the trace anomaly),
\begin{equation}
\label{eq:fN_def_master}
f_N=\sum_{q=u,d,s} f_q^N + \frac{2}{9}\!\left(1-\sum_{q=u,d,s} f_q^N\right) \simeq 0.30\text{--}0.33.
\end{equation}

For phenomenology and numerics we implement each model in {\tt FeynRules}~\cite{Alloul:2013bka} to generate {\tt UFO}~\cite{Degrande:2011ua} and {\tt CalcHEP}~\cite{Belyaev:2012qa} files, interfaced to {\tt MadDM}~\cite{Backovic:2013dpa,Ambrogi:2018jqj,Arina:2021gfn} and {\tt micrOMEGAs}~\cite{Belanger:2006is,Belanger:2013oya,Belanger:2018ccd,Alguero:2023zol} for relic density and indirect detection.

\vspace{0.6em}
\subsection{Dark Matter Simplified Models}
\label{sec:model_simplified}

We consider CP-even/odd spin-0 and spin-1 mediators, and DM with spin \(0,\,\tfrac12,\,1\). 
The low-energy spectrum contains only the DM and one BSM mediator. DM stability follows from an exact \(\mathbb{Z}_2\) symmetry~\cite{ABDALLAH20158,Arina:2018zcq,Chang:2022jgo,Chang:2023cki,Arcadi:2024ukq,DiMauro:2025jia}. 

\paragraph{CP-even scalar mediator \(S\).}
A real CP-even scalar \(S\) couples to SM fermions with MFV and to DM (scalar \(\chi\), Dirac fermion \(\psi\), vector \(V_\mu\)):
\begin{align}
\mathcal{L}_{\rm int}^{\chi} &= \xi\,\mu_\chi \lambda_\chi\, \chi^2 S + \xi\,\lambda_\chi^2\,\chi^2 S^2 
+ \sum_f \frac{g_f}{\sqrt{2}}\frac{m_f}{v_h}\,\bar f f\,S, \\
\mathcal{L}_{\rm int}^{\psi} &= \xi\,\lambda_\psi\,\bar\psi\psi\,S 
+ \sum_f \frac{g_f}{\sqrt{2}}\frac{m_f}{v_h}\,\bar f f\,S, \\
\mathcal{L}_{\rm int}^{V}    &= \mu_V \lambda_V\, V_\mu V^\mu S + \tfrac12 \lambda_V^2 V_\mu V^\mu S^2 
+ \sum_f \frac{g_f}{\sqrt{2}}\frac{m_f}{v_h}\,\bar f f\,S,
\end{align}
with \(\xi=1/2\) for real/Majorana and \(\xi=1\) for complex/Dirac fields. 

\emph{Annihilation and ID.} The $s$-channel mode \({\rm DM\,DM}\to S^\ast\to \bar f f\) is helicity suppressed (\(\propto m_f^2\)); keeping the leading velocity/helicity terms:
\begin{align}
\label{eq:spin0scalar}
\langle\sigma v\rangle_{\bar f f}^{(\chi)} 
&= \frac{N_c^f}{4\pi}\,\frac{\lambda_\chi^2 g_f^2}{2}\,\frac{m_f^2}{v_h^2}\,
\frac{m_\chi^2}{D_{\rm BW}(4m_\chi^2;m_S)}\!\left(1-\frac{m_f^2}{m_\chi^2}\right)^{3/2},\\
\langle\sigma v\rangle_{\bar f f}^{(\psi)} 
&= \frac{N_c^f}{2\pi}\,\lambda_\psi^2 g_f^2\,\frac{m_f^2}{v_h^2}\,
\frac{m_\psi^2 v_{\rm rel}^2}{D_{\rm BW}(4m_\psi^2;m_S)}\!\left(1-\frac{m_f^2}{m_\psi^2}\right)^{3/2},\\
\langle\sigma v\rangle_{\bar f f}^{(V)} 
&= \frac{N_c^f}{12\pi}\,\lambda_V^2 g_f^2\,\frac{m_f^2}{v_h^2}\,
\frac{m_V^2}{D_{\rm BW}(4m_V^2;m_S)}\!\left(1-\frac{m_f^2}{m_V^2}\right)^{3/2},
\end{align}
where $v_{\rm rel}$ is the DM relative velocity.
Scalar/vector DM give $s$-wave (helicity suppressed), while fermion DM is $p$-wave. Therefore, for scalar and vector DM, ID limits can be relevant; for fermion DM, since $v_{\rm rel} \sim\mathcal{O}(10^{-4}\text{--}10^{-5})$ in dSphs, ID bounds are weak.
If kinematically available, \( {\rm DM\,DM}\to SS \) can dominate freeze-out and yields \(4f\) final states via \(S\to \bar f f\)~\cite{Arcadi:2019lka}. In this work we focus on the near-resonant region $m_{\rm DM}\sim m_{\rm med}/2$, where mediator-pair channels are less relevant.

\emph{Direct detection.} In the non-relativistic limit, integrating out \(S\) gives the coherent SI cross section
\begin{equation}
\label{eq:SI_scalar_med}
\sigma_{N}^{\rm SI} \;=\; \frac{\mu_{N}^{2}}{\pi}\,
\left(\frac{g_{\rm DM}}{m_S^{2}}\right)^{2}\left(\frac{m_N}{v_h}\,f_N\right)^{2},
\end{equation}
with $g_{\rm DM}\in\{\lambda_\chi,\,\lambda_\psi,\,\lambda_V\}$. Current SI limits from LZ and XENONnT ~\cite{LZ:2023,Aprile:2023XENONnT,LZ:2024zvo} are very strong and exclude most of the off-resonance parameter space, leaving essentially the $s$-channel funnel~\cite{DiMauro:2025jia}.

\medskip
\paragraph{CP-odd scalar mediator \(A\).}
For a pseudoscalar \(A\) coupled to Dirac DM,
\begin{equation}
\label{eq:pseudoDM_L}
\mathcal{L}^{\psi}_{\rm int} \;=\; i\,\lambda_{\psi}\, \bar{\psi}\gamma_5 \psi\, A 
\;+\; i \sum_f \frac{g_f}{\sqrt{2}} \frac{m_f}{v_h}\, \bar{f}\gamma_5 f\, A.
\end{equation}

\emph{Annihilation and ID.} \(\psi\bar\psi\to A^\ast\to \bar f f\) is $s$-wave but helicity suppressed (\(\propto m_f^2\)), while \(\psi\bar\psi\to AA\) is $p$-wave.

\emph{Direct detection.} The tree-level operator \((\bar\psi i\gamma_5\psi)(\bar q i\gamma_5 q)\) maps to \(\mathcal{O}_6^{\rm NR}\) with \(\mathrm{d}\sigma/\mathrm{d}E_R\propto q^4/m_A^4\)~\cite{Fitzpatrick:2012ix,Arina:2014yna,Dolan:2014ska} and is extremely suppressed.
Loop-induced SI terms, free of the \(q^4\) penalty, can dominate for \(m_A\!\gtrsim\!{\rm GeV}\) and set competitive bounds~\cite{Abe:2018emu,Ertas:2019dew,DiMauro:2025jia}, still typically weaker than tree-level SI in scalar-mediator scenarios.

\medskip
\paragraph{Spin-1 mediator \(Z'\).}
A massive vector \(Z'_\mu\) couples to DM and to SM fermions with generic vector/axial charges:
\begin{align}
\mathcal{L}_{\chi} &= i\,g_{\chi}\,(\chi^* \partial_{\mu}\chi - \chi\,\partial_{\mu}\chi^*)\, Z'^{\mu}
+ g_{\chi}^{2}\,|\chi|^{2}\,Z'_{\mu}Z'^{\mu} \nonumber\\ 
&\quad + g_{f}\,\bar{f}\,\gamma^{\mu}\!\bigl(V_{f}-A_{f}\gamma_{5}\bigr) f\,Z'_{\mu}, 
\label{eq:spin1-Zp_Lchi}\\
\mathcal{L}_{\psi} \!&=\! g_{\psi}\,\bar{\psi}\,\gamma^{\mu}\!\bigl(V_{\psi}\!-\!A_{\psi}\gamma_{5}\bigr) \psi\,Z'_{\mu}
\!+\! g_{f}\,\bar{f}\,\gamma^{\mu}\!\bigl(V_{f} \!-\!A_{f}\gamma_{5}\bigr) f\,Z'_{\mu}.
\label{eq:spin1-Zp_Lpsi}
\end{align}

\emph{Annihilation and ID.} For scalar DM, \(\chi\chi^\ast\!\to\!\bar f f\) is $p$-wave:
\begin{equation}
\langle\sigma v\rangle_{\chi\chi^\ast\to f\bar f}
= \frac{N_c^f\,g_\chi^{2} g_f^{2}}{2\pi}\,
\frac{m_\chi^{2}\,v_{\rm rel}^{2}\,(V_f^{2}+A_f^{2})\,\sqrt{1-\tfrac{m_f^2}{m_\chi^2}}}
{D_{\rm BW}(4m_\chi^2;m_{Z'})}.
\end{equation}
For Dirac DM with pure vector coupling (\(A_\psi=0\)) and generic SM charges:
\begin{equation}
\langle\sigma v\rangle_{\psi\bar\psi\to f\bar f}
= \frac{N_c^f\,g_\psi^{2} g_f^{2}}{2\pi}\,
\frac{ (V_\psi V_f)^2\,m_\psi^{2}\!\left(1+\tfrac{m_f^{2}}{2m_\psi^{2}}\right)\!\sqrt{1-\tfrac{m_f^{2}}{m_\psi^{2}}}}
{D_{\rm BW}(4m_\psi^2;m_{Z'})},
\end{equation}
which is $s$-wave. 
Pure axial--axial is helicity suppressed \(\propto m_f^2\). 
If open, \(Z'Z'\) final states can control freeze-out and lead to \(4f\) signatures after decay.

\emph{Direct detection.} 
Vector quark couplings induce coherent SI scattering,
\begin{eqnarray}
\label{eq:Zp_SI}
\sigma^{\rm SI}_{N}
=\frac{\mu_{N}^{2}}{\pi}\left(\frac{g_\mathrm{DM}\,g_f}{m_{Z'}^{2}}\right)^{\!2}
f_{N}^{2}, \\
f_{p}=2V_{u}+V_{d},\quad f_{n}=V_{u}+2V_{d},
\end{eqnarray}
and are tightly bounded unless isospin-violating cancellations occur. 
Pure axial couplings yield SD, velocity-suppressed scattering,
\begin{eqnarray}
\sigma^{\rm SD}_{N}(v)=\frac{4\,\mu_N^{2}}{\pi}\left(\frac{g_\mathrm{DM}\,g_f}{m_{Z'}^{2}}\right)^{2}
\!\bigl[C_A^{(N)}\bigr]^{2} v^{2},\\
C_A^{(p)}=A_u\Delta u^{(p)}+A_d\Delta d^{(p)}+A_s\Delta s^{(p)},
\end{eqnarray}
and are therefore much less constrained~\cite{Fitzpatrick:2012ix,Arcadi:2024ukq}. 
In all cases, the near-resonant strip \(m_{Z'}\simeq 2m_{\rm DM}\) permits the thermal relic with smaller couplings~\cite{DiMauro:2025jia}.

\vspace{0.6em}
\subsection{Higgs Portal Models}
\label{sec:higgsportal}

We also consider SM-singlet DM coupled to the Higgs doublet \(H\) for three spins: scalar \(\chi\), vector \(V_\mu\), and Dirac \(\psi\)
~\cite{Silveira1985136,Beniwal:2015sdl,Arcadi:2019lka,DiMauro:2023tho}. 
Before EWSB:
\begin{align}
\mathcal{L}_{\chi} &= \tfrac{1}{2} (\partial_\mu \chi)^2 - \tfrac{1}{2} \mu_\chi^2 \chi^2 - \tfrac{1}{4!}\lambda_\chi \chi^4
 - \tfrac{1}{2} \lambda_{h\chi} \chi^2 H^\dagger H, \\
\mathcal{L}_{V} &= -\tfrac{1}{4} W_{\mu\nu} W^{\mu\nu} + \tfrac{1}{2} \mu_V^2 V_\mu V^\mu -\tfrac{1}{4!} \lambda_{hV} (V_\mu V^\mu)^2 \nonumber\\
 &\quad + \tfrac{1}{2} \lambda_{hV} V_\mu V^\mu H^\dagger H, \\
\mathcal{L}_{\psi} &= \bar{\psi} (i \slashed{\partial} - \mu_\psi) \psi
 - \frac{\lambda_{h\psi}}{\Lambda_\psi}\,\bar{\psi}\psi\, H^\dagger H ,
\end{align}
where $\Lambda_{\psi}$ represents the new physics scale that we fix to 1 TeV \cite{Arcadi:2024ukq}.
After EWSB, with \(H^\dagger H=\tfrac12 v_h^2+v_h h+\tfrac12 h^2\), the DM masses and Higgs couplings are:
\begin{align}
m_\chi^2&=\mu_\chi^2+\tfrac12\lambda_{h\chi}v_h^2,\ \ & g_{\chi}=\lambda_{h\chi}v_h, \\
m_V^2&=\mu_V^2+\tfrac12\lambda_{hV}v_h^2,\ \ & g_{V}=\lambda_{hV}v_h,\\
m_\psi&=\mu_\psi+\tfrac12\frac{\lambda_{h\psi}}{\Lambda_\psi}v_h^2,\ \ & g_{\psi}=\frac{\lambda_{h\psi}}{\Lambda_\psi}v_h.
\end{align}

\emph{Annihilation and ID.} The dominant modes are \(s\)-channel \(h^\ast\to f\bar f,\,W^+W^-,\,ZZ,\,gg\) and, if open, \(hh\). 
A compact expression valid for any spin is
\begin{equation}
\label{eq:HP_factorized}
\sigma v \;\simeq\;
\frac{4\,g_{\rm DM}^{2}}{D_{\rm BW}(4m_{\rm DM}^2;m_h)}
\;\frac{\Gamma_h^{\rm SM}(s=4m_{\rm DM}^{2})}{2m_{\rm DM}},
\end{equation}
with ${\rm DM}\in\{\chi,V,\psi\}$. Equation~\eqref{eq:HP_factorized} resums thresholds and captures the enhancement near \(m_{\rm DM}\simeq m_h/2\).
For \(\chi,V\) the \(\bar f f\) mode is $s$-wave but helicity suppressed; for Dirac \(\psi\) (scalar portal) it is $p$-wave and thus ID-negligible today. 
Above the $WW/ZZ$ thresholds, bosonic final states often dominate; for $m_{\rm DM}>m_h$, the $hh$ channel (plus contact/$t$/$u$ for $\chi,V$) can be important.
Eq.~\ref{eq:HP_factorized} represents a way to factorize the decay width into the annihilation cross section and it is not valid for production of $hh$.

\emph{Direct detection.} Integrating out \(h\) yields the coherent SI rate
\begin{equation}
\label{eq:HP_SI}
\sigma_{N}^{\rm SI}
=\frac{\mu_N^{2}}{\pi}\left(\frac{g_{\rm DM}}{m_h^{2}}\right)^{\!2}
\left(\frac{m_N}{v_h}\,f_N\right)^{\!2}.
\end{equation}
Away from resonance this enforces strong upper bounds on \(g_{\rm DM}\). For \(m_{\rm DM}<m_h/2\),
invisible Higgs decays further restrict the portal:
\begin{align}
\Gamma(h\to \chi\chi) &= \frac{g_{\chi}^{2}}{32\pi m_h}\,\sqrt{1-\frac{4m_\chi^{2}}{m_h^{2}}}, \\
\Gamma(h\to V V) &= \frac{g_{V}^{2} m_h^{3}}{128\pi m_V^{4}} \left(1-4r_V+12 r_V^{2}\right)\sqrt{1-4 r_V}, \\
\Gamma(h\to \psi\bar\psi) &= \frac{g_{\psi}^{2} m_h}{8\pi}\left(1-\frac{4m_\psi^{2}}{m_h^{2}}\right)^{3/2},
\end{align}
with \(r_V=m_V^{2}/m_h^{2}\).
The invisible branching ratio,
\(
\mathcal{B}_{h,\rm{inv}}= \Gamma_{h,\rm{inv}}/(\Gamma_{h,\rm{inv}}+\Gamma_{h,\rm{SM}}),
\)
is currently constrained at the \(\sim 0.1\) level~\cite{ATLAS-CONF-2020-008,CMS:2018yfx}, typically weaker than SI DD bounds; we therefore do not include it in the baseline constraints.

\emph{Fermion portal with a pseudoscalar admixture.} 
If a CP-odd component is allowed after EWSB, we write
\begin{equation}
\mathcal{L}\supset -\,\bar{\psi}\,(c_S^{\psi}+i c_P^{\psi}\gamma_5)\psi\,h,
\end{equation}
with $c_S^{\psi} = (\lambda_{h\psi}/\Lambda_\psi)\,v_h \cos\xi$ and
$c_P^{\psi} = (\lambda_{h\psi}/\Lambda_\psi)\,v_h \sin\xi$. 
Here $c_P^\psi$ yields unsuppressed $s$-wave annihilation but negligible coherent SI scattering at tree level.

\vspace{0.6em}
\subsection{UV–Complete Vector Higgs Portal}
\label{sec:VDM_UV}

We summarize the main ingredients of the model (see \cite{Farzan:2012hh,Ko:2014gha,Baek:2014goa,Duch:2015jta,Arcadi:2020jqf} for details).
We add an Abelian gauge symmetry $U(1)_X$ with gauge boson $A_\mu^X$ and a $U(1)_X$–charged complex scalar $S$ (SM singlet).  
All SM fields are neutral under $U(1)_X$. A discrete $\mathbb{Z}_2$
\begin{equation}
A_\mu^X\to- A_\mu^X,\qquad S\to S^{*}
\end{equation}
(for $S=\phi\,e^{i\sigma}$ this is $\phi\!\to\!\phi$, $\sigma\!\to\!-\sigma$) forbids renormalizable $U(1)_X$–$U(1)_Y$ kinetic mixing and stabilizes the vector boson, which is our DM candidate.  
The covariant derivative on $S$ is
\begin{equation}
D_\mu S=\big(\partial_\mu+i g_X q_X^S A_\mu^X\big)S,\qquad |q_X^S|=1,
\end{equation}
and $X_{\mu\nu}=\partial_\mu A_\nu^X-\partial_\nu A_\mu^X$.

\paragraph{Lagrangian and symmetry breaking.}
Writing the Higgs portal uniformly with the rest of the paper,
\begin{eqnarray}
&&\mathcal{L}\supset |D_\mu H|^2+|D_\mu S|^2
-\mu_H^2|H|^2+\lambda_H|H|^4 \nonumber \\
&&-\mu_S^2|S|^2+\lambda_S|S|^4 +\kappa |S|^2|H|^2 \nonumber \\
&&- \frac14 B_{\mu\nu}B^{\mu\nu}-\frac14 W^a_{\mu\nu}W^{a\,\mu\nu}-\frac14 X_{\mu\nu}X^{\mu\nu}.
\label{eq:VDM_Lagrangian}
\end{eqnarray}
EWSB and $U(1)_X$ breaking occur for
\(
\langle H\rangle=\tfrac{1}{\sqrt2}(0,v_h)^T,\;
\langle S\rangle=\tfrac{1}{\sqrt2}v_r
\)
with $v_h=246.22$~GeV and $v_r>0$.  Expanding the CP–even fluctuations,
\(
H^0=\tfrac{1}{\sqrt2}(v_h+\phi_H+i\sigma_H),\;
S=\tfrac{1}{\sqrt2}(v_r+\phi_S+i\sigma_S),
\)
the CP–even mass matrix in the $(\phi_H,\phi_S)$ basis is
\begin{equation}
\mathcal{M}^2=
\begin{pmatrix}
2\lambda_H v_h^2 & \kappa v_h v_r \\
\kappa v_h v_r & 2\lambda_S v_r^2
\end{pmatrix},
\qquad
\tan 2\alpha=\frac{\kappa v_h v_r}{\lambda_S v_r^2-\lambda_H v_h^2}.
\label{eq:mixing}
\end{equation}
Defining mass eigenstates \(H\) (SM–like) and \(H_p\) (portal scalar) via
\begin{equation}
\begin{pmatrix} H \\[2pt] H_p \end{pmatrix}
=
\begin{pmatrix}
\cos\alpha & \ \ \sin\alpha \\
-\sin\alpha & \ \ \cos\alpha
\end{pmatrix}
\begin{pmatrix} \phi_H \\[2pt] \phi_S \end{pmatrix},
\label{eq:mixing-HP}
\end{equation}
their masses are
\begin{equation}
m_{H,H_p}^2=\lambda_H v_h^2+\lambda_S v_r^2
\mp\sqrt{(\lambda_S v_r^2-\lambda_H v_h^2)^2+(\kappa v_h v_r)^2},
\end{equation}
with $m_h \simeq 125.7~\mathrm{GeV}$.
From $|D_\mu S|^2$ one gets the DM mass $m_X=g_X v_r$ and the scalar–vector couplings
\begin{equation}
g_{XXH}=-\frac{2m_X^2}{v_r}\sin\alpha,\qquad
g_{XXH_p}=+\frac{2m_X^2}{v_r}\cos\alpha.
\label{eq:VDM_gXXH}
\end{equation}
Couplings of the scalars to SM fields are rescaled by $\cos\alpha$ for $H$ and by $\sin\alpha$ for $H_p$.

\paragraph{Annihilation (relic density) and ID.}
Annihilation proceeds via $s$–channel $H,H_p$ into SM states.  A compact expression that resums thresholds is
\begin{equation}
\sigma v \simeq
\sum_{i\in\{H,H_p\}}
\frac{4\,g_{XXi}^2}{D_{\rm BW}(4m_X^2;m_i)}\;
\frac{\Gamma_i^{\rm SM}(s=4m_X^2)}{2m_X},
\label{eq:VDM_sigv}
\end{equation}
where $\Gamma_i^{\rm SM}(s)$ includes the appropriate mixing rescaling ($\propto \cos^2\alpha$ for $H$, $\propto \sin^2\alpha$ for $H_p$).
If kinematically open, $XX\to HH,\,HH_p,\,H_pH_p$ via contact and $t/u$ diagrams can impact freeze‐out.
Breit–Wigner enhancement occurs near $2m_X\simeq m_h$ or $m_{H_p}$; above the $WW/ZZ$ thresholds these channels typically dominate.

\paragraph{Direct detection.}
Elastic scattering arises from $t$–channel $H,H_p$ exchange and is SI and coherent.
For $|q|\ll m_i$ the per–nucleon cross section reads
\begin{equation}
\sigma_N^{\rm SI}
=\frac{\mu_N^2}{\pi}\left(\frac{m_N f_N}{v_h}\right)^2
\left|
\frac{g_{XXH}\cos\alpha}{m_{H}^2}
+\frac{g_{XXH_p}\sin\alpha}{m_{H_p}^2}
\right|^2,
\label{eq:VDM_SI}
\end{equation}
with $\mu_N=m_X m_N/(m_X+m_N)$ and $f_N\simeq 0.30$–$0.33$.
Because $g_{XXH}\propto -\sin\alpha$ and $g_{XXH_p}\propto \cos\alpha$, destructive $H$–$H_p$ interference can appreciably weaken SI limits for suitable $(\alpha,m_{H_p})$, whereas in the small–mixing limit $\sigma_N^{\rm SI}\propto \sin^2\alpha$ and is dominated by $H$ exchange.

\vspace{0.6em}
\subsection{\texorpdfstring{$Z$}{Z} Portal}
\label{sec:Zportal}

As a complementary benchmark we take the SM \(Z\) boson as mediator~\cite{Arcadi:2014lta}. 
For Dirac DM \(\psi\),
\begin{equation}
\label{eq:ZportalLag}
\mathcal{L}_{Z\text{-portal}} \;=\; \bar{\psi}\gamma^\mu \bigl(g_V - g_A \gamma^5\bigr)\psi\, Z_\mu.
\end{equation}

\emph{Annihilation and ID.} The process \(\psi\bar\psi\to f\bar f\) via $s$-channel \(Z\) is $s$-wave if \(g_V\neq 0\), and helicity/$p$-wave suppressed for purely axial couplings. The resonant region \(m_{\rm DM}\simeq m_Z/2\) allows the thermal target with small couplings. 

\emph{Direct detection.} Nonzero \(g_V\) induces coherent SI $Z$ exchange and is therefore very strongly constrained; viable regions typically prefer axial portals (SD, velocity suppressed). 
For \(m_{\rm DM}<m_Z/2\) the invisible $Z$ width limits \(\Gamma(Z\!\to\!\psi\bar\psi)\); above threshold, EW precision and LHC mono-$X$ searches provide complementary bounds~\cite{Arcadi:2014lta,Arcadi:2019lka,Arcadi:2024ukq}.

\vspace{0.6em}
\subsection{\texorpdfstring{$U(1)_{L_i-L_j}$}{Li-Lj} and \texorpdfstring{$U(1)_{B-L}$}{B-L}}
\label{sec:LiLjBL}

We study anomaly-free gauge extensions \(U(1)_{L_i-L_j}\) and \(U(1)_{B-L}\) with a new vector \(X_\mu\) and Dirac DM \(\chi\)~\cite{Foot:1990mn,1991PhRvD..43...22H,Heeck:2011wj,Bauer:2018onh,Koechler:2025ryv}. 
Before kinetic/mass diagonalization:
\begin{eqnarray}
\mathcal{L}&=&\mathcal{L}_{\rm SM}-\tfrac14 X_{\mu\nu}X^{\mu\nu}-\tfrac{\epsilon}{2}F_{\mu\nu}X^{\mu\nu}
+\tfrac12 m_X^2 X_\mu X^\mu \nonumber\\ 
&\quad& -\, g_X J_X^\mu X_\mu - m_\chi \bar\chi\chi - g_X q_X \bar\chi\gamma_\mu\chi X^\mu.
\end{eqnarray}
In \(L_i-L_j\) only families \(i,j\) carry \(\pm1\) charge (family \(k\) is neutral); in \(B\!-\!L\) one has \(q_X=1/3\) for quarks and \(-1\) for leptons. 
Kinetic mixing \(\epsilon F_{\mu\nu}X^{\mu\nu}\) is loop-generated even if absent at tree level~\cite{Burgess:2008ri,Bauer:2018onh}. 
For small \(\epsilon\), after diagonalization,
\begin{eqnarray}
\mathcal{L}_{\rm int}&\simeq& -eA_\mu J_{\rm EM}^\mu - Z_\mu\bigl[g_ZJ_Z^\mu+g_X\sin\xi\,J_X^\mu\bigr] \\
&\quad& -\, A'_\mu\bigl[g_X J_X^\mu - e\,\epsilon\cos\theta_W J_{\rm EM}^\mu\bigr], \nonumber
\end{eqnarray}
with \(\xi\simeq \epsilon \sin\theta_W/(1-m_X^2/m_Z^2)\).
Thus the $A'$ coupling to DM is controlled by $g_X$, while the coupling to quarks relevant for DD is \(\propto \epsilon\).

\emph{Annihilation and ID.} The relic abundance is typically set by \(\chi\bar\chi\to\ell_i\bar\ell_i,\nu_i\bar\nu_i,\ell_j\bar\ell_j,\nu_j\bar\nu_j\) controlled by \(g_X\); the resonant corridor \(m_{A'}\simeq 2m_\chi\) allows the thermal target with smaller couplings. 

\emph{Direct detection.} In \(L_i-L_j\) scattering on quarks proceeds via kinetic mixing and is suppressed by \(\epsilon^2\); in \(B\!-\!L\) it occurs at tree level and is much more constrained. 
The momentum-dependent loop correction from \((\ell_i,\ell_j)\),
\begin{equation}
\label{eq:epsQ}
\epsilon_\textrm{TOT}(Q) = \epsilon - \frac{e \, g_{X}}{2\pi^2} \int_0^1 \!dx\, x(1-x)\, 
\log \!\left( \frac{m_j^2 + Q^2 x(1 - x)}{m_i^2 + Q^2 x(1 - x)} \right),
\end{equation}
interpolates between an IR plateau and the UV value \(\epsilon\), and can be tuned in the WIMP recoil window (\(|\vec q|\sim 10\text{--}400\) MeV) without conflicting with collider bounds~\cite{Bauer:2018onh}. 
For a nuclear target \(N=(A,Z)\) the vector-portal cross section reads~\cite{Evans:2017kti,Arcadi:2018tly}
\begin{widetext}
\begin{align}
\sigma_{N}(Q)
&= \frac{\mu_{\chi N}^{2}}{\pi}\,q_X^2g_X^{2}\,\epsilon_{\rm{TOT}}^{2}(Q)\,F^2_\text{Helm}(Q)\,
\left|
    \frac{f_{N}^{(A')}}{m_{A'}^{2}+Q^2} - 
    \frac{\sin\theta_W\, f_{N}^{(Z)}}{(m_{Z}^{2}-m_{A'}^{2})}
\right|^{2},\\
f_{N}^{(X)} &= \frac{1}{A}\!\left[ Z\bigl(2g_{uX}+g_{dX}\bigr)
      + (A-Z)\bigl(g_{uX}+2g_{dX}\bigr)\right],\quad X=A',Z,
\end{align}
\end{widetext}
with \(F_\text{Helm}(Q)\) the Helm form factor~\cite{Helm:1956zz,Duda:2006uk}.


\begin{table*}[t]
\centering
\setlength{\tabcolsep}{5.5pt}
\renewcommand{\arraystretch}{1.18}
\begin{tabular}{|l|l|l|c|c|c|l|c|}
\hline
\textbf{Model} & \textbf{Med} & \textbf{DM} & \textbf{Type} & \textbf{DD: SI} & \textbf{DD: SD} & \textbf{ID, $v\!\to\!0$} & \textbf{GCE fit} \\
\hline
\multicolumn{8}{|c|}{\emph{Higgs portals (simplified, SM $h$)}}\\
\hline
Higgs portal & $h$  & $S$ & hadronic & YES & NO & $s$-wave, helicity & \textbf{Yes}\\
Higgs portal & $h$  & $\psi$ (scalar) & hadronic & YES & NO & $p$-wave & \textbf{No} \\
Higgs portal & $h$  & $V_\mu$ & hadronic & YES & NO & $s$-wave, helicity & \textbf{Yes}\\
\hline
\multicolumn{8}{|c|}{\emph{$Z$ portal}}\\
\hline
$Z$ portal & $Z$  & $\psi$ (vector) & mixed & YES & NO & $s$-wave & \textbf{No}\\
$Z$ portal & $Z$  & $\psi$ (axial) & mixed & NO &  YES & helicity or $p$-wave & \textbf{No}\\
\hline
\multicolumn{8}{|c|}{\emph{UV–complete Vector Higgs portal ($H$–$H_p$ mixing)}}\\
\hline
UV VHP & $H,H_p$& $X_\mu$ & mixed & YES & NO & $s$-wave \\
\hline
\multicolumn{8}{|c|}{\emph{Simplified scalar mediator (mass $m_S$ fixed in figures)}}\\
\hline
Scalar med. & $S$  & $\chi$ & hadronic & YES & NO & $s$-wave, helicity & \textbf{Yes}\\
Scalar med. & $S$ & $\psi$ (scalar) & hadronic & YES & NO & $p$-wave & \textbf{No}\\
Scalar med. & $S$  & $V_\mu$ & hadronic & YES & NO & $s$-wave, helicity & \textbf{Yes}\\
\hline
\multicolumn{8}{|c|}{\emph{Simplified pseudoscalar mediator}}\\
\hline
Pseudoscalar & $P$  & $\psi$ (pseudo) & hadronic & loop$^{\ast}$ & SD, $\propto q^4$$^{\ast}$ & $s$-wave & \textbf{Yes}\\
\hline
\multicolumn{8}{|c|}{\emph{Simplified vector mediator ($Z'$)}}\\
\hline
Vector med. & $Z'$  & $\psi$ (vector) & mixed & YES & NO & $s$-wave & \textbf{Marginal}\\
Vector med. & $Z'$  & $\psi$ (axial) & mixed & NO & YES & helicity or $p$-wave & \textbf{No}\\
\hline
\multicolumn{8}{|c|}{\emph{Leptophilic gauge bosons}}\\
\hline
$U(1)_{L_\mu-L_e}$ & $A'$  &  $\chi$ & leptonic & YES (via $\epsilon^2$) & SD (subleading) & $s$-wave;  & \textbf{Yes}\\
$U(1)_{L_e-L_\tau}$ & $A'$  &  $\chi$ & leptonic & YES (via $\epsilon^2$) & SD (subleading) & $s$-wave;  & \textbf{Yes}\\
$U(1)_{L_\mu-L_\tau}$ & $A'$  &  $\chi$ & leptonic & YES (via $\epsilon^2$) & SD (subleading) & $s$-wave;  & \textbf{Borderline}\\
$U(1)_{B-L}$ & $A'$ &  $\chi$ & mixed & YES & SD (subleading) & $s$-wave;  & \textbf{Yes}\\
\hline
\end{tabular}
\caption{Summary of all benchmark models considered. Column~1 lists the model name; 
Columns~2–3 give the mediator and DM species. ``Type'' classifies the mediator’s dominant SM couplings (hadronic, leptonic, or mixed). The DD columns indicate whether coherent spin-independent (SI) or spin-dependent (SD) scattering arises at leading order. 
The ID column describes $s$-channel annihilation in the $v\!\to\!0$ limit—relevant for present-day indirect searches—and specifies whether it is $s$- or $p$-wave and whether it is
helicity suppressed ($\propto m_f^2$). 
The last column states whether an overlap region exists that satisfies the relic-density target, DD/ID limits, and our GCE flux fit (typically near the $s$-channel resonance).
\newline
$^{\ast}$~For a pseudoscalar mediator, tree-level nuclear scattering maps onto
momentum-suppressed non-relativistic operators (e.g.\ $\propto q^4$); SI arises only at loop
level and is typically much weaker than tree-level scalar exchange.
}
\label{tab:all_models_summary}
\end{table*}

\section{Results}
\label{sec:results}

We present the viable parameter space for physically motivated BSM DM models, considering the DM relic density observation, DD/ID bounds and a fit to the GCE flux data. 
For each model we show (i) a global scan in the range \(m_{\rm DM}\in[1,10^3]~\mathrm{GeV}\) and (ii) a zoom around the $s$-channel resonance \(m_{\rm med}\simeq 2m_{\rm DM}\), where annihilation rates are Breit–Wigner enhanced and small couplings can satisfy the relic target while easing DD limits.
The results are expressed in terms of a single coupling \(\lambda_{\rm portal}\).
For Higgs portals this is \(\lambda_{h\chi}\), \(\lambda_{hV}\), or \(\lambda_{h\psi}\);
for the \(Z\) portal we set \(\lambda_{\rm portal}=g_V\) (pure vector) or \(g_A\) (pure axial);
for simplified models we take \(g_f\equiv g_{\rm DM}\equiv\lambda_{\rm portal}\);
for \(L_i-L_j\), \(B-L\), and the UV–complete Vector Higgs portal we choose \(g_X=\lambda_{\rm portal}\).

The constraints used in the analysis are:
\begin{itemize}
\item \textbf{Cosmology.} The thermal relic abundance matches \(\Omega_{\rm DM} h^2 \simeq 0.120\) within uncertainties~\cite{Planck:2018vyg}.
\item \textbf{Direct detection.} We compare SI and SD DM–nucleus cross sections with the current LZ limits~\cite{LZ:2024zvo} and with the projected DARWIN/XLZD limits~\cite{DARWIN}.
\item \textbf{Indirect detection.} We impose the combined dSph \(\gamma\)-ray limits on \(\langle\sigma v\rangle\) as compiled in~\cite{McDaniel:2023bju}, evaluating channel-by-channel constraints according to the predicted final-state composition at each parameter point.
\item \textbf{GCE fit.} We fit the GCE spectra from Ref.~\cite{Cholis:2021rpp}, profiling over the channel mixture fixed by each model and scanning the overall normalization via \(\langle\sigma v\rangle\).
We include the systematic uncertainties from interstellar emission modeling as reported in~\cite{Cholis:2021rpp}. We also take into account in the results the possible uncertainties for the geometrical factor in the GC region. In particular, considering the results in Refs.~\cite{DiMauro:2021qcf,Koechler:2025ryv} we assume an average geometrical factor $J$ of 100 and a variation between 40 and 300 considering different DM density profiles and local DM density values. Therefore, the best-fit value for $\langle \sigma v \rangle$ obtained through the fit to the GCE flux inherits an uncertainty of about a factor of 2.5 in the lower side and 3 in the upper side. Since, $\langle \sigma v \rangle \propto \lambda_{\rm{portal}}^2$ or $\lambda_{\rm{portal}}^4$ this translates into a factor less than two in the coupling.
The prompt component is used for all models; for \(L_i-L_j\) and \(B-L\) we also include ICS following~\cite{Koechler:2025ryv}. 
Fitting the Di~Mauro et al.~(2021) dataset~\cite{DiMauro:2021raz} yields consistent results.
\end{itemize}

We do not include a dedicated collider recast in our baseline plots, for the following reasons. 
(i) For models with \emph{dominant SI} scattering (e.g., vector mediators with vector quark couplings, scalar Higgs portals), present DD bounds already dominate over mono-$X$ collider searches for \(m_{\rm DM}\gtrsim\) few GeV (see, e.g.,~\cite{ATLAS:2024kpy}). 
(ii) For a \emph{pseudoscalar} mediator, tree-level scattering is SD and momentum suppressed (\(\propto q^4\)); loop–induced SI terms exist but, near the $s$-channel pole, collider null results translate only weakly into limits on \(\langle\sigma v\rangle\), especially under the “minimal width’’ and benchmark choices \(g_q=g_\chi\sim 1\) used in~\cite{ATLAS:2024kpy}. In the resonant regime, the ID limits we include are typically more constraining. 
(iii) For an \emph{axial–vector} mediator with Dirac DM, collider bounds can be stronger than current SD limits, but public results are given at discrete mediator masses and benchmark widths that preclude a faithful recast across our continuous scans. 
(iv) In \(U(1)_{L_i-L_j}\) models, collider constraints on the dark photon–SM coupling via kinetic mixing are generically weaker than the DD limits once the loop- and momentum-dependent kinetic mixing relevant for scattering is accounted for; in \(U(1)_{B-L}\) tree-level quark couplings make DD even more restrictive. 
(v) For Higgs portals, limits on \(\mathcal{B}(h\to \mathrm{inv})\) are typically weaker than DD in the mass range of interest and do not modify our conclusions~\cite{Beniwal:2015sdl,Arcadi:2019lka,DiMauro:2023tho}.

\noindent
In summary, our figures display the interplay of relic density, DD, ID, and the GCE fit, with dedicated insets around the resonant corridor where the phenomenology changes most sharply. Collider results are discussed qualitatively above and do not affect the main viable regions under our assumptions.

\subsection{Higgs Portal Models}
\label{sec:results_Higgs}

\begin{figure*}
  \centering
  \includegraphics[width=\columnwidth]{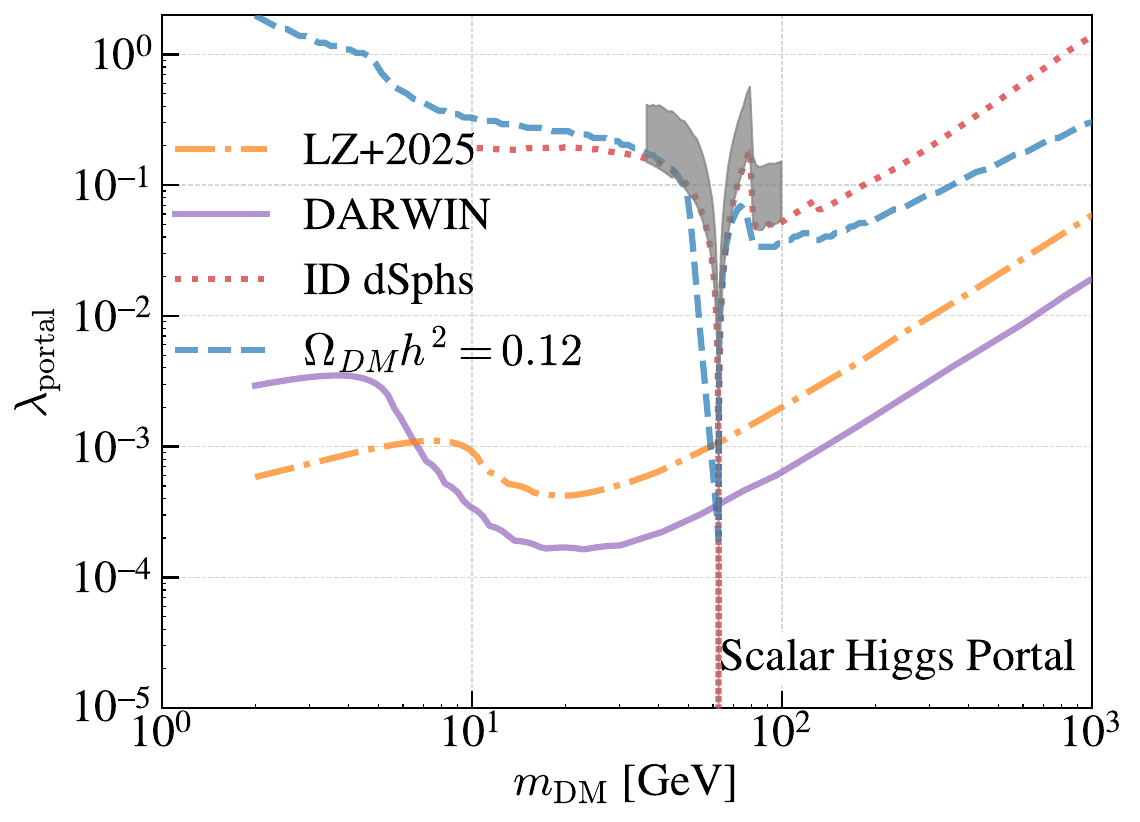}
  \includegraphics[width=\columnwidth]{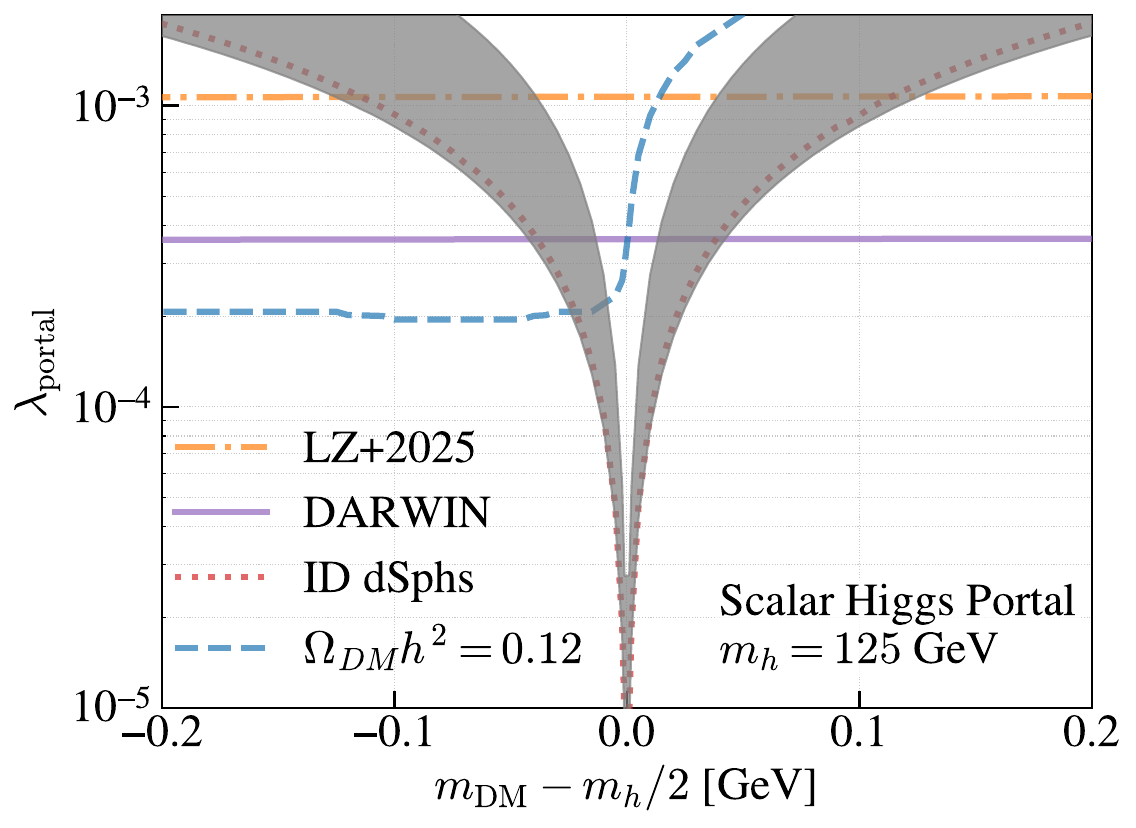}
  \includegraphics[width=\columnwidth]{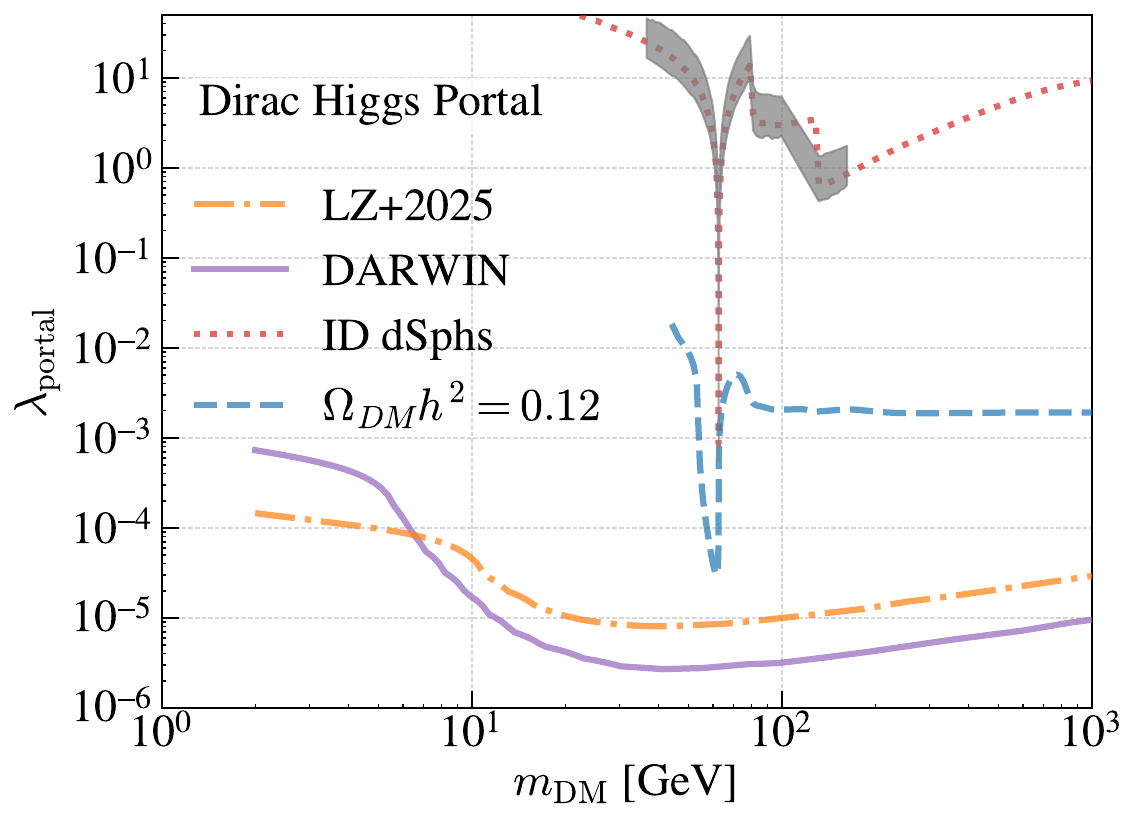}
  \includegraphics[width=\columnwidth]{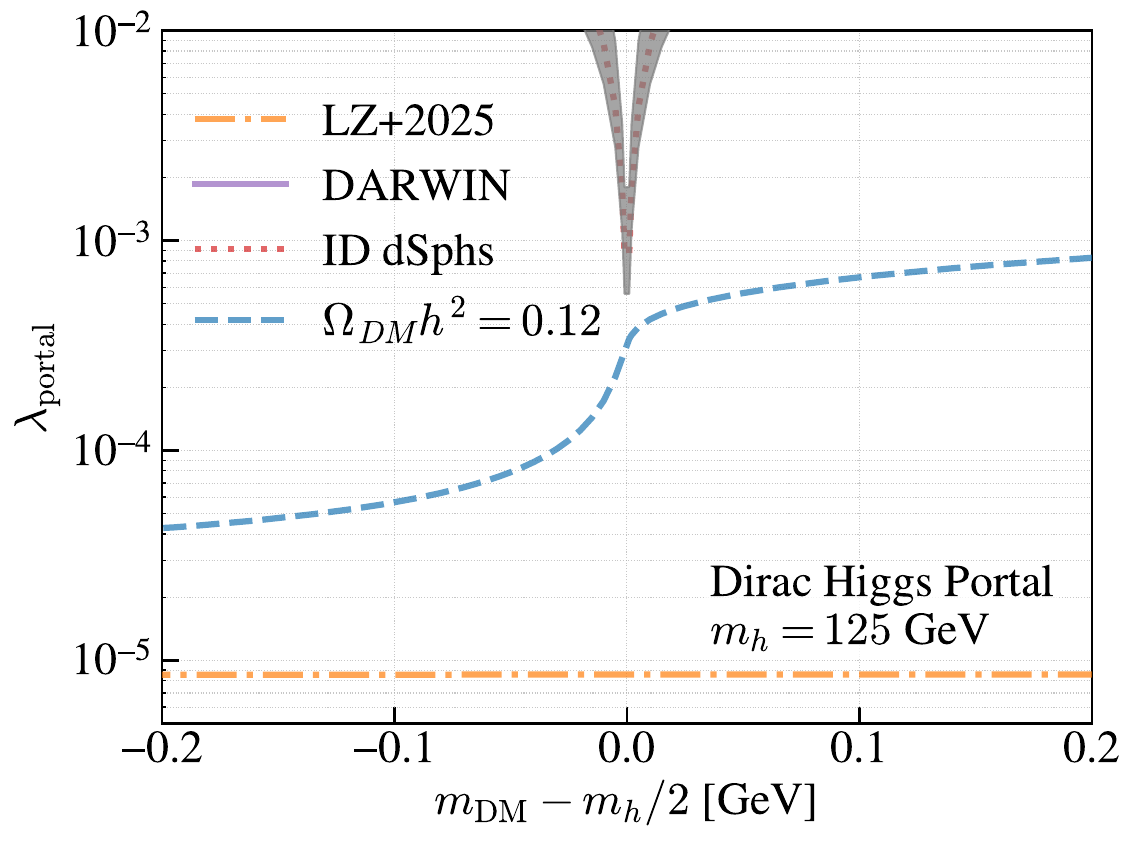}
  \includegraphics[width=\columnwidth]{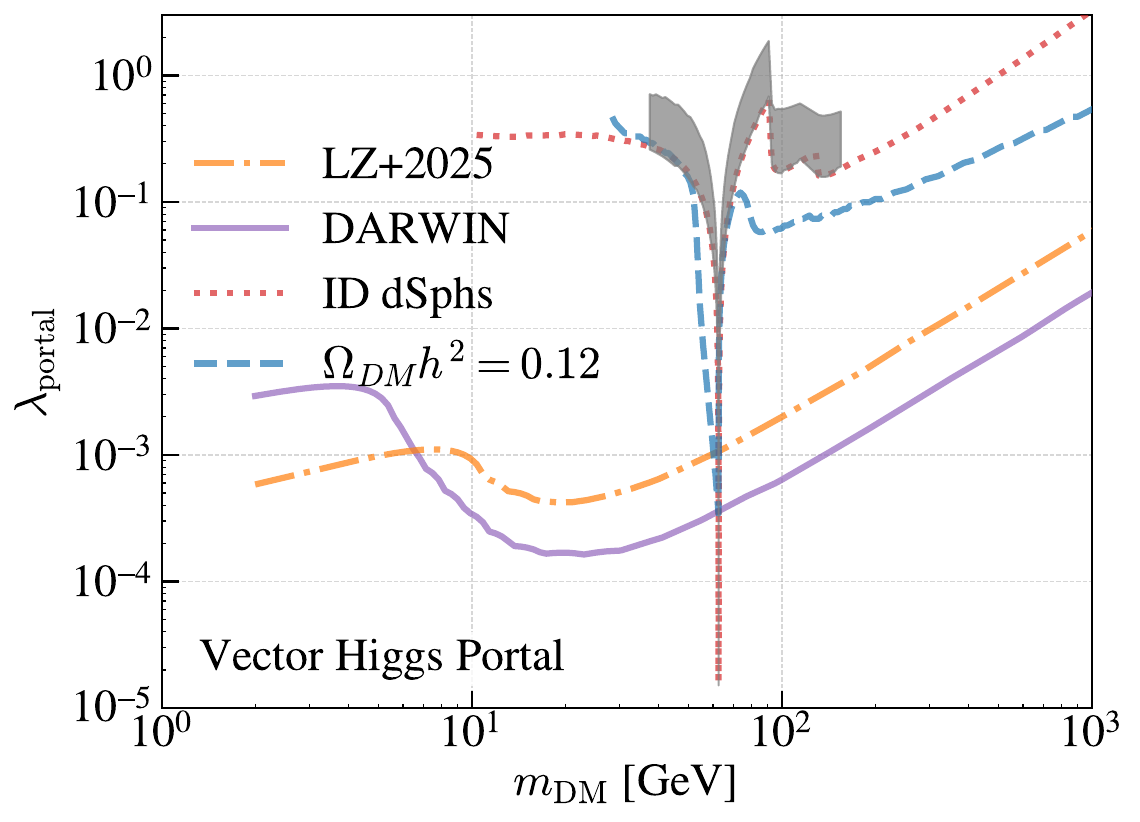}
  \includegraphics[width=\columnwidth]{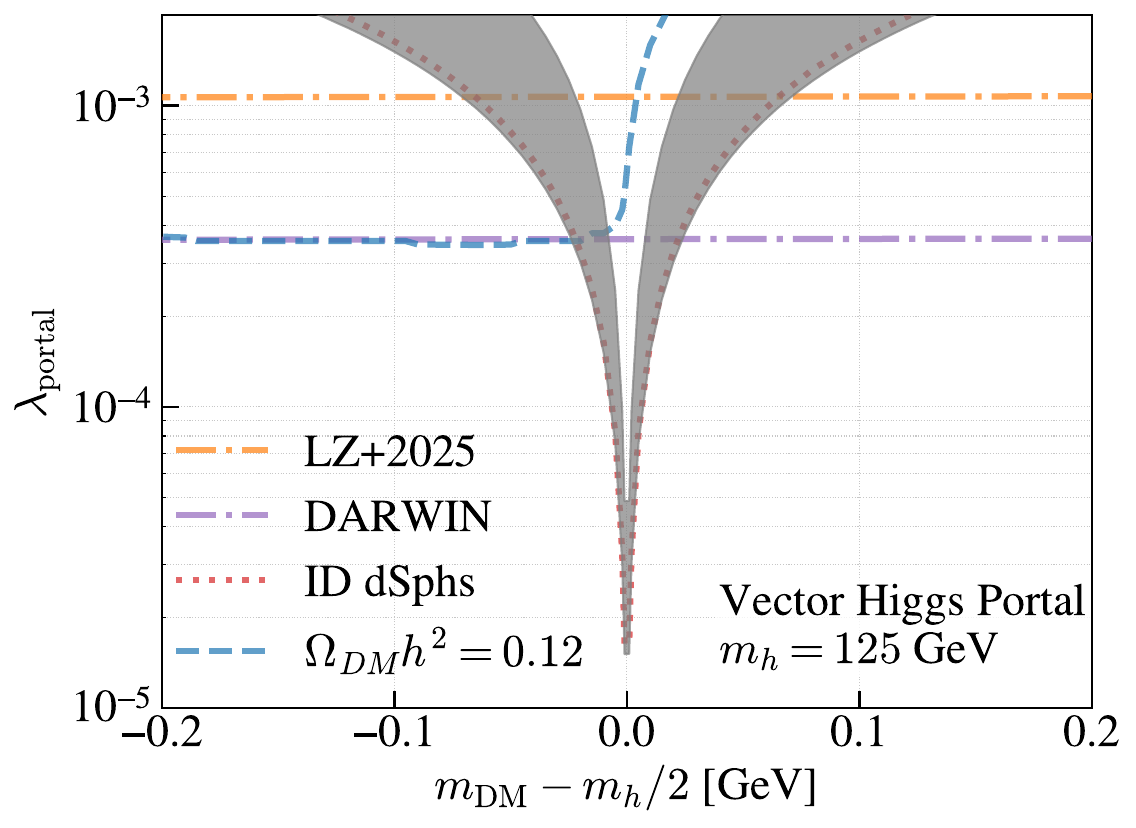}
  \caption{Constraints in the $(m_{\rm DM},\lambda_{\rm portal})$ plane for three Higgs portal models.
  The blue dashed line indicates the parameters for which $\Omega_{\rm DM} h^2=0.12$.
  The red dotted line shows SI ID upper limits. The orange dot-dashed line labeled and the purple solid line labeled show SI DD upper limits for LZ+2025 and DARWIN/XLZD.
  The gray band marks the region favored by the GCE flux fit with $\Delta\chi^2\le 9.21$ (2 d.o.f.) relative to the best fit, corresponding to $\approx99\%$ C.L.
  }
  \label{fig:higgs_portals}
\end{figure*}

Fig.~\ref{fig:higgs_portals} summarizes the results for the \emph{scalar}, \emph{Dirac}, and \emph{vector} Higgs portals (top to bottom). 
For each case we show a wide scan in \(m_{\rm DM}\in[2,10^{3}]~\mathrm{GeV}\) (left) and a zoom around the $s$–channel resonance (right), where the horizontal axis is shifted so that the origin marks the Higgs pole, \(m_{\rm DM}=m_h/2\simeq62.5~\mathrm{GeV}\).
The blue dashed line indicates the coupling that reproduces \(\Omega_{\rm DM}h^2\simeq0.12\); the orange dot–dashed and red dotted lines show, respectively, the LZ SI limit and the joint dSph \(\gamma\)–ray limit mapped onto \(\lambda_{\rm portal}\); the grey band denotes the region compatible with our GCE fit.

\paragraph{Scan procedure.}
At fixed \(m_{\rm DM}\) we sample \(\lambda_{\rm portal}\in[10^{-6},1]\), compute \(\Omega_{\rm DM}h^2(\lambda_{\rm portal})\), and interpolate to the value that matches 0.12. 
If \(\Omega_{\rm DM}h^2\) does not cross 0.12 within the scanned coupling range, no solution is recorded at that mass.\footnote{This happens only far from the resonance when the entire scanned interval either overcloses or undercloses the Universe. It occurs, in particular, for Dirac and vector Higgs portals below a few tens of GeV.}

\paragraph{Relic density.}
The relic–matching curve shows the expected narrow dip at \(m_{\rm DM}\simeq m_h/2\) in all three portals due to the Breit–Wigner enhancement. 
Away from resonance, viable relic solutions typically require \(\lambda_{\rm portal}\sim10^{-1}\text{--}10^{0}\), as expected in the WIMP regime.\footnote{In the WIMP case the couplings that reproduce the thermal cross section are roughly of order the electroweak couplings \(g,g'\sim\mathcal{O}(0.1\text{--}1)\).} 
For the Dirac case, the derived \(\lambda_{\rm portal}\) is affected by the choice of the new–physics scale \(\Lambda_\psi\) (here fixed to 1 TeV). 
Within the resonance funnel the relic abundance is achieved with couplings smaller by several orders of magnitude.

\paragraph{Direct detection.}
Higgs exchange induces coherent SI scattering in all three portals. 
For the \emph{Dirac} portal the current LZ limit lies \emph{below} the relic–density line essentially across the full mass range, leaving no region that simultaneously satisfies \(\Omega_{\rm DM}h^2\) and SI bounds.
By contrast, for the \emph{scalar} and \emph{vector} portals a thin resonant strip survives: in the neighborhood of \(m_h/2\) the relic curve dips beneath the LZ constraint, whereas most off–resonance solutions are excluded.

\paragraph{Indirect detection.}
In the Dirac case, annihilation through a scalar Higgs portal is $p$–wave, so present–day rates in dSphs are velocity suppressed and ID limits are weak compared to SI bounds. 
For scalar and vector portals the dominant channels contain $s$–wave components (helicity suppressed for light fermions but unsuppressed for bosonic final states above threshold), so dSph constraints remove portions of the parameter space, mainly off resonance and at the edges of the funnel. 
Overall, SI direct detection remains the leading constraint.

\paragraph{Net outcome and detuning.}
Combining relic density and SI/ID bounds, we find:
(i) the \emph{Dirac Higgs portal} is excluded across our scanned masses by SI limits, despite weak ID;
(ii) the \emph{scalar} and \emph{vector} portals retain a robust resonant corridor around \(m_{\rm DM}\simeq m_h/2\) that also accommodates the GCE flux fit. 
The resonance requires a modest mass detuning,
\begin{equation}
\Delta \;\equiv\; \frac{\lvert m_{\rm DM}-m_h/2\rvert}{m_h/2}\,,
\end{equation}
at the few-percent level once relic, DD, and ID are imposed. 
From our scans we extract \(\Delta \simeq 6\%\) (scalar) and \(\Delta \simeq 4\%\) (vector).\footnote{These estimates do not include the GCE fit in the definition of \(\Delta\); adding it selects the same corridor but further narrows the preferred band.}

\begin{figure}
  \centering
  \includegraphics[width=\columnwidth]{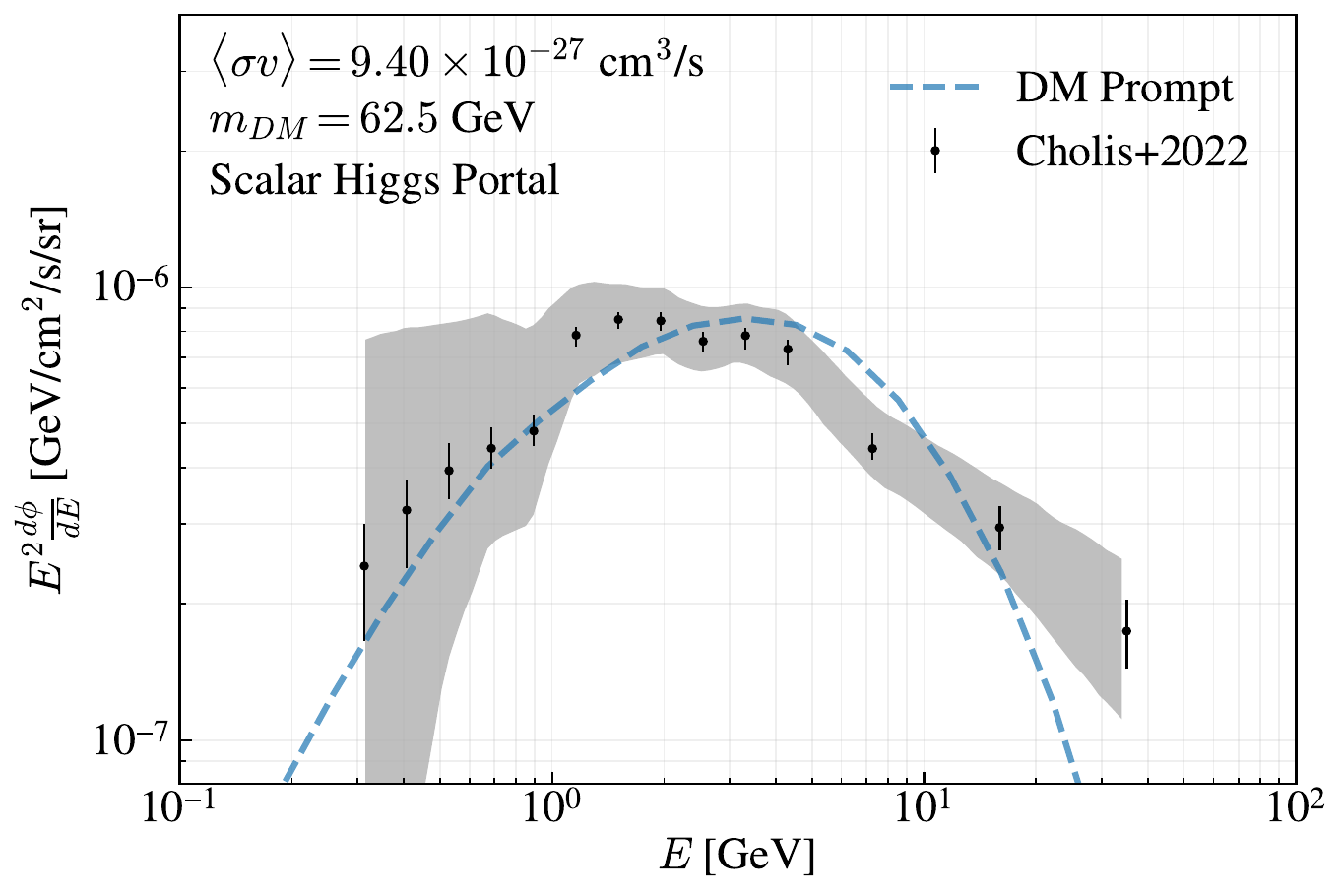}
  \caption{Best–fit of the scalar Higgs portal to the Cholis+22 GCE datasets described in Section \ref{sec:results_Higgs}. We show the GCE data and the theoretical predictions for the DM prompt flux (blue dashed curve).}
  \label{fig:SHP_flux}
\end{figure}

\paragraph{Benchmark points near the funnel that fit the GCE.}
For the \emph{scalar} portal, a representative point satisfying relic density, DD, ID, and the GCE fit lies at 
\(m_{\rm DM}-m_h/2 \simeq -0.025~\mathrm{GeV}\) with \(\lambda_{\rm portal} \simeq 2\times10^{-4}\). 
For the \emph{vector} portal, a compatible point is found instead with \(\lambda_{\rm portal} \simeq 3.5\times10^{-4}\). 
The exact values depend mildly on the adopted $J$–factor and on the treatment of sub-threshold channels via the off-shell Higgs width.
Fig.~\ref{fig:SHP_flux} compares the best-fit scalar-portal prompt spectrum with the GCE data, showing good agreement within the reported systematics and a best-fit \(\langle\sigma v\rangle\) of order \(10^{-26}\,\mathrm{cm^3\,s^{-1}}\).

\subsection{UV–Complete Vector Higgs Portal: Results}
\label{sec:results_UV_VHP}

\begin{figure*}[htbp]
  \centering
  \includegraphics[width=\columnwidth]{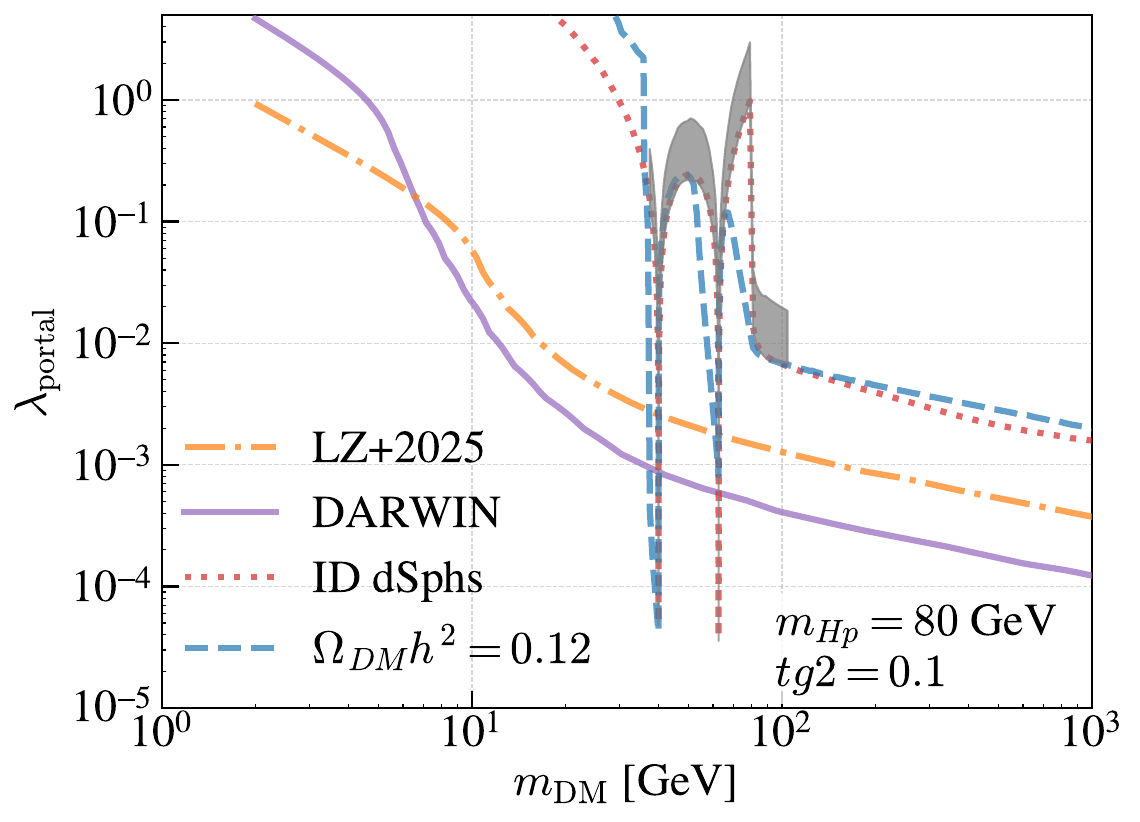}
  \includegraphics[width=\columnwidth]{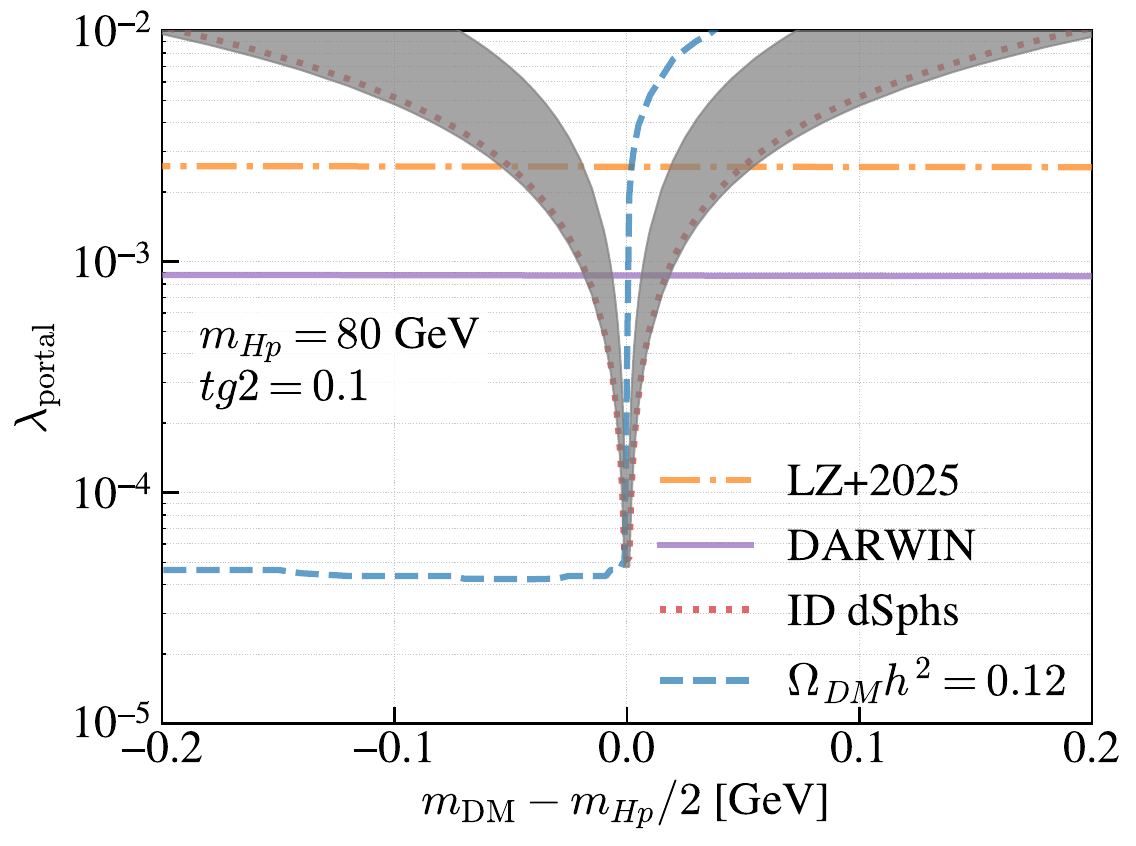}
  \includegraphics[width=\columnwidth]{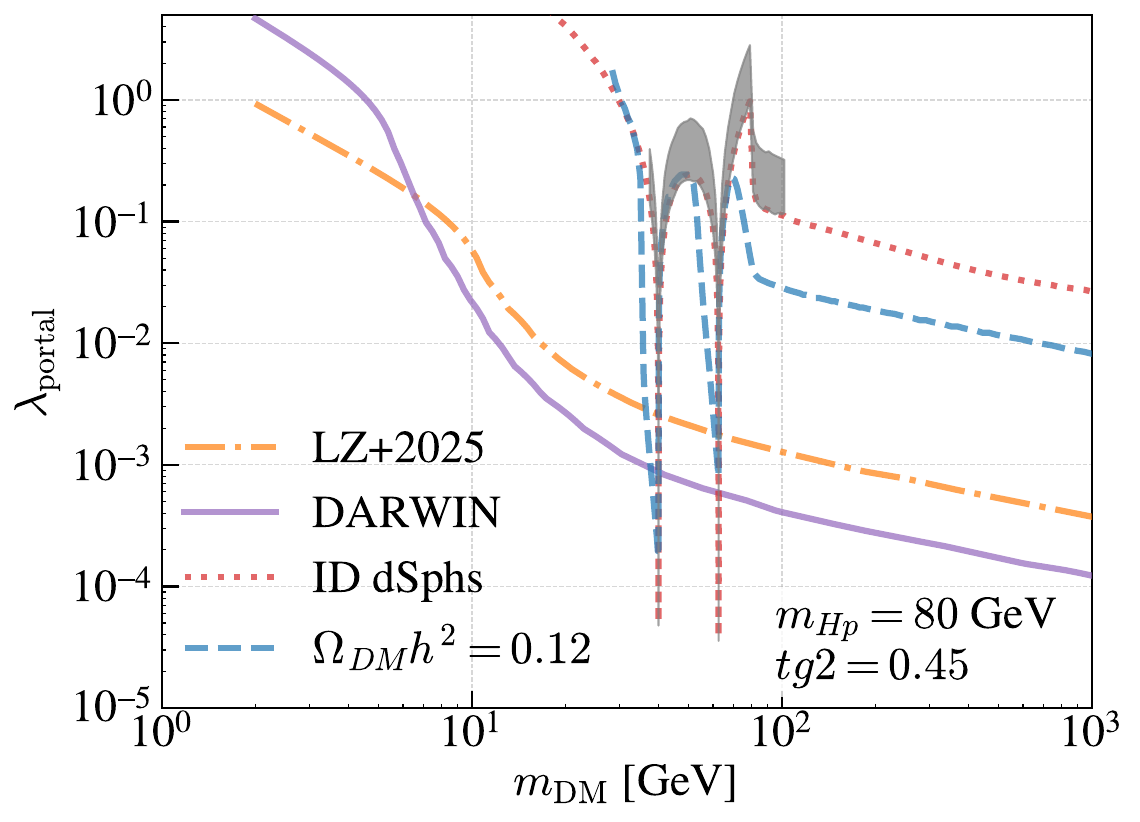}
  \includegraphics[width=\columnwidth]{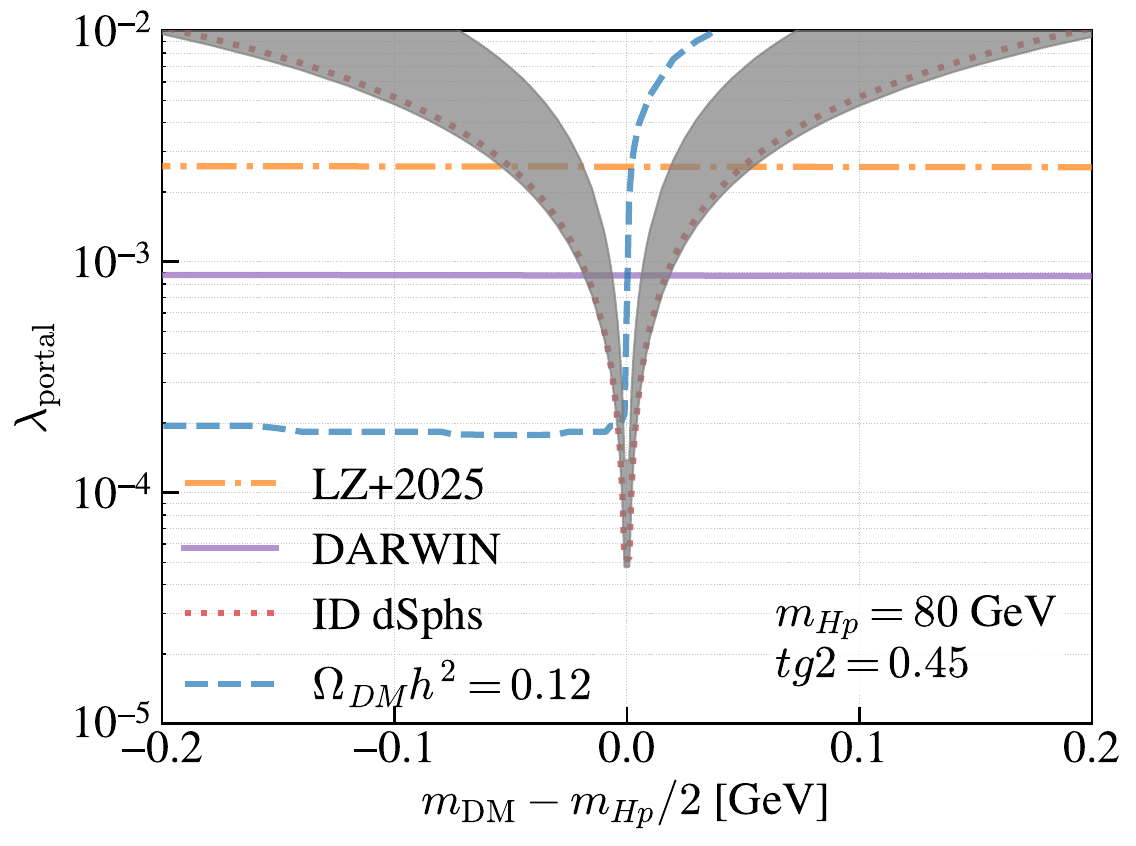}
  \caption{Constraints in the $(m_{\rm DM},\,\lambda_{\rm portal})$ plane for the UV-complete Higgs-portal model under two parameter settings. In both cases we fix $m_{Hp}=80~\mathrm{GeV}$ and take the mixing to be $\tan{2\alpha}\equiv tg2=0.1$ or $0.45$. The curves and shaded bands follow the same definitions as in Fig.~\ref{fig:higgs_portals}.}
  \label{fig:UV_VHP_MHp_80}
\end{figure*}

\begin{figure*}[htbp]
  \centering
  \includegraphics[width=\columnwidth]{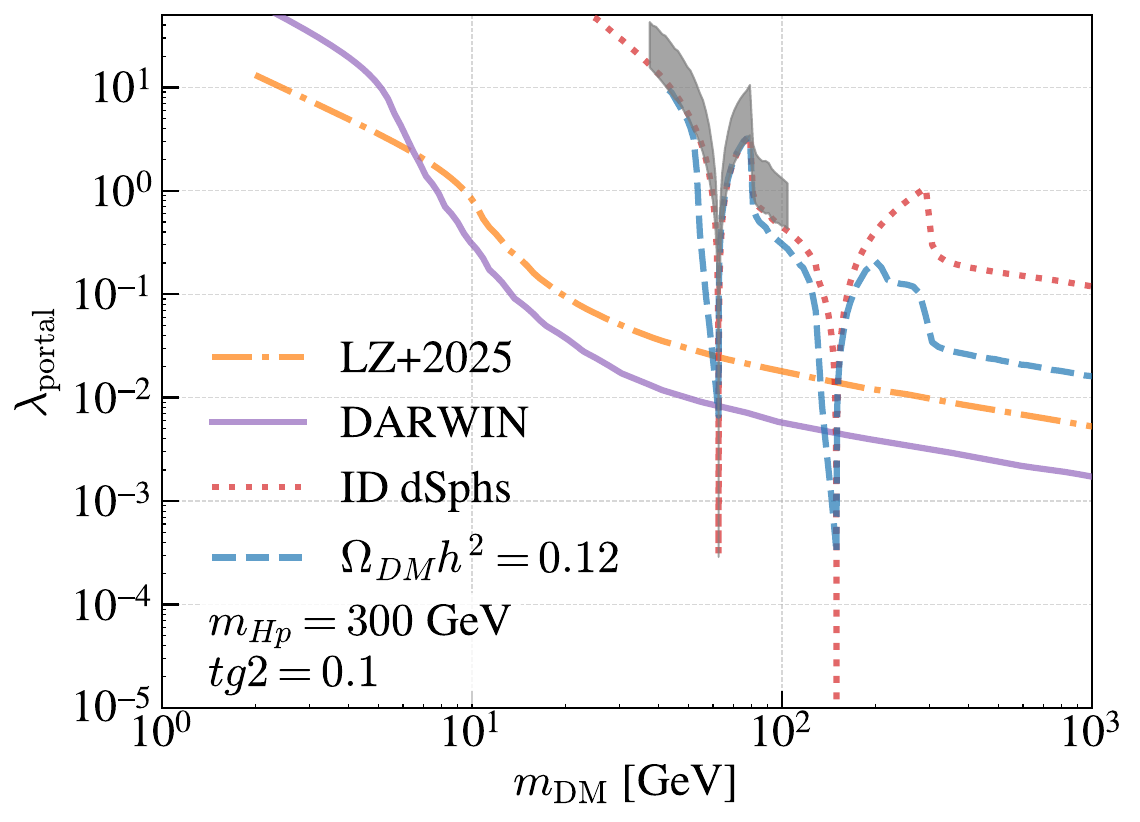}
  \includegraphics[width=\columnwidth]{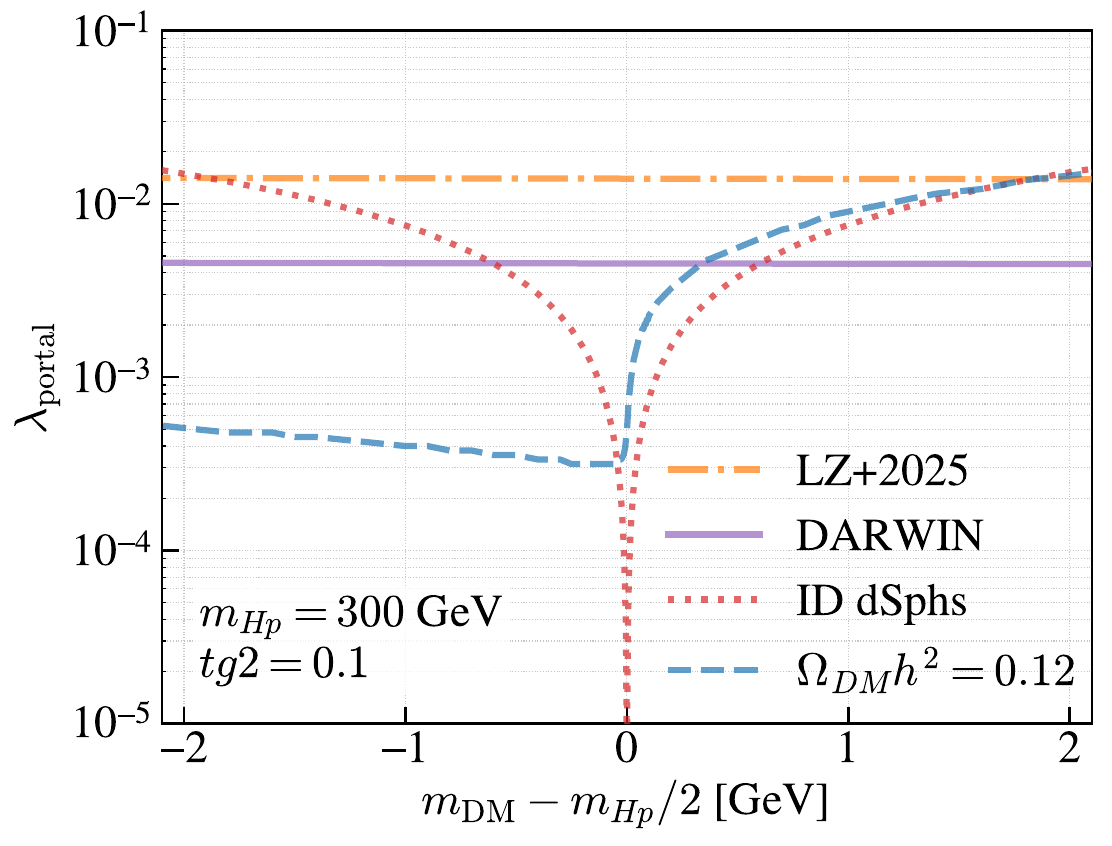}
  \includegraphics[width=\columnwidth]{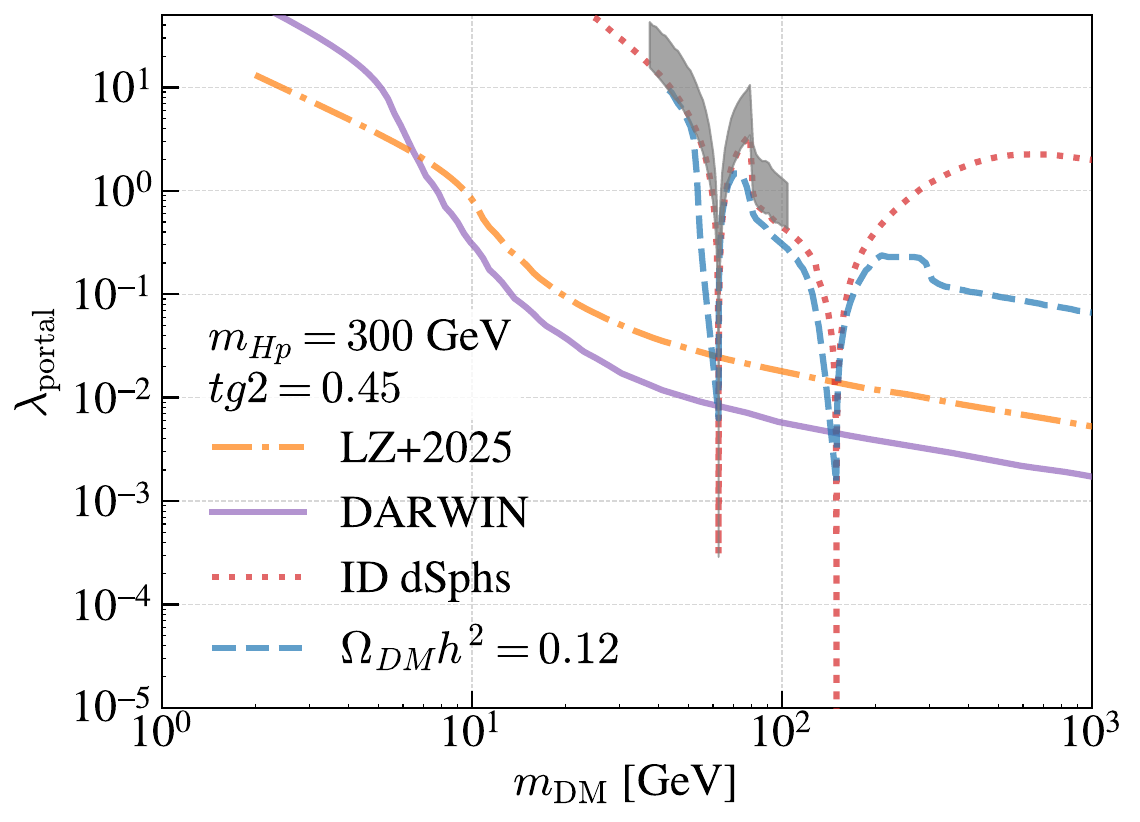}
  \includegraphics[width=\columnwidth]{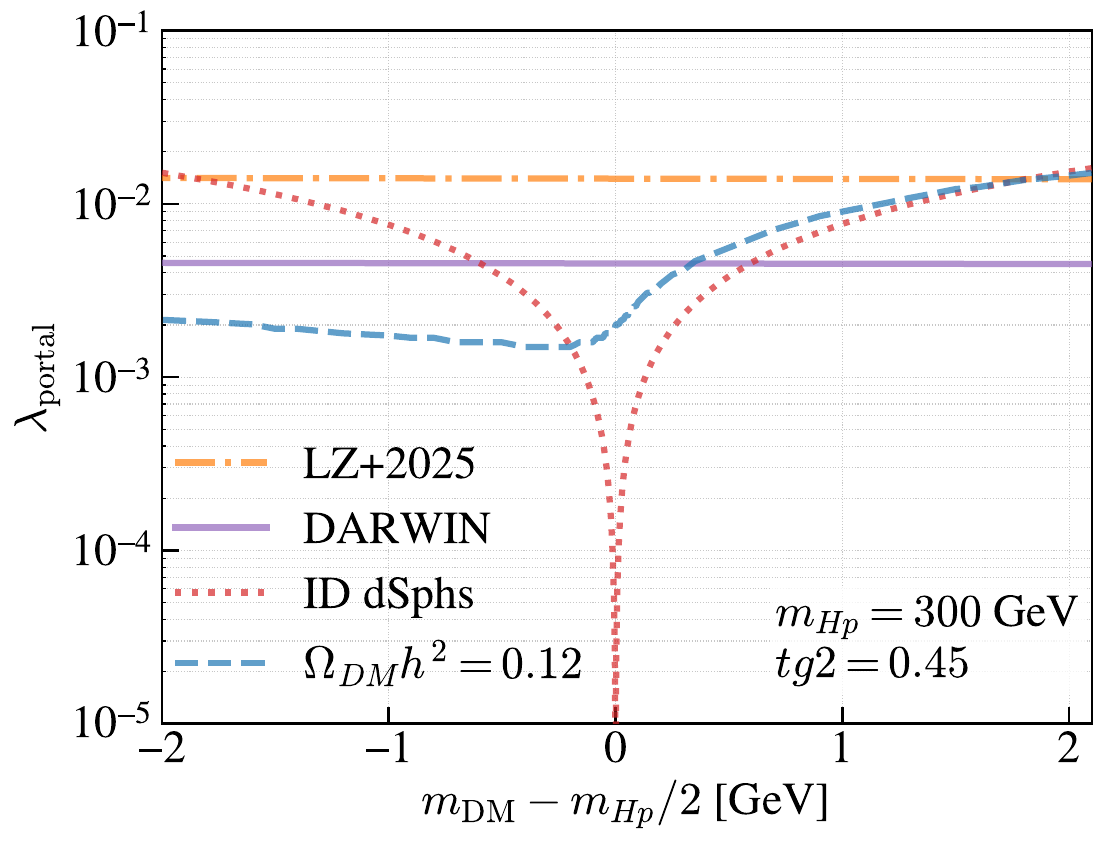}
\caption{Constraints in the $(m_{\rm DM},\,\lambda_{\rm portal})$ plane for the UV-complete Higgs-portal model under two parameter settings. Same as Fig.~\ref{fig:UV_VHP_MHp_80}, except that $m_{Hp}$ is set to $300~\mathrm{GeV}$.}
  \label{fig:UV_VHP_MHp_300}
\end{figure*}

Fig.~\ref{fig:UV_VHP_MHp_80} (for $m_{H_p}=80~\mathrm{GeV}$) and Fig.~\ref{fig:UV_VHP_MHp_300} (for $m_{H_p}=300~\mathrm{GeV}$) show our scans of the UV–complete vector Higgs–portal model for two benchmark mixings, $\tan(2\alpha)\!\equiv\!\mathrm{tg2}=0.1$ and $0.45$.  
Left panels display the full range $m_{\rm DM}\!\in\![1,10^3]~\mathrm{GeV}$; right panels zoom on the new scalar $H_p$ resonance with the horizontal axis shifted so that the origin is $m_{\rm DM}=m_{H_p}/2$.  
In each panel we overlay the coupling that reproduces the relic abundance (blue dashed, $\Omega_{\rm DM}h^2\simeq0.12$), the LZ SI limit (orange dot–dash), the DARWIN/XLZD SI bounds (purple solid), the dSph ID limit (red dotted), and the region preferred by our GCE flux fit (grey band).

\paragraph{Relic density.}
As expected for scalar mediation, the relic–density curve exhibits sharp dips at the $s$–channel poles. Besides the SM–Higgs funnel at $m_{\rm DM}\simeq m_h/2$, a second—typically deeper—dip appears at $m_{\rm DM}\simeq m_{H_p}/2$.  
Increasing the mixing from $\mathrm{tg2}=0.1$ to $0.45$ raises the relic–density line across masses, thereby shrinking the width of the $H_p$ funnel.  
Numerically, for $m_{H_p}=80~\mathrm{GeV}$ the relic–density requirement near the $H_p$ pole moves from $\lambda_{\rm portal}\sim 5\times10^{-5}$ to $\sim 2\times10^{-4}$ when $\mathrm{tg2}$ increases from $0.1$ to $0.45$.  
For $m_{H_p}=300~\mathrm{GeV}$, the same change shifts the resonant coupling from $\sim 5\times10^{-4}$ to $\sim 2\times10^{-3}$.

\paragraph{Direct detection.}
The LZ (orange) and DARWIN (purple) bounds are nearly insensitive to the change in $\mathrm{tg2}$ for the benchmarks considered and sits around $\lambda_{\rm portal}\sim 10^{-3}$ near the $m_{H_p}=80~\mathrm{GeV}$ resonance and $\sim 10^{-2}$ for $m_{H_p}=300~\mathrm{GeV}$.  
As a result, for $m_{H_p}=80~\mathrm{GeV}$ a narrow strip around the $H_p$ resonance remains allowed (especially at smaller mixing), whereas for $m_{H_p}=300~\mathrm{GeV}$ the permitted band is much broader.

\paragraph{Indirect detection.}
Away from the poles, the dSph limit (red) is weaker than LZ for the parameter ranges shown. Close to the $H_p$ funnel (right panels) ID trims the edges of the resonant band but does not dominate over SI bounds; the surviving region is therefore primarily set by the interplay of the relic–density requirement and LZ.

\paragraph{Net outcome and GCE fit.}
The UV vector–portal retains a viable resonant corridor around $m_{\rm DM}\simeq m_{H_p}/2$ in all cases, with the width controlled by $\mathrm{tg2}$.  
Taking into account relic density, DD, and ID, the model remains viable in the resonant region for a detuning $\Delta\!\sim\!6\%$ for $m_{H_p}=80~\mathrm{GeV}$, and over a broader range just below $m_{H_p}/2$ for $m_{H_p}=300~\mathrm{GeV}$.  
When the GCE flux fit is included (grey band), only the $m_{H_p}=80~\mathrm{GeV}$ benchmarks overlap the preferred band, whereas for $m_{H_p}=300~\mathrm{GeV}$ the resonant strip no longer intersects the GCE best-fit region.  
Consequently, within these benchmarks the $m_{H_p}=80~\mathrm{GeV}$ cases provide the strongest combined agreement among relic density, DD, ID, and the GCE, with $\lambda_{\rm portal} \simeq (4\text{–}15)\times 10^{-5}$ depending on $\mathrm{tg2}$.

\subsection{\texorpdfstring{$Z$}{Z} Portal Model}
\label{sec:results_Zportal}

\begin{figure*}[htbp]
  \centering
  \includegraphics[width=\columnwidth]{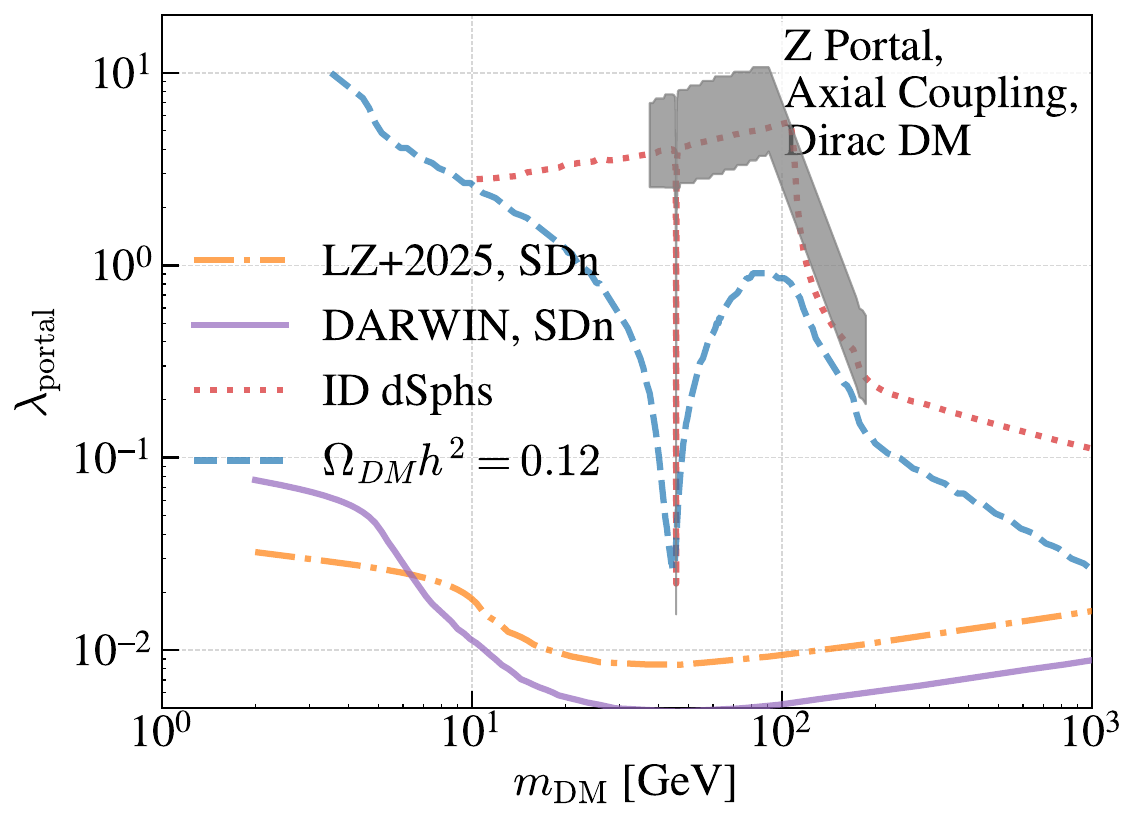}
  \includegraphics[width=\columnwidth]{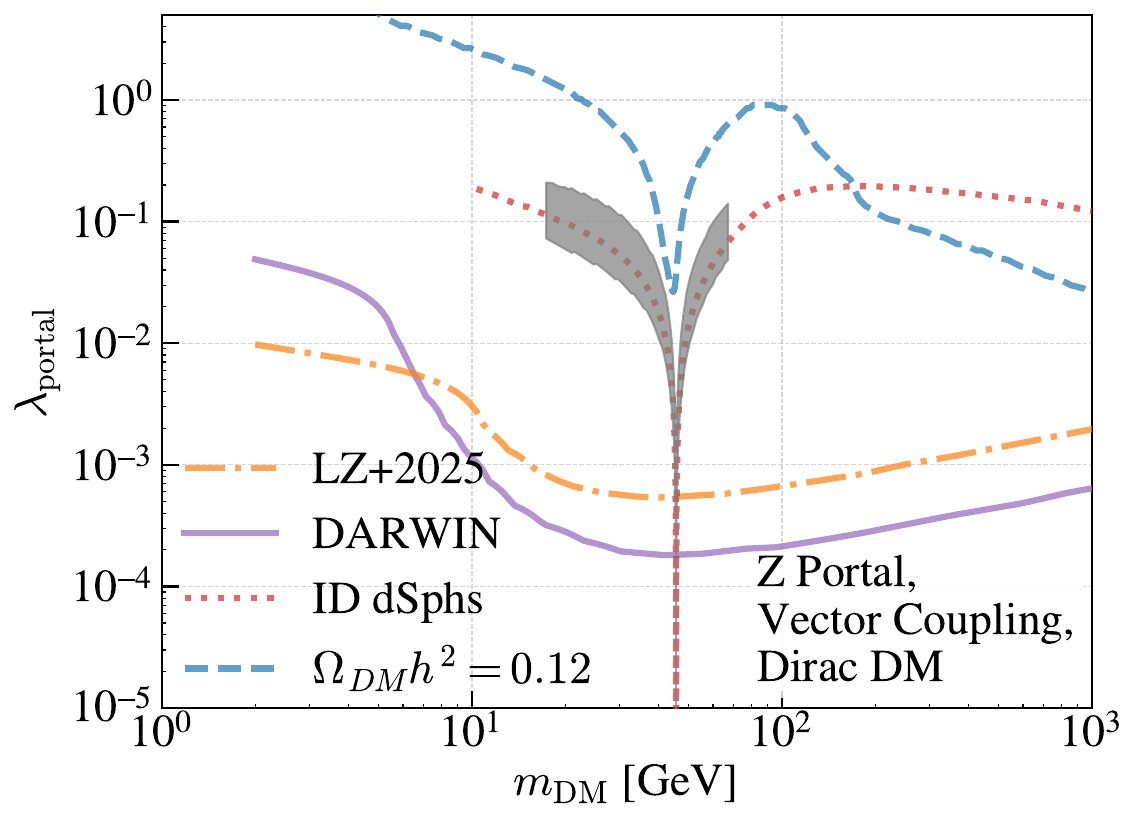}
  \caption{Constraints in the $(m_{\rm DM},\,\lambda_{\rm portal})$ plane for the $Z$-portal with Dirac dark matter under two coupling choices. The left panel shows the axial coupling, where we adopt SD neutron limits (orange dot–dashed and purple solid lines); the right panel shows the vector coupling, where SI limits are used. The curves and shaded bands follow the same definitions as in Fig.~\ref{fig:higgs_portals}.}

  \label{fig:Z_Portal}
\end{figure*}

Fig.~\ref{fig:Z_Portal} summarizes scans for a Dirac fermion coupled to the SM $Z$ boson with either \emph{axial} (left) or \emph{vector} (right) interactions. 
As in the Higgs–portal plots, the blue dashed curve shows the coupling that reproduces $\Omega_{\rm DM}h^2\simeq 0.12$, the orange dot–dashed curve the relevant DD bound, and the red dotted curve the combined dSph limit; the gray band indicates the region preferred by the GCE flux fit.

\paragraph{Axial coupling.}
For purely axial DM–$Z$ interactions, present–day annihilation is helicity/$p$–wave suppressed, so ID limits are comparatively weak. 
However, SD scattering on nuclei provides strong constraints: adopting the tighter SD–neutron bound, the LZ curve already lies below the relic–density requirement across the full mass range, including the $Z$–pole funnel at $m_{\rm DM}\simeq m_Z/2$. 
Therefore, no simultaneous solution to relic density and DD is found.

\paragraph{Vector coupling.}
With a vector interaction the annihilation is $s$–wave, but coherent SI scattering via $Z$ exchange is extremely constrained. 
The SI limit excludes the relic–density curve by a wide margin for all masses, again including the resonant region.

\paragraph{Outcome.}
For both coupling choices, the $Z$–portal with Dirac DM is excluded once relic density and DD limits are combined. 
The GCE–favored band does not reopen viable parameter space and is shown only for completeness.

\subsection{Simplified Models: Scalar (and Pseudoscalar) Mediators}
\label{sec:results_scalar_simplified}

\begin{figure*}[htbp]
  \centering
  \includegraphics[width=\columnwidth]{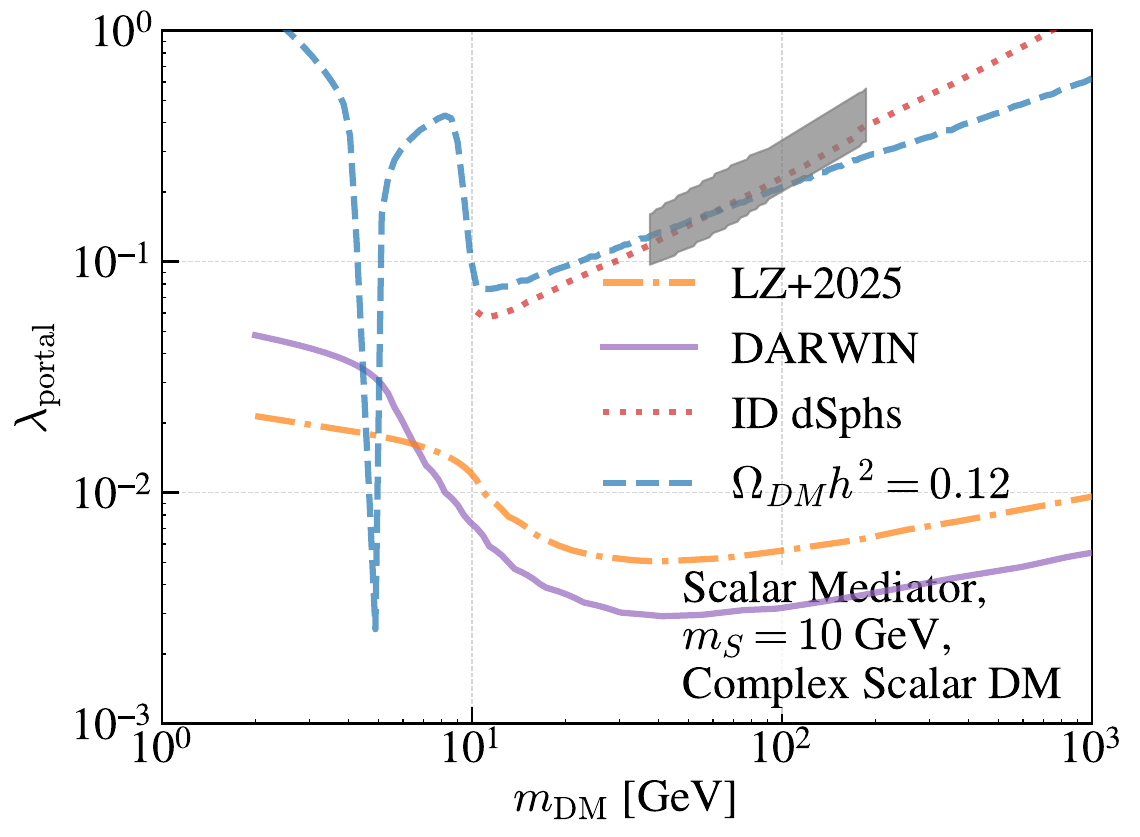}
  \includegraphics[width=\columnwidth]{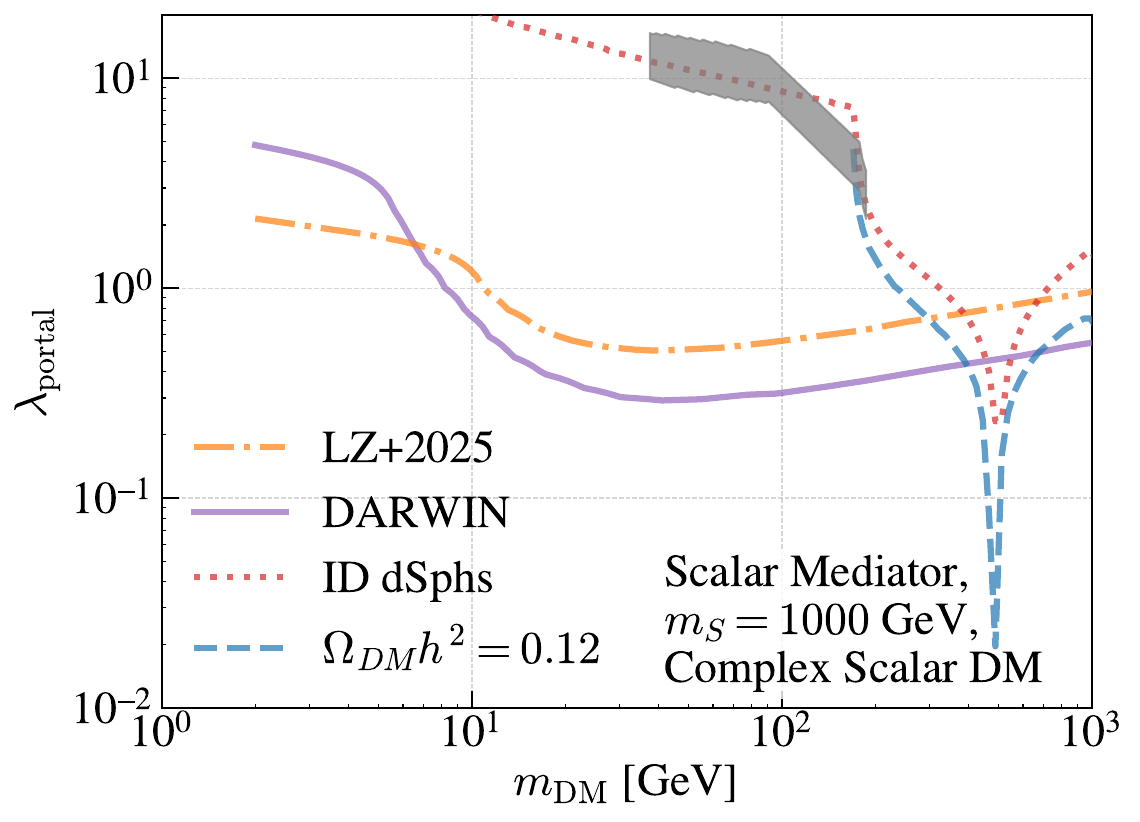}
  \caption{Constraints in the $(m_{\rm DM},\,\lambda_{\rm portal})$ plane for the simplified scalar-mediator model with complex-scalar dark matter under two mediator-mass benchmarks. We fix $m_{S}=10~\mathrm{GeV}$ (left) and $m_{S}=1000~\mathrm{GeV}$ (right). Here, $m_{y}$ denotes the scalar-mediator mass. The curves and shaded bands follow the same definitions as in Fig.~\ref{fig:higgs_portals}.}
  \label{fig:scalar_mediator_complex_mass_selection}
\end{figure*}

\paragraph{Mediator–mass benchmarks ($m_{S}=10$ and $1000$ GeV).}
We first consider mediator masses that are far from the DM mass values needed to fit the GCE.
Fig.~\ref{fig:scalar_mediator_complex_mass_selection} illustrates two benchmarks cases with $m_{S}=10$ and $1000$ GeV in the
\emph{complex–scalar DM + scalar mediator} setup.
For a light mediator ($m_{S}=10~\mathrm{GeV}$, left) the relic–density curve develops a narrow
\emph{s}–channel funnel at $m_{\rm DM}\simeq m_{S}/2$, well below the DM masses preferred by the GCE fit;
away from resonance, the couplings required by $\Omega_{\rm DM}h^2\simeq0.12$ lie above the SI limit, so
no parameter space with the observed relic density survives once DD is applied.  
For a heavy mediator ($m_{S}=1000~\mathrm{GeV}$, right), the resonance shifts to
$m_{\rm DM}\simeq 500~\mathrm{GeV}$, but the SI bound again lies below the relic curve near the dip in the GCE-favoured mass range, excluding any funnel solution compatible with the GCE,
excluding the funnel.  
In both cases the GCE–favoured band (grey) does not simultaneously satisfy relic density and DD, so in the rest we
focus on $m_{S}=120~\mathrm{GeV}$ that is twice the best-fit DM mass of the GCE and for which a viable resonant overlap exists.

\begin{figure*}[htbp]
  \centering
  \includegraphics[width=\columnwidth]{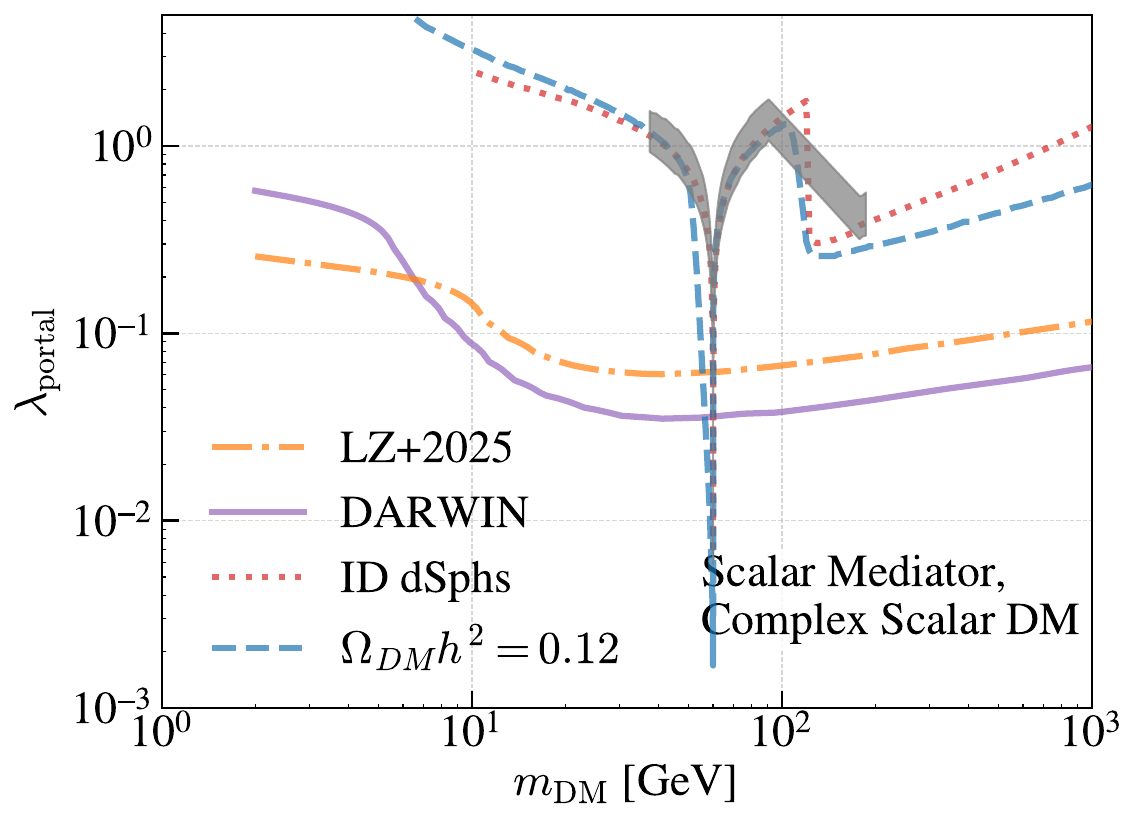}
  \includegraphics[width=\columnwidth]{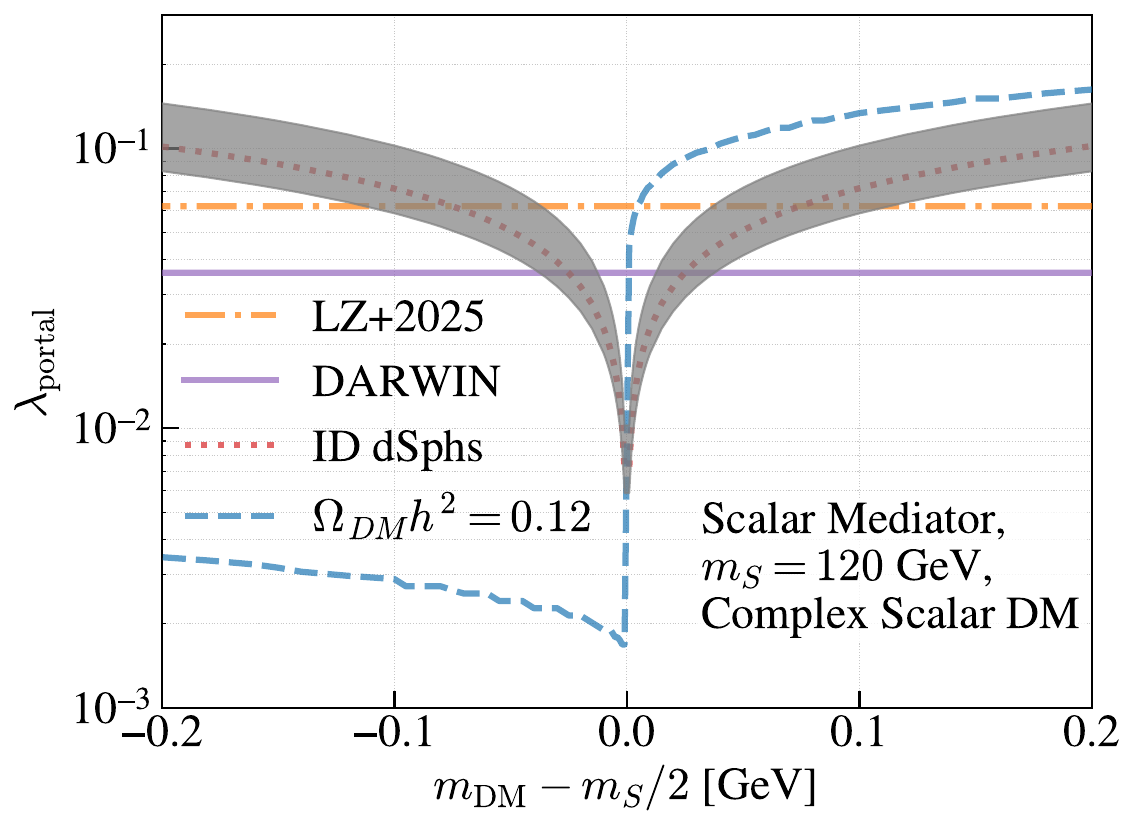}
  \includegraphics[width=\columnwidth]{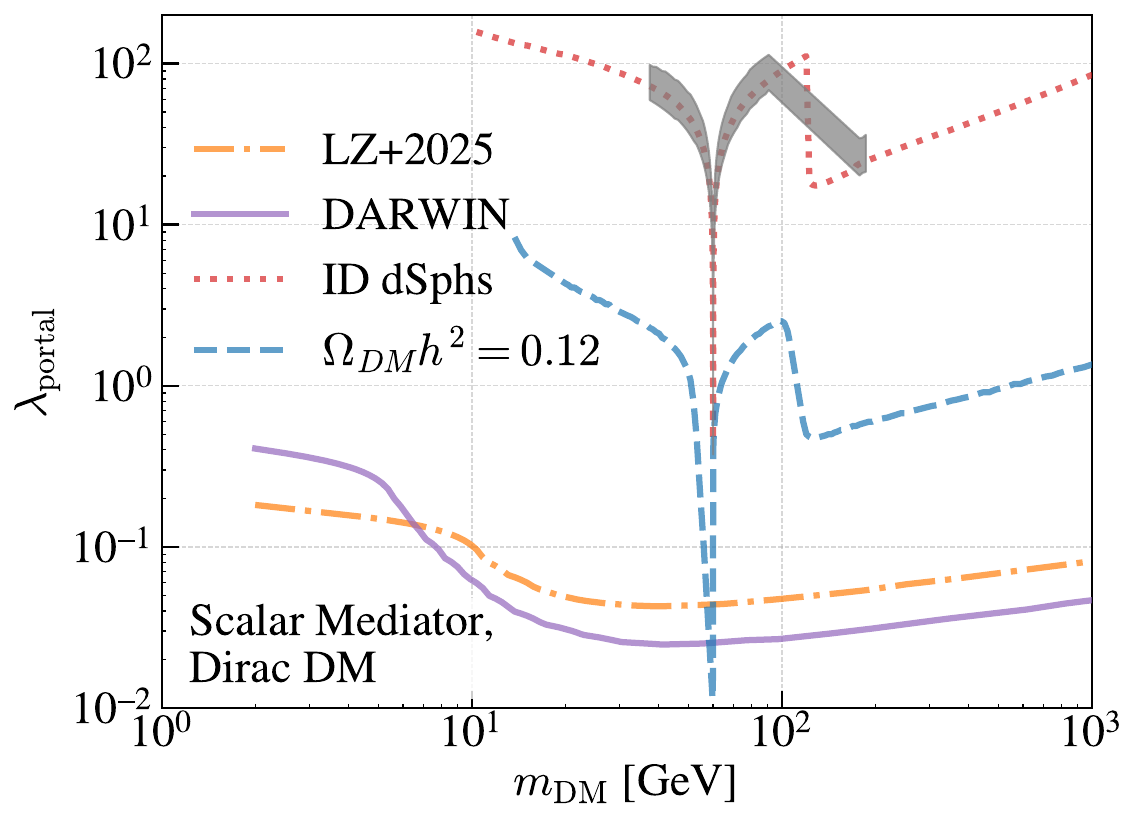}
  \includegraphics[width=\columnwidth]{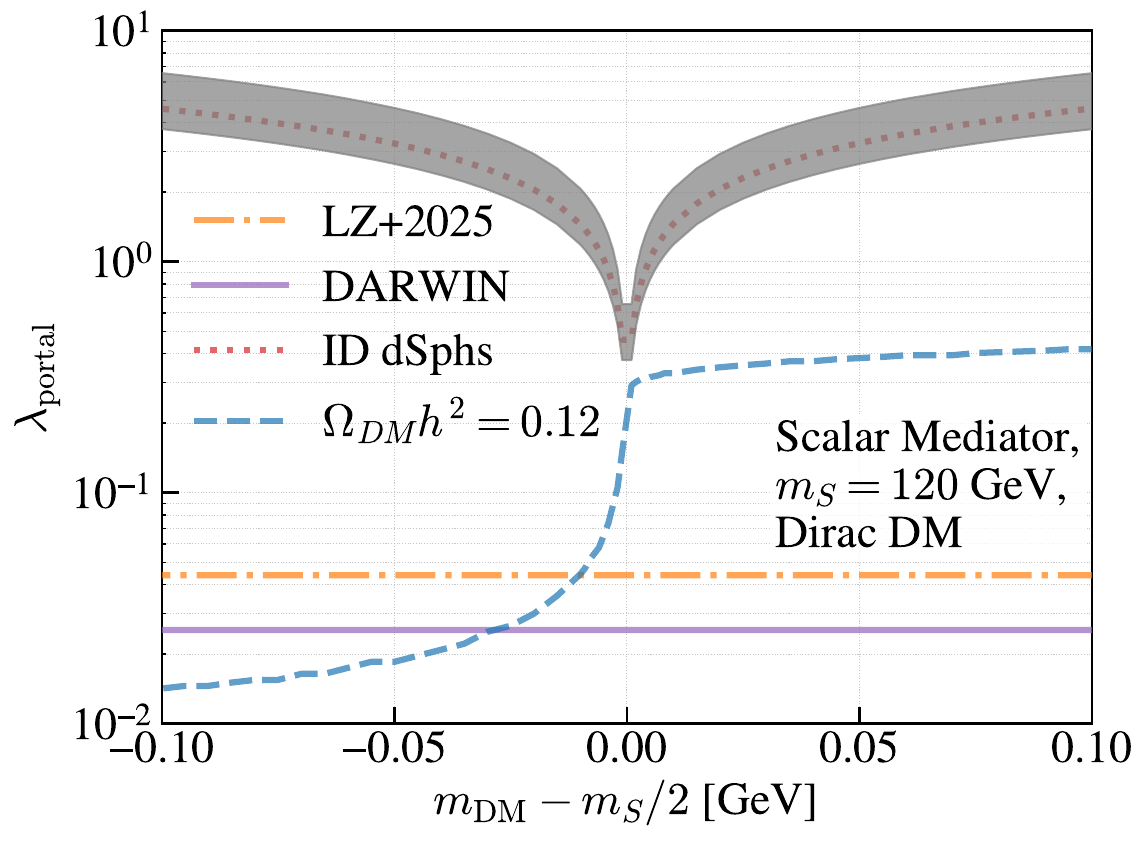}
  \includegraphics[width=\columnwidth]{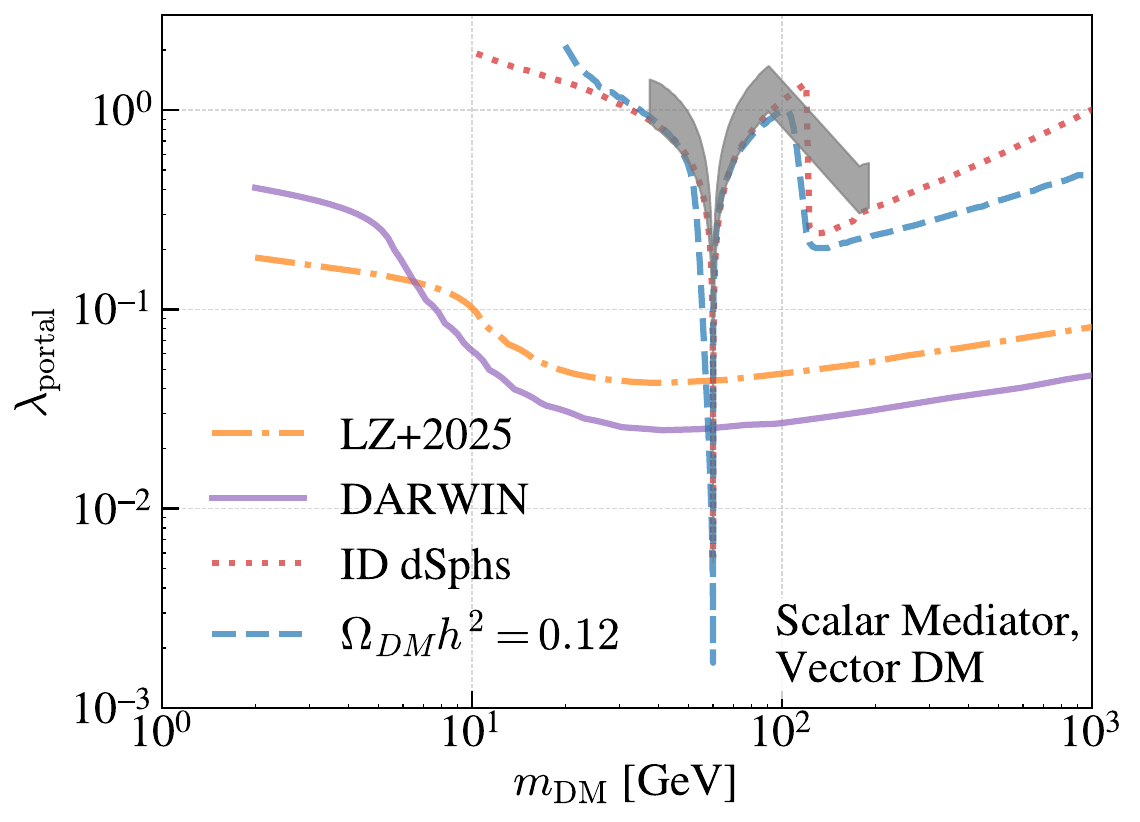}
  \includegraphics[width=\columnwidth]{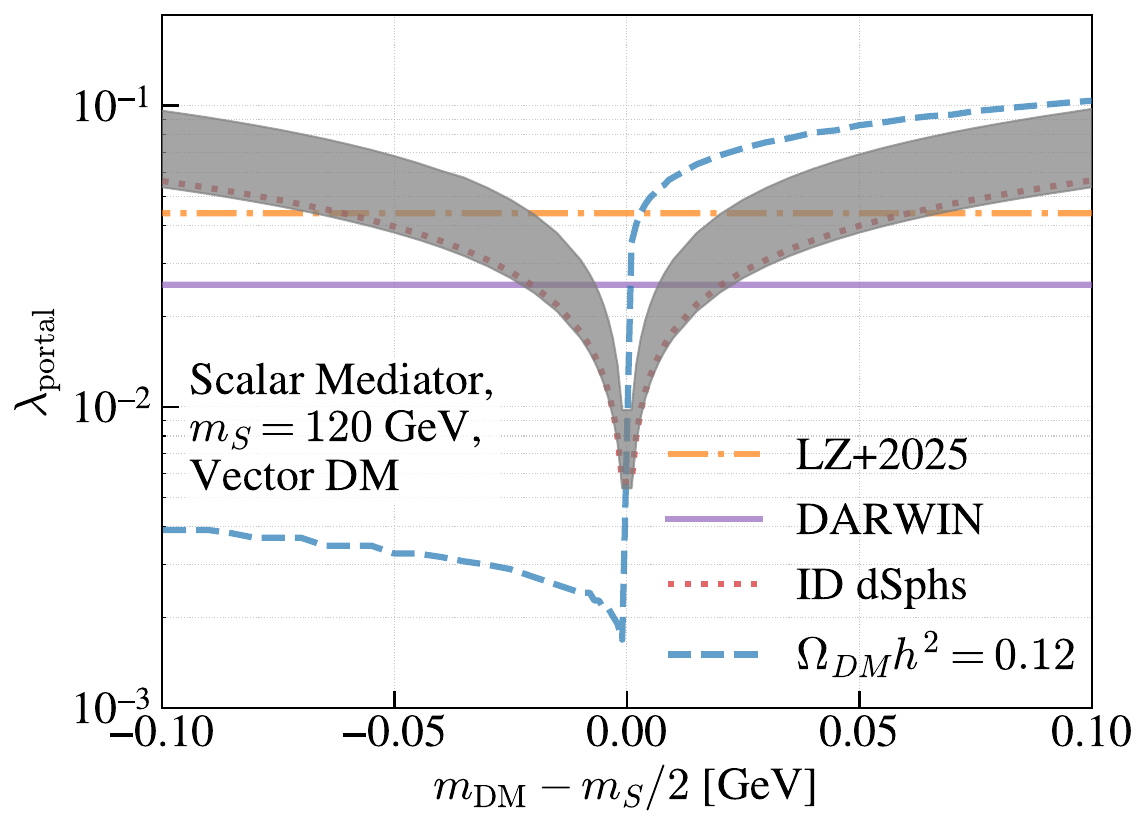}
  \caption{Constraints in the $(m_{\rm DM},\,\lambda_{\rm portal})$ plane for the simplified scalar-mediator model with complex-scalar, Dirac, and vector dark matter at fixed $m_S=120~\mathrm{GeV}$. Here, $m_{y}$ denotes the scalar-mediator mass. The curves and shaded bands follow the same definitions as in Fig.~\ref{fig:higgs_portals}.}
  \label{fig:Scalar_Mediator_120}
\end{figure*}

\begin{figure*}[htbp]
  \centering
  \includegraphics[width=\columnwidth]{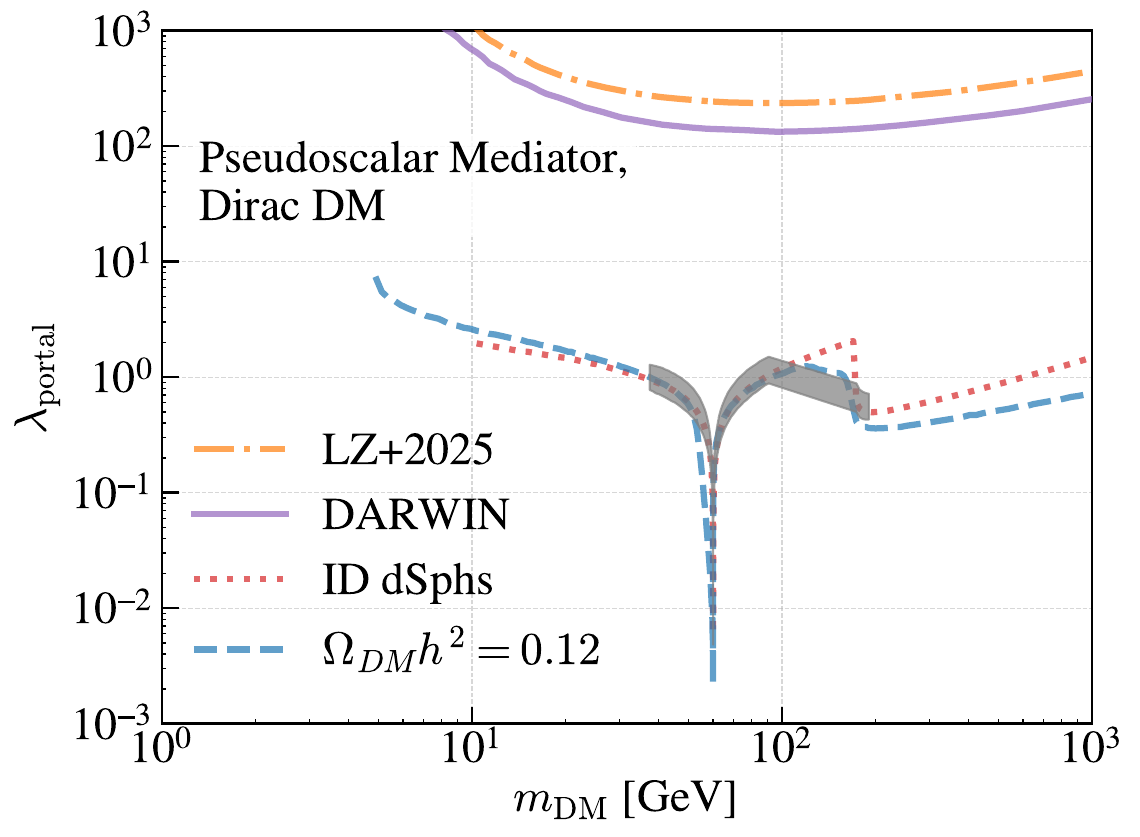}
  \includegraphics[width=\columnwidth]{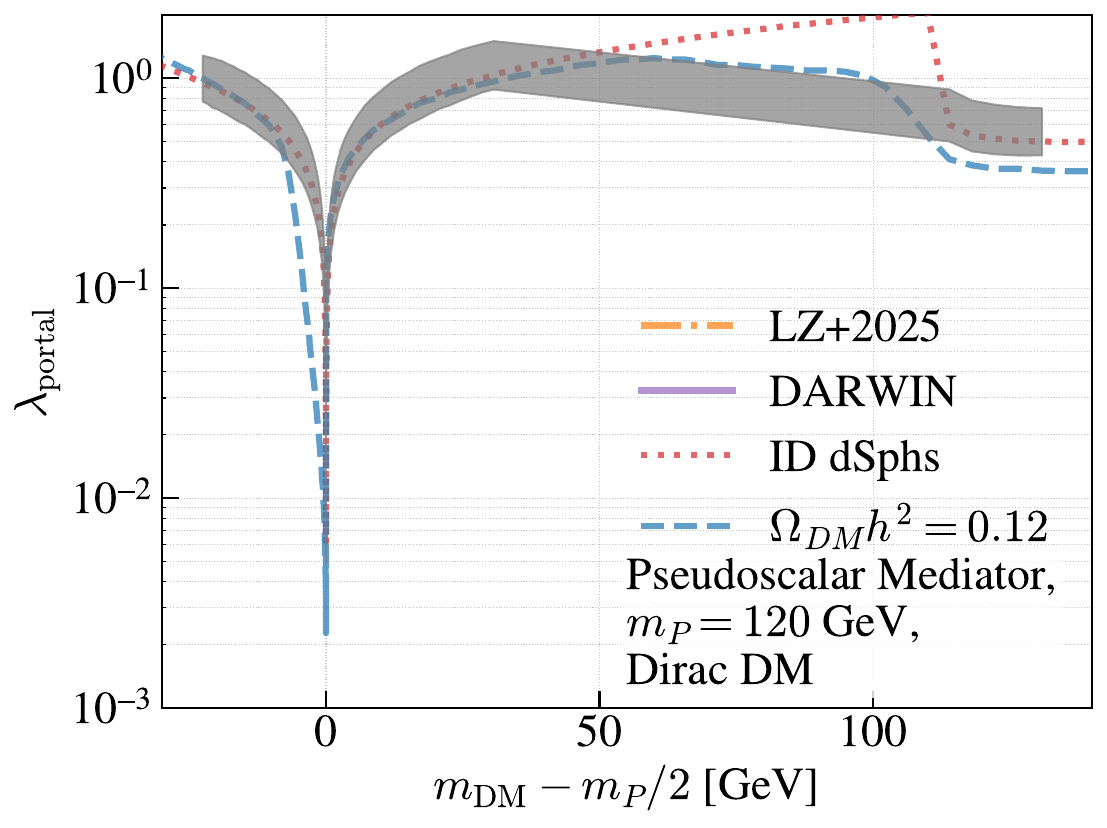}
 \caption{Constraints in the $(m_{\rm DM},\,\lambda_{\rm portal})$ plane for the simplified pseudoscalar-mediator model with Dirac dark matter at fixed $m_P=120~\mathrm{GeV}$. where $P$ denotes the pseudoscalar-mediator. The curves and shaded bands follow the same definitions as in Fig.~\ref{fig:higgs_portals}.}
 \label{fig:pseudoscalar_mediator_dirac_120}
\end{figure*}

\paragraph{Scan setup at $m_{S}=120$ GeV.}
Fig.~\ref{fig:Scalar_Mediator_120}–\ref{fig:pseudoscalar_mediator_dirac_120} summarize the scans for a
\emph{scalar} mediator with three DM spins (complex scalar, Dirac, vector), and for a \emph{pseudoscalar}
mediator with Dirac DM.  For each case we show a global $(m_{\rm DM},\lambda_{\rm portal})$ scan (left)
and a resonance zoom (right), where the horizontal axis is shifted so that the origin corresponds to
$m_{\rm DM}=m_S/2=60~\mathrm{GeV}$.  

\paragraph{Complex–scalar DM + scalar mediator.}
The relic target produces the expected, very narrow funnel centered at
$m_{\rm DM}\simeq m_S/2$.
Away from this resonance, the couplings required to achieve
$\Omega_{\rm DM}h^2\simeq0.12$ are firmly excluded by SI direct detection,
which therefore controls the off–resonance parameter space.
Inside the funnel, however, the relic curve plunges below the LZ and DARWIN lines, leaving a thin but viable strip.
This strip overlaps the gray GCE band in the zoomed panel, so a consistent
solution exists only in the immediate neighborhood of the pole.

\paragraph{Dirac DM + scalar mediator.}
Here annihilation into SM fermions is $p$-wave, so present–day rates are
velocity suppressed and dSph limits are comparatively weak.
The allowed region is thus determined almost entirely by the crossing of the
relic curve with the SI limit.
Only a very tight resonant corridor survives; once the GCE flux fit is
imposed, this corridor no longer overlaps the preferred band in our scan.

\paragraph{Vector DM + scalar mediator.}
As in the complex–scalar case, annihilation contains an $s$-wave component.
Consequently, SI direct detection excludes essentially all off–resonance
solutions.
Near $m_{\rm DM}\simeq m_S/2$ the relic curve again dips below the LZ bound,
leaving a narrow strip where all requirements—relic density, DD, and the GCE
fit—can be met simultaneously.
Outside that funnel, the model is ruled out by DD.

\paragraph{Dirac DM + pseudoscalar mediator.}
For $m_P=120~\mathrm{GeV}$ (Fig.~\ref{fig:pseudoscalar_mediator_dirac_120}),
tree–level scattering is momentum/velocity suppressed and SI bounds are weak; loop–induced SI is subleading in the ranges of interest.
The surviving parameter space is therefore set mainly by the relic target and the dSph constraint.
A comparatively broad resonant band persists around $m_{\rm DM}\sim m_P/2$ and, importantly, it continues to overlap the GCE–preferred region in the zoomed panel.

\medskip
\noindent\textbf{Takeaway.}
With scalar mediators, SI direct detection removes the entire off–resonance
plane; only a narrow resonance funnel remains.
That funnel overlaps the GCE band for complex–scalar and vector DM, but not
for Dirac DM.
For a pseudoscalar mediator with Dirac DM, the DD suppression leaves a broader
near–resonant region that survives all constraints and still provides a good
GCE fit.

\subsection{Simplified Models: Vector Mediator}
\label{sec:results_vector_simplified}

\begin{figure*}[htbp]
  \centering
  \includegraphics[width=\columnwidth]{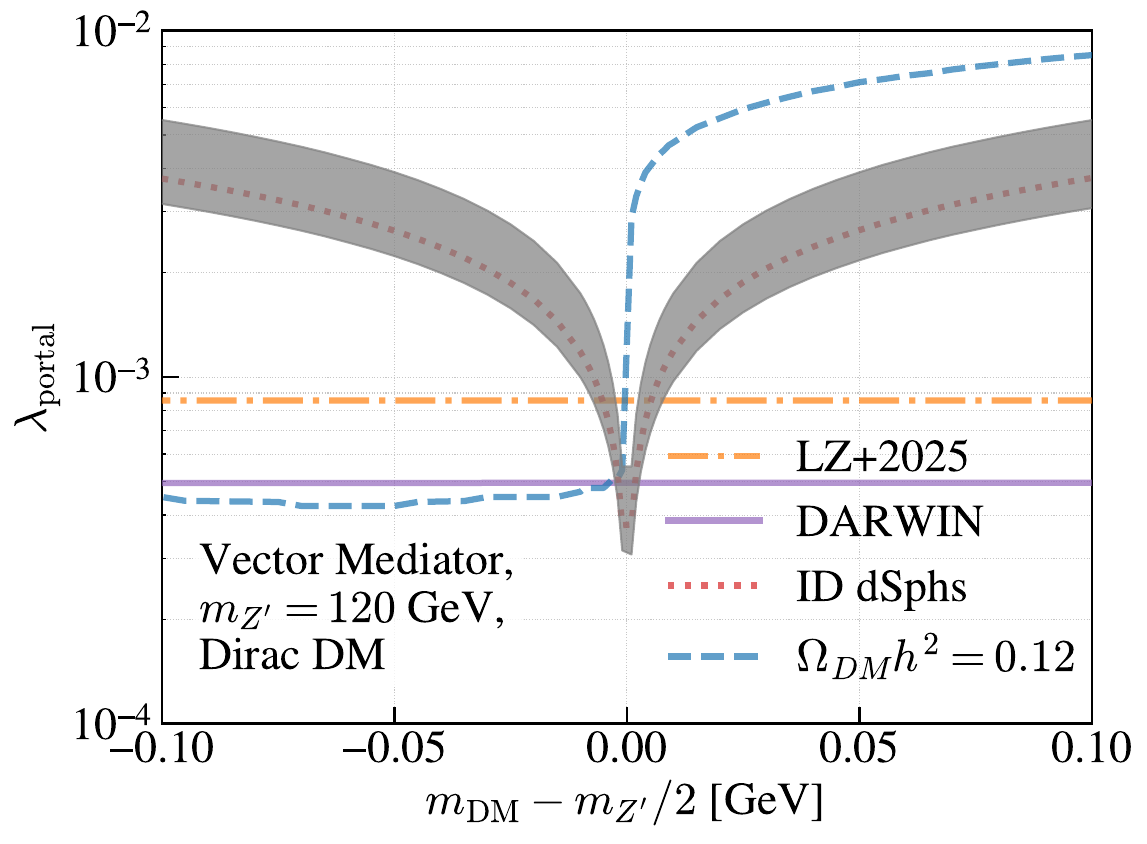}
  \includegraphics[width=\columnwidth]{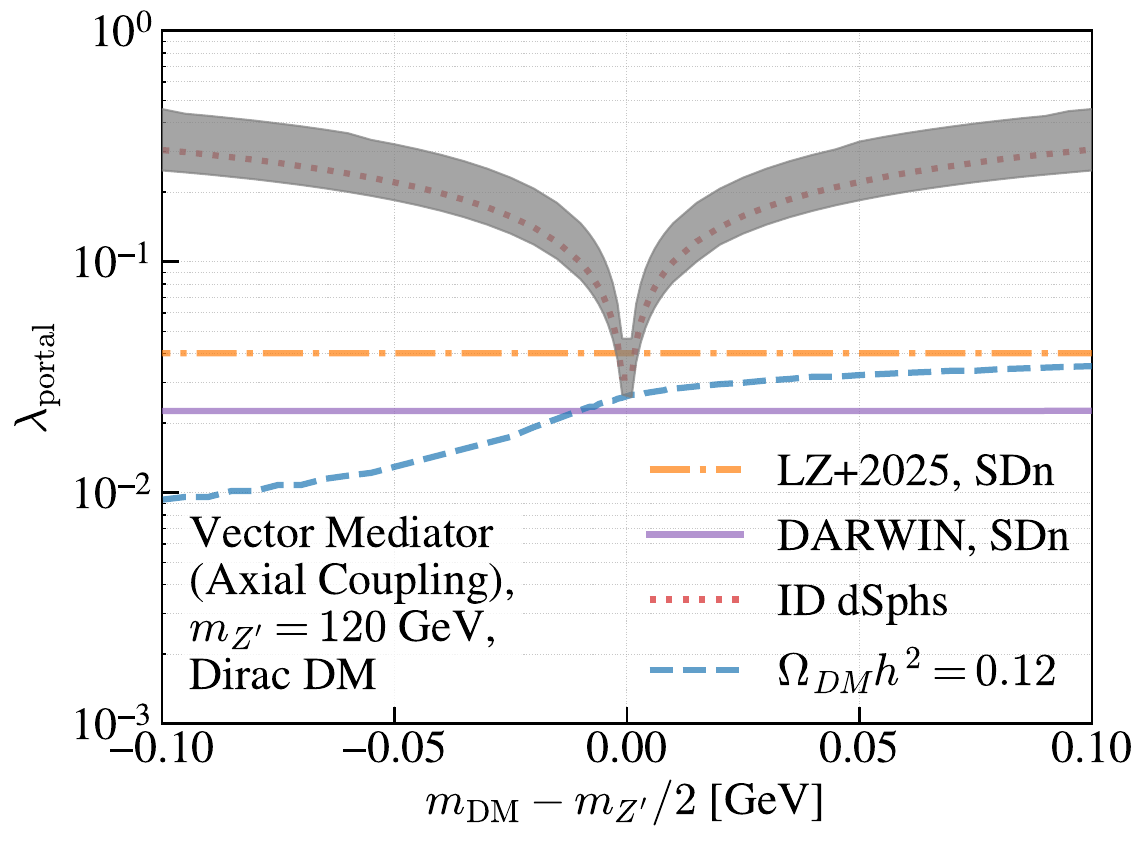}
  \caption{Constraints in the $(m_{\rm DM},\,\lambda_{\rm portal})$ plane for the simplified vector/axial-vector model with Dirac dark matter at fixed $m_{Z'}=120~\mathrm{GeV}$. Here, $m_{Z'}$ denotes the vector-mediator mass. The curves and shaded bands follow the same definitions as in Fig.~\ref{fig:higgs_portals}.}
  \label{fig:Vector_Mediator}
\end{figure*}

We study a spin–1 mediator $Z'$ with Dirac DM under two coupling structures to the dark current:
(i) \emph{vector} and (ii) \emph{axial–vector}. Guided by the GCE fit—which prefers resonant solutions—we
fix the mediator mass to $m_{Z'}=120~\mathrm{GeV}$ and focus on the resonance region
$m_{\rm DM}\simeq m_{Z'}/2=60~\mathrm{GeV}$. Fig.~\ref{fig:Vector_Mediator} shows zoom–in panels with the
horizontal axis shifted to $m_{\rm DM}-m_{Z'}/2$. Curves and bands follow the same conventions as in
Fig.~\ref{fig:higgs_portals}: the blue dashed line matches $\Omega_{\rm DM}h^2\simeq0.12$, the orange
dot–dashed line is the direct–detection bound (SI or SD as appropriate), the red dotted line gives the
combined dSph limit, and the grey band denotes the region preferred by our GCE flux fit.

\paragraph{Vector coupling.}
The $s$–channel funnel is narrow and centered at $m_{\rm DM}\simeq m_{Z'}/2$. Away from the pole, the
spin–independent LZ limit lies well below the relic–density line and excludes the parameter space.
Near the pole, Breit–Wigner enhancement lowers the coupling required for the thermal target, producing
a thin corridor where the relic curve dips beneath the SI bound and overlaps the GCE–preferred band,
with typical values $\lambda_{\rm portal}\sim 5\times10^{-4}$. Indirect limits (dSphs) modestly trim
the edges of this corridor but remain subdominant to SI constraints.

\paragraph{Axial–vector coupling.}
Here the leading nuclear response is spin–dependent; we therefore compare to the LZ SD–neutron bound. We also display the bounds obtained with the sensitivity forecast from the DARWIN detector for SD–neutron bound.
The relic–density curve is flatter across the resonance than in the pure–vector case, but SD constraints
still restrict viability to a very small neighborhood around the pole. After imposing the GCE fit,
only a narrow overlap remains, with representative couplings
$\lambda_{\rm portal}\sim 2.5\times10^{-2}$. Away from the resonance, the combination of relic density
and SD direct–detection limits excludes the model.

For both couplings, the surviving parameter space is confined to a narrow resonant corridor around
$m_{\rm DM}\simeq m_{Z'}/2$.  The pure vector case retains a small but finite overlap between relic density,
SI limits, and the GCE band, while the axial–vector case leaves only a marginal sliver once SD and GCE
constraints are applied.

\subsection{$U(1)_{L_i-L_j}$ and $U(1)_{B-L}$}
\label{sec:results_LiLj_BL}

\begin{figure*}[htbp]
  \centering
  \includegraphics[width=\columnwidth]{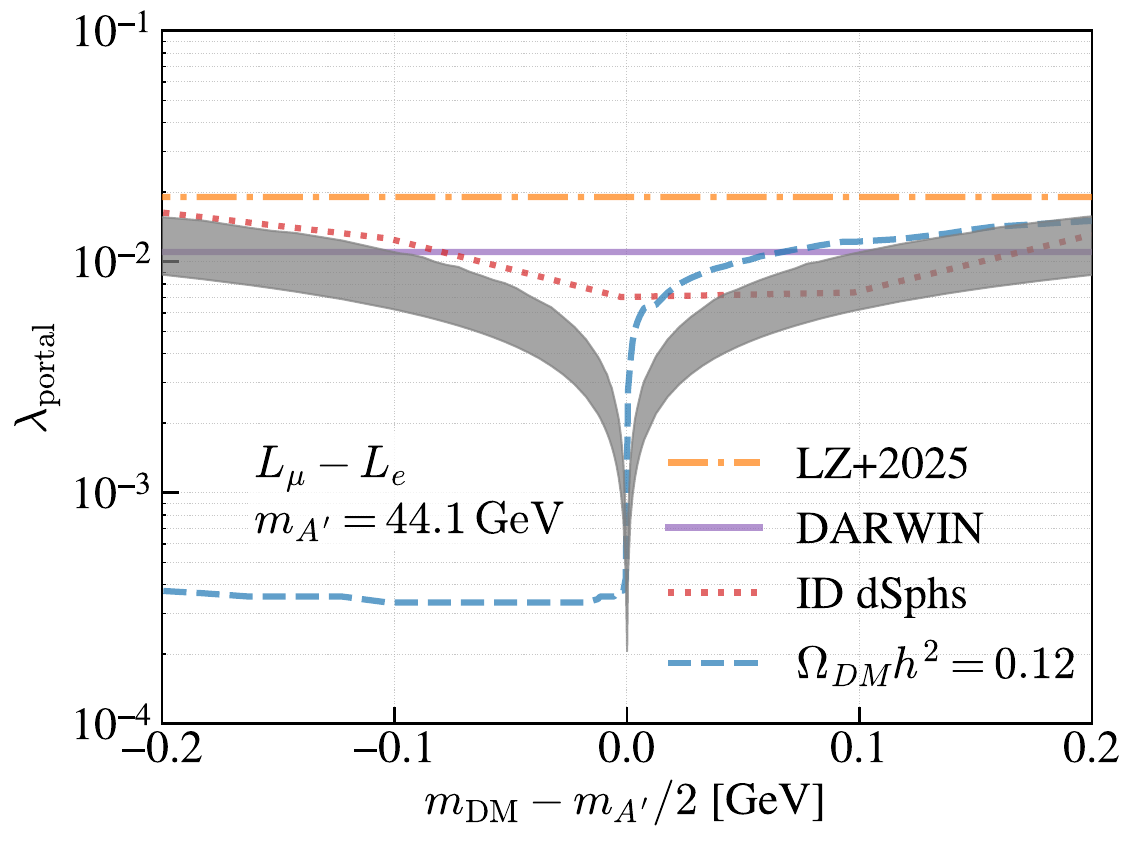}
  \includegraphics[width=\columnwidth]{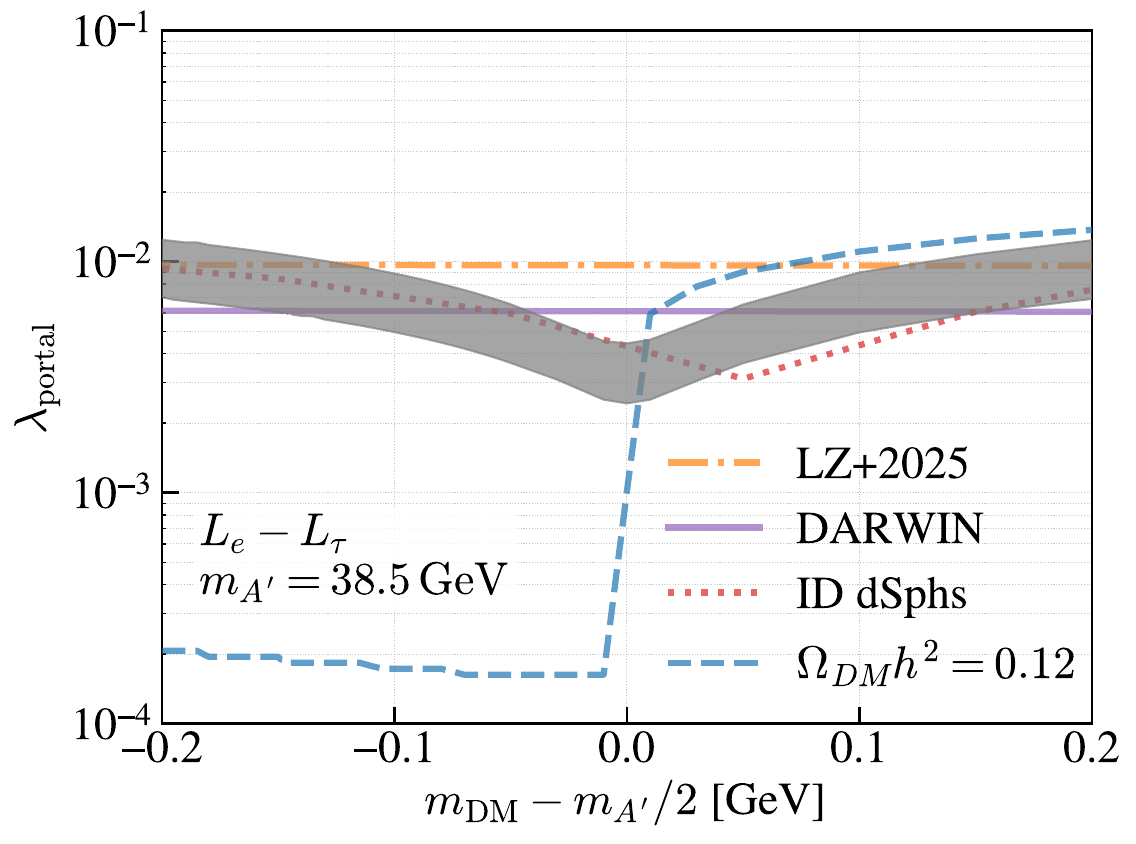}
  \includegraphics[width=\columnwidth]{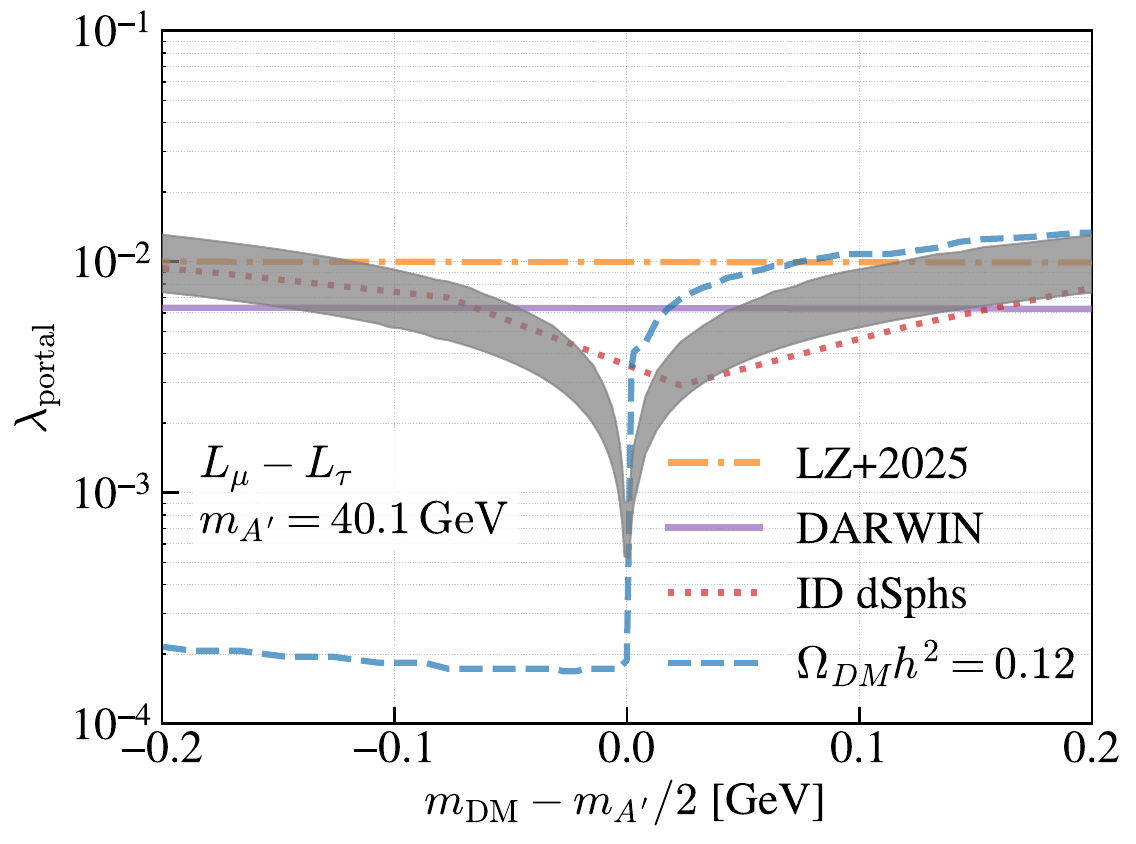}
  \includegraphics[width=\columnwidth]{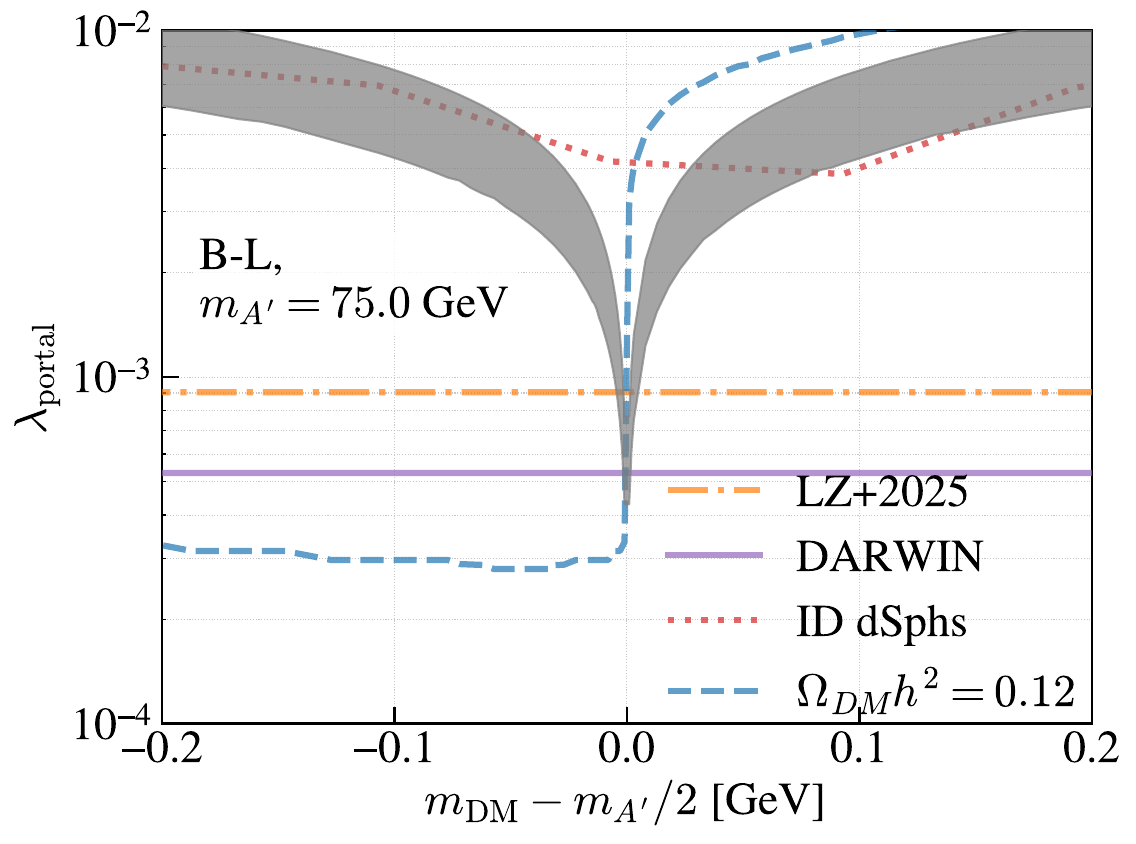}
  \caption{Constraints in the $(m_{\rm DM},\,\lambda_{\rm portal})$ plane for the $B\!-\!L$ and $L_i\!-\!L_j$ models with different vector mediator masses. Here, $m_{A'}$ denotes the vector–mediator mass. The curves and shaded bands follow the same definitions as in Fig.~\ref{fig:higgs_portals}.}
  \label{fig:LiLj_zoomin}
\end{figure*}

Fig.~\ref{fig:LiLj_zoomin} shows the zoom–in constraints in the $(m_{\rm DM},\lambda_{\rm portal})$ plane for the
$U(1)_{L_i-L_j}$ ($L_\mu\!-\!L_e$, $L_e\!-\!L_\tau$, $L_\mu\!-\!L_\tau$) and $U(1)_{B-L}$ models.  
For each panel the horizontal axis is shifted so that the origin corresponds to the $s$–channel pole, $m_{\rm DM}=m_{A'}/2$.  

\paragraph{Constraints and viable regions.}
All models exhibit a narrow resonant funnel around $m_{\rm DM}\simeq m_{A'}/2$.  
Because $B\!-\!L$ couples to quarks at tree level, its SI limit is comparatively strong; only a thin portion of the
resonant strip lies below LZ+2025. The remaining portion will be even narrower with the future DARWIN bounds.
In $L_i\!-\!L_j$ the scattering on nuclei proceeds via kinetic mixing, so the SI bound is weaker (at the $\sim10^{-2}$ level in the figure), allowing a somewhat wider overlap with the relic line.  
Among these, $L_\mu\!-\!L_e$ leaves the broadest allowed strip, while $L_\mu\!-\!L_\tau$ is the most constrained once the GCE
preference is imposed.  
Away from the pole, all models are excluded by DD.

\begin{figure*}[htbp]
  \centering
  \includegraphics[width=\columnwidth]{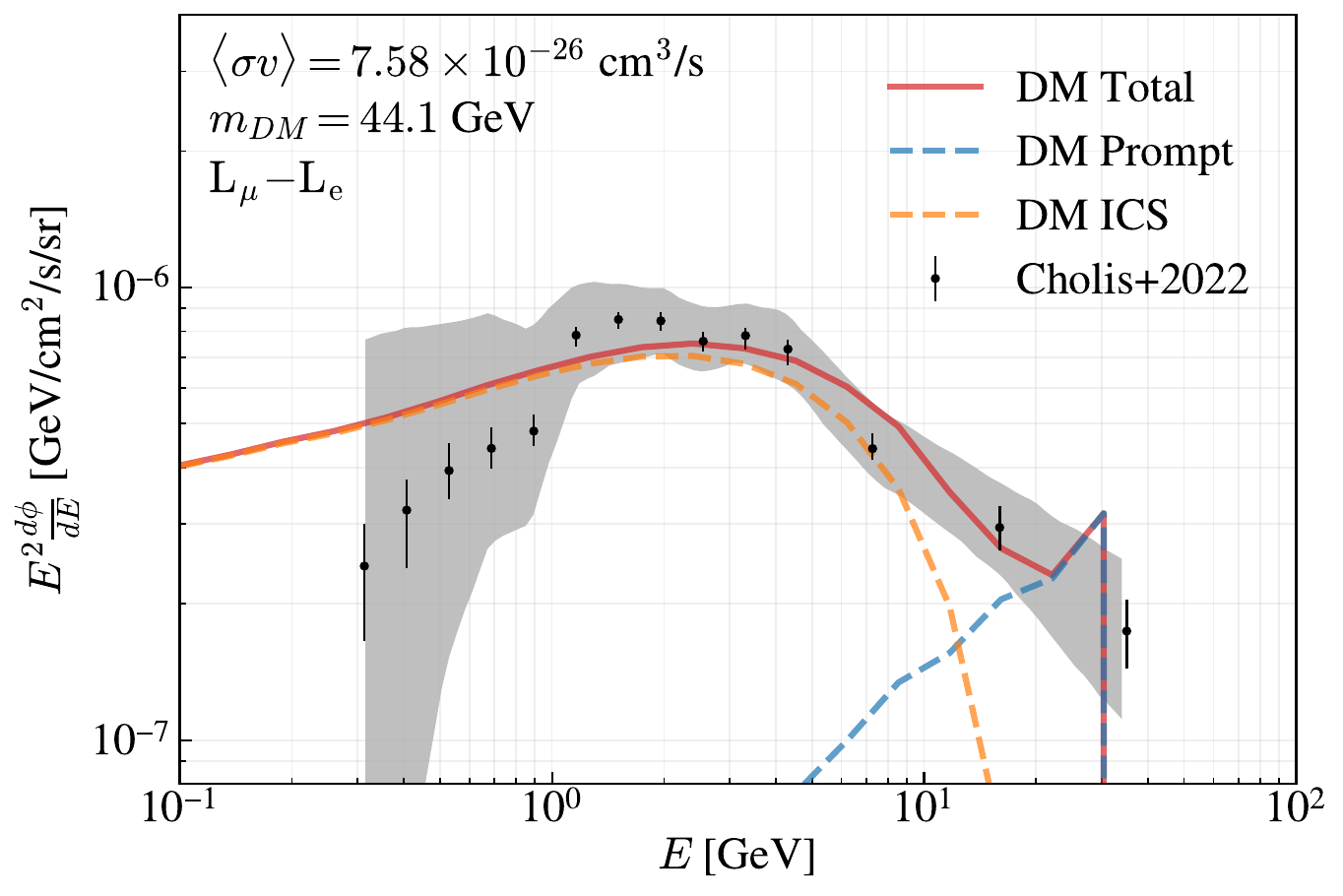}
  \includegraphics[width=\columnwidth]{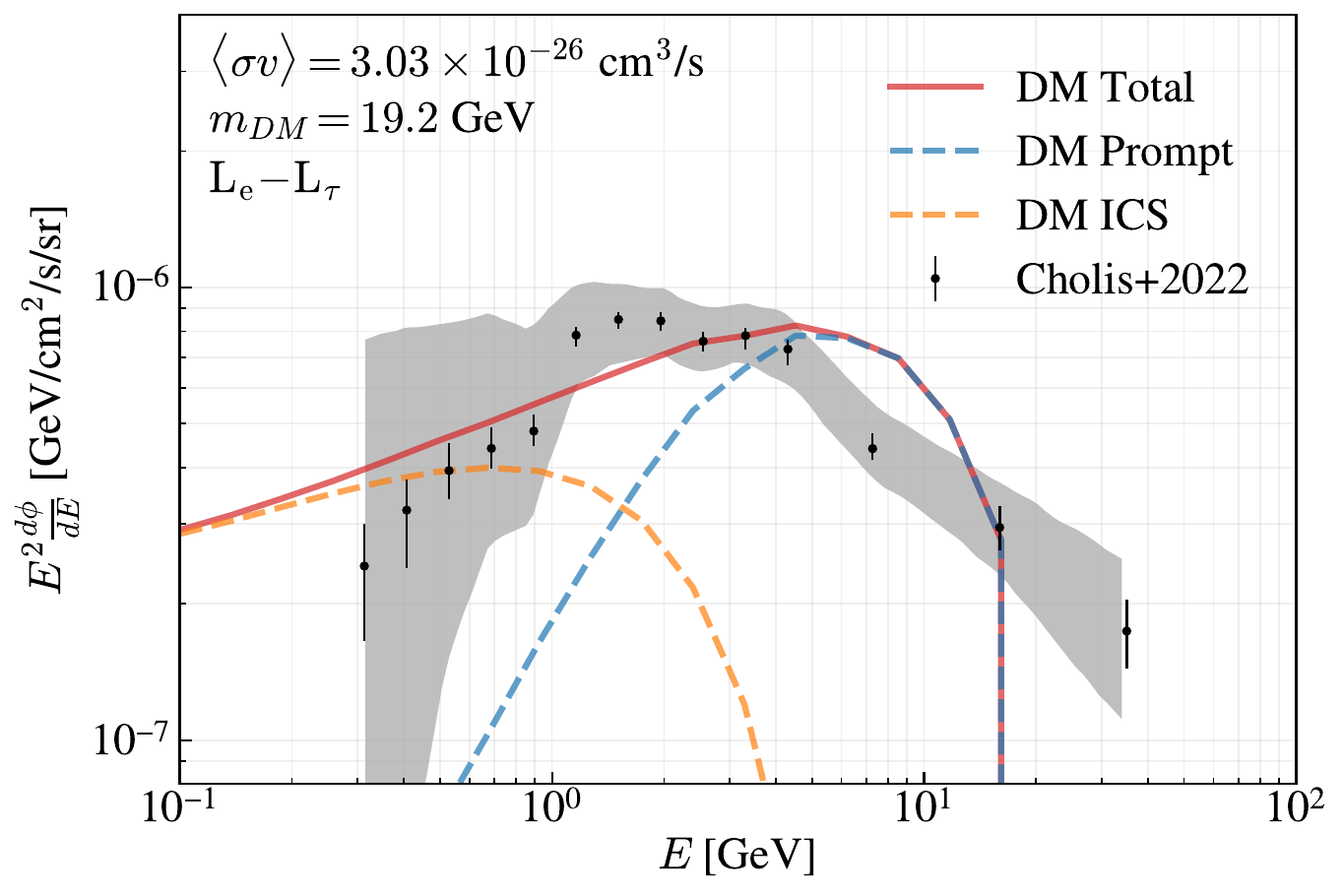}
  \includegraphics[width=\columnwidth]{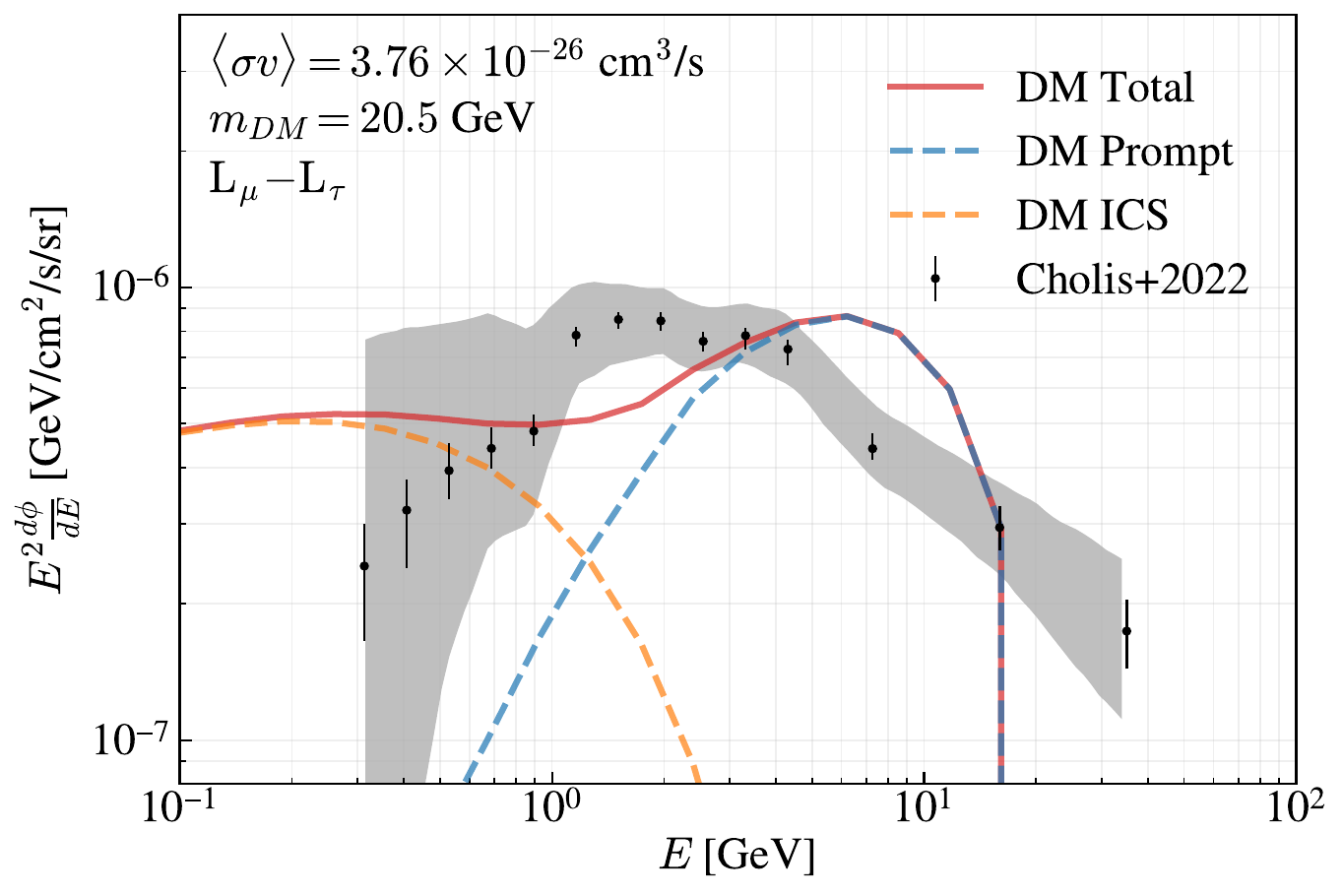}
  \includegraphics[width=\columnwidth]{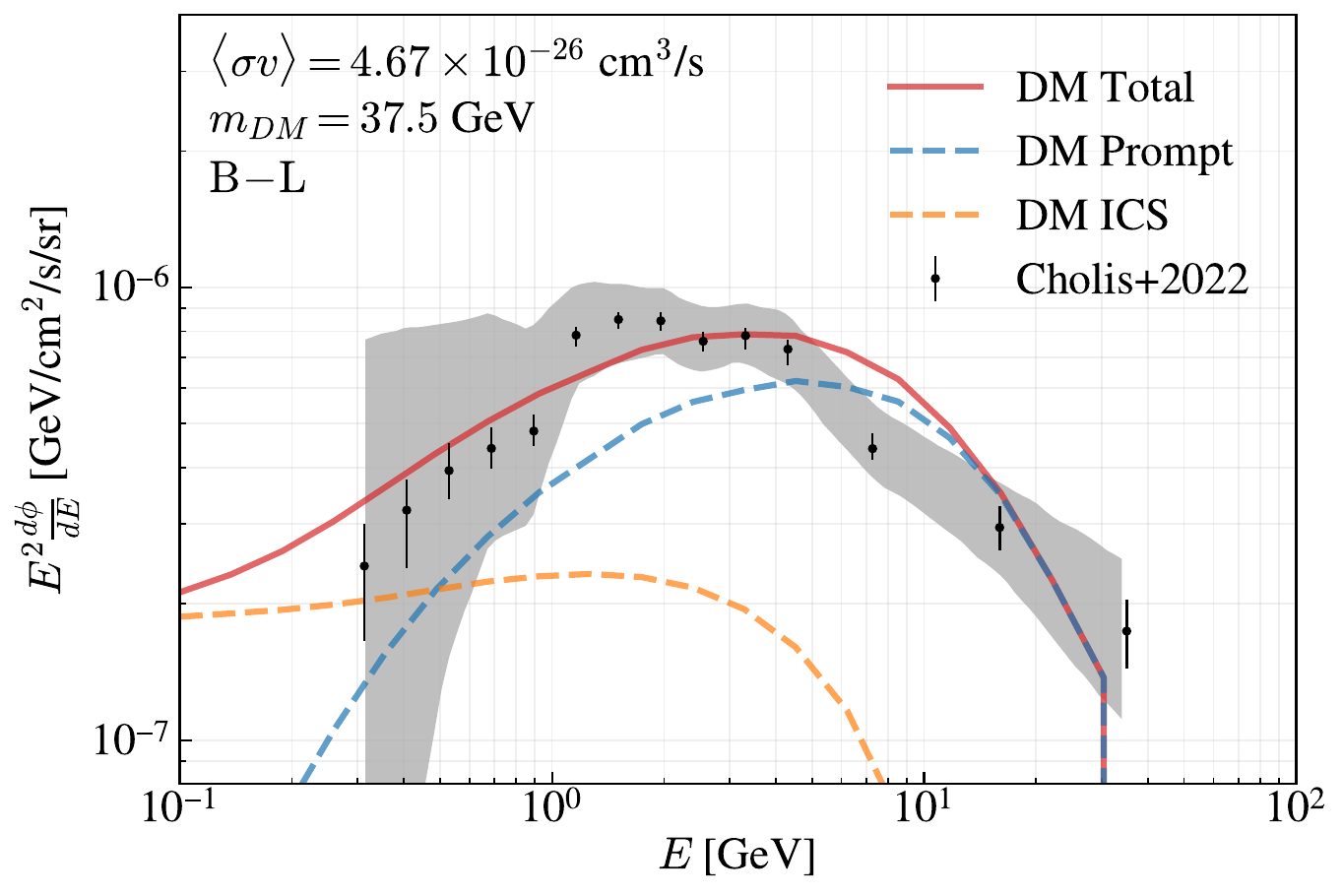}
  \caption{Best–fit of the four models considered in section~\ref{sec:results_LiLj_BL} to the Cholis+22 GCE datasets. In each panel, we show the GCE data and the theoretical predictions for the DM prompt (blue), ICS (orange), and total $\gamma$-ray flux (red).}
  \label{fig:LiLj_flux}
\end{figure*}

\paragraph{GCE flux fits.}
Fig.~\ref{fig:LiLj_flux} displays the best–fit spectra to the Cholis+22 dataset for the four models, decomposed into prompt and ICS.  
The ICS contribution governs the sub–GeV to few–GeV rise, while the prompt part dominates toward higher energies.  
$L_\mu\!-\!L_e$ provides the best overall match (lowest $\chi^2$), $B\!-\!L$ performs comparably well, $L_e\!-\!L_\tau$ yields an
acceptable fit with slightly larger $\chi^2$, and $L_\mu\!-\!L_\tau$ is disfavored.  
The corresponding best–fit parameters $(m_{\rm DM},\langle\sigma v\rangle)$ are summarized in
Table~\ref{tab:gce_bestfits} and are used to set $m_{A'}\simeq 2\,m_{\rm DM}$ in the constraint panels.
In particular, we can see that the DM mass needed to fit the GCE flux is between 20 to 45 GeV and the cross section is around the thermal one considering the uncertainties due to the geometrical factor.
We also report in the same table the best-fit results for the GCE obtained in the Singlet Dark Matter model with a Higgs portal (SHP; see Sec.~\ref{sec:higgsportal}), which we take as representative of a purely hadronic scenario. We find that the goodness of fit is comparable among the \(L_\mu-L_e\), \(B-L\), and purely hadronic models. The best-fit DM masses for these three model classes are also similar, spanning \(m_{\rm DM}\simeq 40\text{--}60~\mathrm{GeV}\). By contrast, the preferred annihilation cross section is smaller for the purely hadronic SHP scenario, \(\langle\sigma v\rangle \sim 10^{-26}\,\mathrm{cm}^3\,\mathrm{s}^{-1}\), while it is larger for the purely leptonic case, \(\langle\sigma v\rangle \sim 7\times 10^{-26}\,\mathrm{cm}^3\,\mathrm{s}^{-1}\). This difference reflects the dominant emission mechanism: in purely hadronic models the prompt component dominates, whereas in purely leptonic models most of the flux arises from inverse-Compton scattering.

\begin{table}[t] 
\centering 
\caption{Best-fit parameters for the $L_i-L_j$, $B-L$ and Singlet DM model with the Higgs portal (SHP, see Sec.~\ref{sec:higgsportal}) models to the Cholis+22 GCE dataset. The last column represents the best-fit $\chi^2$ value.} \label{tab:gce_bestfits} 
\begin{tabular}{lccc} \hline Model & $m_{\rm DM}$ [GeV] & $\langle\sigma v\rangle$ [$10^{-26}\,\mathrm{cm^3\,s^{-1}}$] & $\chi^2$ \\ 
\hline 
$L_\mu - L_e$ & 44.1 & 7.58 & 5.2 \\ 
$L_e - L_\tau$ & 19.2 & 3.03 & 16.5 \\ 
$L_\mu - L_\tau$ & 20.5 & 3.76 & 28.5 \\
\hline
$B - L$ & 37.5 & 4.67 & 6.2 \\ 
\hline
SHP & 62.5 & 0.94 & 11.8 \\
\hline
\end{tabular} \end{table}

Once relic density, dSphs, LZ+2025, and the GCE fit are combined, only narrow resonant bands remain viable.  
$L_\mu\!-\!L_e$ retains the largest overlap with the GCE–preferred region, $B\!-\!L$ is more tightly constrained by SI scattering but still admits a small window, and $L_\mu\!-\!L_\tau$ is strongly disfavored by the flux fit despite satisfying the resonant requirements.

\section{Conclusions}
\label{sec:conclusions}

Across all frameworks we examined, current DD and ID constraints squeeze viable WIMP solutions into \emph{resonant funnels} at $m_{\rm DM}\simeq m_{\rm med}/2$. 
We use the factor $\Delta$ to describe the degree of fine-tuning of the DM and mediator masses
\begin{equation}
\Delta \equiv \frac{|\,m_{\rm DM}-m_{\rm med}/2\,|}{m_{\rm med}/2}\,.
\end{equation}
Away from resonance, achieving the relic density generally requires couplings
$g_{\rm DM}\sim 0.1$--$1$, which are typically excluded by present SI/SD DD limits.
Near the pole, the annihilation rate is Breit--Wigner enhanced, so much smaller portals suffice and SI
bounds can be evaded. 
In this regime, ID limits for $s$-wave models also begin to be relevant.

\medskip
\noindent\textbf{Models that remain compatible.}
Collecting the scenarios that survive all bounds \emph{and} admit a fit to the GCE data, we find:
\begin{itemize}

\item \emph{Higgs portals (scalar and vector DM):} narrow corridors around the Higgs pole with
$\Delta \sim 4$--$6\%$ survives to DD/ID and relic density and
$\lambda_{\rm portal}\sim (2$--$4)\times 10^{-4}$ are preferred by the GCE flux data.

\item \emph{UV--complete Vector Higgs portal ($H$--$H_p$ mixing):} There is a viable strip around the $H_p$ pole with
$\Delta\sim 6$--$11\%$ and the GCE selects $\lambda_{\rm portal}\sim (0.4$--$2)\times 10^{-4}$ (mixing angle dependent).

\item \emph{Simplified scalar mediator with complex--scalar or vector DM:} The resonance windows are open at $\sim m_S/2$ with $\Delta\sim 5$--$8\%$ and $\lambda_{\rm portal}$ in the $10^{-3}$--$10^{-1}$ range (model-- and channel--dependent);
GCE--favored points sit near the lower end of this interval for $\lambda_{\rm portal}$.

\item \emph{Simplified vector mediator with Dirac DM (vector coupling):} a thin region around $m_{Z'}/2$ with
$\Delta\sim 2\%$ and $\lambda_{\rm portal}\sim 5\times 10^{-4}$ survives to all the observations.

\item \emph{Leptonic vectors $U(1)_{L_i-L_j}$ and $U(1)_{B-L}$ (prompt $+$ ICS):} We find good fits to the GCE with
$\langle\sigma v\rangle$ close to the thermal one and DM masses between 20-50 GeV. For these models we have included both the prompt and ICS $\gamma$-ray emission. The allowed corridors have $\Delta\sim 14$--$26\%$ for $L_i-L_j$ and $\Delta\sim 4\%$ for $B\!-\!L$, with portal strengths
$\lambda_{\rm portal}\sim 10^{-3}$--$10^{-2}$. Among these, $L_\mu\!-\!L_e$ and $B\!-\!L$ yield the best overall fits,
while $L_\mu\!-\!L_\tau$ is disfavored by the spectral fit.
\item \emph{Pseudoscalar mediator with Dirac DM:} a broader resonant band (loop induced SI suppressed) remains,
with $\lambda_{\rm portal}\sim 5\times 10^{-2}$--$1$ and $\Delta$ larger than in the scalar cases (up to the
few$\times 10\%$ level), still compatible with the GCE.
\end{itemize}

\noindent\textbf{Models that are effectively ruled out.}
\begin{itemize}
\item \emph{Dirac Higgs portal (scalar operator):} excluded across the scanned masses by SI direct detection,
despite $p$-wave ID suppression.
\item \emph{$Z$ portal with Dirac DM:} excluded for \emph{vector} couplings by SI limits and for \emph{axial}
couplings by SD (neutron) bounds; the $Z$-pole funnel does not reopen viable space.
\item \emph{Simplified scalar mediator with Dirac DM:} ruled out by SI limits away from resonance and lacks a
simultaneous solution once the GCE fit is imposed since the annihilation cross section into fermions is p-wave.
\end{itemize}

\medskip
\noindent\textbf{Outlook.}
If the GCE is of particle origin, present DD/ID constraints and precise relic density measurements favor models where $m_{\rm{DM}}\sim m_{\rm{med}}/2$ with a fine tune at the level of $\Delta\!\sim\!{\rm few}\%$ and portal couplings $\lambda_{\rm portal}$ as small as ${\sim}10^{-4}$ in hadronic portals and
${\sim}10^{-3}$--$10^{-2}$ in leptonic vectors. Next--generation SI (and SD--neutron) sensitivities from the DARWIN experiment \cite{DARWIN}, improved dSph/GC $\gamma$-ray analyses, and targeted collider searches for mediators near the pole masses will be
decisive in testing these last viable corridors.

\medskip

\begin{acknowledgments} 
M.D.M. acknowledges support from the research grant {\sl TAsP (Theoretical Astroparticle Physics)} funded by Istituto Nazionale di Fisica Nucleare (INFN) and from the Italian Ministry of University and Research (MUR), PRIN 2022 ``EXSKALIBUR – Euclid-Cross-SKA: Likelihood Inference Building for Universe’s Research'', Grant No. 20222BBYB9, CUP I53D23000610 0006, and from the European Union -- Next Generation EU. The code for this work is publicly available at \url{https://github.com/KK-cloudhub/GCE_analysis}.
\end{acknowledgments}

\bibliography{paper}

\providecommand{\noopsort}[1]{}\providecommand{\singleletter}[1]{#1}%
\begin{thebibliography}{90}%
\makeatletter
\providecommand \@ifxundefined [1]{%
 \@ifx{#1\undefined}
}%
\providecommand \@ifnum [1]{%
 \ifnum #1\expandafter \@firstoftwo
 \else \expandafter \@secondoftwo
 \fi
}%
\providecommand \@ifx [1]{%
 \ifx #1\expandafter \@firstoftwo
 \else \expandafter \@secondoftwo
 \fi
}%
\providecommand \natexlab [1]{#1}%
\providecommand \enquote  [1]{``#1''}%
\providecommand \bibnamefont  [1]{#1}%
\providecommand \bibfnamefont [1]{#1}%
\providecommand \citenamefont [1]{#1}%
\providecommand \href@noop [0]{\@secondoftwo}%
\providecommand \href [0]{\begingroup \@sanitize@url \@href}%
\providecommand \@href[1]{\@@startlink{#1}\@@href}%
\providecommand \@@href[1]{\endgroup#1\@@endlink}%
\providecommand \@sanitize@url [0]{\catcode `\\12\catcode `\$12\catcode
  `\&12\catcode `\#12\catcode `\^12\catcode `\_12\catcode `\%12\relax}%
\providecommand \@@startlink[1]{}%
\providecommand \@@endlink[0]{}%
\providecommand \url  [0]{\begingroup\@sanitize@url \@url }%
\providecommand \@url [1]{\endgroup\@href {#1}{\urlprefix }}%
\providecommand \urlprefix  [0]{URL }%
\providecommand \Eprint [0]{\href }%
\providecommand \doibase [0]{http://dx.doi.org/}%
\providecommand \selectlanguage [0]{\@gobble}%
\providecommand \bibinfo  [0]{\@secondoftwo}%
\providecommand \bibfield  [0]{\@secondoftwo}%
\providecommand \translation [1]{[#1]}%
\providecommand \BibitemOpen [0]{}%
\providecommand \bibitemStop [0]{}%
\providecommand \bibitemNoStop [0]{.\EOS\space}%
\providecommand \EOS [0]{\spacefactor3000\relax}%
\providecommand \BibitemShut  [1]{\csname bibitem#1\endcsname}%
\let\auto@bib@innerbib\@empty
\bibitem [{\citenamefont {Silk}\ \emph {et~al.}(2010)\citenamefont {Silk} \emph
  {et~al.}}]{Bertone:2010zza}%
  \BibitemOpen
  \bibfield  {author} {\bibinfo {author} {\bibfnamefont {J.}~\bibnamefont
  {Silk}} \emph {et~al.},\ }\href {\doibase 10.1017/CBO9780511770739} {\emph
  {\bibinfo {title} {{Particle Dark Matter: Observations, Models and
  Searches}}}},\ edited by\ \bibinfo {editor} {\bibfnamefont {G.}~\bibnamefont
  {Bertone}}\ (\bibinfo  {publisher} {Cambridge Univ. Press},\ \bibinfo
  {address} {Cambridge},\ \bibinfo {year} {2010})\BibitemShut {NoStop}%
\bibitem [{\citenamefont {Bertone}\ and\ \citenamefont
  {Hooper}(2018)}]{Bertone:2016nfn}%
  \BibitemOpen
  \bibfield  {author} {\bibinfo {author} {\bibfnamefont {G.}~\bibnamefont
  {Bertone}}\ and\ \bibinfo {author} {\bibfnamefont {D.}~\bibnamefont
  {Hooper}},\ }\href {\doibase 10.1103/RevModPhys.90.045002} {\bibfield
  {journal} {\bibinfo  {journal} {Rev. Mod. Phys.}\ }\textbf {\bibinfo {volume}
  {90}},\ \bibinfo {pages} {045002} (\bibinfo {year} {2018})},\ \Eprint
  {http://arxiv.org/abs/1605.04909} {arXiv:1605.04909 [astro-ph.CO]}
  \BibitemShut {NoStop}%
\bibitem [{\citenamefont {Cirelli}\ \emph {et~al.}(2024)\citenamefont
  {Cirelli}, \citenamefont {Strumia},\ and\ \citenamefont
  {Zupan}}]{Cirelli:2024ssz}%
  \BibitemOpen
  \bibfield  {author} {\bibinfo {author} {\bibfnamefont {M.}~\bibnamefont
  {Cirelli}}, \bibinfo {author} {\bibfnamefont {A.}~\bibnamefont {Strumia}}, \
  and\ \bibinfo {author} {\bibfnamefont {J.}~\bibnamefont {Zupan}},\
  }\href@noop {} {\  (\bibinfo {year} {2024})},\ \Eprint
  {http://arxiv.org/abs/2406.01705} {arXiv:2406.01705 [hep-ph]} \BibitemShut
  {NoStop}%
\bibitem [{\citenamefont {Aghanim}\ \emph {et~al.}(2018)\citenamefont {Aghanim}
  \emph {et~al.}}]{Aghanim:2018eyx}%
  \BibitemOpen
  \bibfield  {author} {\bibinfo {author} {\bibfnamefont {N.}~\bibnamefont
  {Aghanim}} \emph {et~al.} (\bibinfo {collaboration} {Planck}),\ }\href@noop
  {} {\  (\bibinfo {year} {2018})},\ \Eprint {http://arxiv.org/abs/1807.06209}
  {arXiv:1807.06209 [astro-ph.CO]} \BibitemShut {NoStop}%
\bibitem [{\citenamefont {Schumann}(2019)}]{Schumann:2019eaa}%
  \BibitemOpen
  \bibfield  {author} {\bibinfo {author} {\bibfnamefont {M.}~\bibnamefont
  {Schumann}},\ }\href {\doibase 10.1088/1361-6471/ab2ea5} {\bibfield
  {journal} {\bibinfo  {journal} {J. Phys. G}\ }\textbf {\bibinfo {volume}
  {46}},\ \bibinfo {pages} {103003} (\bibinfo {year} {2019})},\ \Eprint
  {http://arxiv.org/abs/1903.03026} {arXiv:1903.03026 [astro-ph.CO]}
  \BibitemShut {NoStop}%
\bibitem [{\citenamefont {Boveia}\ and\ \citenamefont
  {Doglioni}(2018)}]{Boveia:2018yeb}%
  \BibitemOpen
  \bibfield  {author} {\bibinfo {author} {\bibfnamefont {A.}~\bibnamefont
  {Boveia}}\ and\ \bibinfo {author} {\bibfnamefont {C.}~\bibnamefont
  {Doglioni}},\ }\href {\doibase 10.1146/annurev-nucl-101917-021008} {\bibfield
   {journal} {\bibinfo  {journal} {Ann. Rev. Nucl. Part. Sci.}\ }\textbf
  {\bibinfo {volume} {68}},\ \bibinfo {pages} {429} (\bibinfo {year} {2018})},\
  \Eprint {http://arxiv.org/abs/1810.12238} {arXiv:1810.12238 [hep-ex]}
  \BibitemShut {NoStop}%
\bibitem [{\citenamefont {Gaskins}(2016)}]{Gaskins:2016cha}%
  \BibitemOpen
  \bibfield  {author} {\bibinfo {author} {\bibfnamefont {J.~M.}\ \bibnamefont
  {Gaskins}},\ }\href {\doibase 10.1080/00107514.2016.1175160} {\bibfield
  {journal} {\bibinfo  {journal} {Contemp. Phys.}\ }\textbf {\bibinfo {volume}
  {57}},\ \bibinfo {pages} {496} (\bibinfo {year} {2016})},\ \Eprint
  {http://arxiv.org/abs/1604.00014} {arXiv:1604.00014 [astro-ph.HE]}
  \BibitemShut {NoStop}%
\bibitem [{\citenamefont {Pieri}\ \emph {et~al.}(2011)\citenamefont {Pieri},
  \citenamefont {Lavalle}, \citenamefont {Bertone},\ and\ \citenamefont
  {Branchini}}]{Pieri:2009je}%
  \BibitemOpen
  \bibfield  {author} {\bibinfo {author} {\bibfnamefont {L.}~\bibnamefont
  {Pieri}}, \bibinfo {author} {\bibfnamefont {J.}~\bibnamefont {Lavalle}},
  \bibinfo {author} {\bibfnamefont {G.}~\bibnamefont {Bertone}}, \ and\
  \bibinfo {author} {\bibfnamefont {E.}~\bibnamefont {Branchini}},\ }\href
  {\doibase 10.1103/PhysRevD.83.023518} {\bibfield  {journal} {\bibinfo
  {journal} {Phys. Rev. D}\ }\textbf {\bibinfo {volume} {83}},\ \bibinfo
  {pages} {023518} (\bibinfo {year} {2011})},\ \Eprint
  {http://arxiv.org/abs/0908.0195} {arXiv:0908.0195 [astro-ph.HE]} \BibitemShut
  {NoStop}%
\bibitem [{\citenamefont {Goodenough}\ and\ \citenamefont
  {Hooper}(2009)}]{Goodenough:2009gk}%
  \BibitemOpen
  \bibfield  {author} {\bibinfo {author} {\bibfnamefont {L.}~\bibnamefont
  {Goodenough}}\ and\ \bibinfo {author} {\bibfnamefont {D.}~\bibnamefont
  {Hooper}},\ }\href@noop {} {\  (\bibinfo {year} {2009})},\ \Eprint
  {http://arxiv.org/abs/0910.2998} {arXiv:0910.2998 [hep-ph]} \BibitemShut
  {NoStop}%
\bibitem [{\citenamefont {Hooper}\ and\ \citenamefont
  {Goodenough}(2011)}]{Hooper:2010mq}%
  \BibitemOpen
  \bibfield  {author} {\bibinfo {author} {\bibfnamefont {D.}~\bibnamefont
  {Hooper}}\ and\ \bibinfo {author} {\bibfnamefont {L.}~\bibnamefont
  {Goodenough}},\ }\href {\doibase 10.1016/j.physletb.2011.02.029} {\bibfield
  {journal} {\bibinfo  {journal} {Phys. Lett.}\ }\textbf {\bibinfo {volume}
  {B697}},\ \bibinfo {pages} {412} (\bibinfo {year} {2011})},\ \Eprint
  {http://arxiv.org/abs/1010.2752} {arXiv:1010.2752 [hep-ph]} \BibitemShut
  {NoStop}%
\bibitem [{\citenamefont {Boyarsky}\ \emph {et~al.}(2011)\citenamefont
  {Boyarsky}, \citenamefont {Malyshev},\ and\ \citenamefont
  {Ruchayskiy}}]{Boyarsky:2010dr}%
  \BibitemOpen
  \bibfield  {author} {\bibinfo {author} {\bibfnamefont {A.}~\bibnamefont
  {Boyarsky}}, \bibinfo {author} {\bibfnamefont {D.}~\bibnamefont {Malyshev}},
  \ and\ \bibinfo {author} {\bibfnamefont {O.}~\bibnamefont {Ruchayskiy}},\
  }\href {\doibase 10.1016/j.physletb.2011.10.014} {\bibfield  {journal}
  {\bibinfo  {journal} {Phys. Lett.}\ }\textbf {\bibinfo {volume} {B705}},\
  \bibinfo {pages} {165} (\bibinfo {year} {2011})},\ \Eprint
  {http://arxiv.org/abs/1012.5839} {arXiv:1012.5839 [hep-ph]} \BibitemShut
  {NoStop}%
\bibitem [{\citenamefont {Hooper}\ and\ \citenamefont
  {Linden}(2011)}]{Hooper:2011ti}%
  \BibitemOpen
  \bibfield  {author} {\bibinfo {author} {\bibfnamefont {D.}~\bibnamefont
  {Hooper}}\ and\ \bibinfo {author} {\bibfnamefont {T.}~\bibnamefont
  {Linden}},\ }\href {\doibase 10.1103/PhysRevD.84.123005} {\bibfield
  {journal} {\bibinfo  {journal} {Phys. Rev.}\ }\textbf {\bibinfo {volume}
  {D84}},\ \bibinfo {pages} {123005} (\bibinfo {year} {2011})},\ \Eprint
  {http://arxiv.org/abs/1110.0006} {arXiv:1110.0006 [astro-ph.HE]} \BibitemShut
  {NoStop}%
\bibitem [{\citenamefont {Abazajian}\ and\ \citenamefont
  {Kaplinghat}(2012)}]{Abazajian:2012pn}%
  \BibitemOpen
  \bibfield  {author} {\bibinfo {author} {\bibfnamefont {K.~N.}\ \bibnamefont
  {Abazajian}}\ and\ \bibinfo {author} {\bibfnamefont {M.}~\bibnamefont
  {Kaplinghat}},\ }\href {\doibase 10.1103/PhysRevD.86.083511,
  10.1103/PhysRevD.87.129902} {\bibfield  {journal} {\bibinfo  {journal} {Phys.
  Rev.}\ }\textbf {\bibinfo {volume} {D86}},\ \bibinfo {pages} {083511}
  (\bibinfo {year} {2012})},\ \bibinfo {note} {[Erratum: Phys.
  Rev.D87,129902(2013)]},\ \Eprint {http://arxiv.org/abs/1207.6047}
  {arXiv:1207.6047 [astro-ph.HE]} \BibitemShut {NoStop}%
\bibitem [{\citenamefont {Gordon}\ and\ \citenamefont
  {Macias}(2013)}]{Gordon:2013vta}%
  \BibitemOpen
  \bibfield  {author} {\bibinfo {author} {\bibfnamefont {C.}~\bibnamefont
  {Gordon}}\ and\ \bibinfo {author} {\bibfnamefont {O.}~\bibnamefont
  {Macias}},\ }\href {\doibase 10.1103/PhysRevD.88.083521,
  10.1103/PhysRevD.89.049901} {\bibfield  {journal} {\bibinfo  {journal} {Phys.
  Rev.}\ }\textbf {\bibinfo {volume} {D88}},\ \bibinfo {pages} {083521}
  (\bibinfo {year} {2013})},\ \bibinfo {note} {[Erratum: Phys.
  Rev.D89,no.4,049901(2014)]},\ \Eprint {http://arxiv.org/abs/1306.5725}
  {arXiv:1306.5725 [astro-ph.HE]} \BibitemShut {NoStop}%
\bibitem [{\citenamefont {Abazajian}\ \emph {et~al.}(2014)\citenamefont
  {Abazajian}, \citenamefont {Canac}, \citenamefont {Horiuchi},\ and\
  \citenamefont {Kaplinghat}}]{Abazajian:2014fta}%
  \BibitemOpen
  \bibfield  {author} {\bibinfo {author} {\bibfnamefont {K.~N.}\ \bibnamefont
  {Abazajian}}, \bibinfo {author} {\bibfnamefont {N.}~\bibnamefont {Canac}},
  \bibinfo {author} {\bibfnamefont {S.}~\bibnamefont {Horiuchi}}, \ and\
  \bibinfo {author} {\bibfnamefont {M.}~\bibnamefont {Kaplinghat}},\ }\href
  {\doibase 10.1103/PhysRevD.90.023526} {\bibfield  {journal} {\bibinfo
  {journal} {Phys. Rev.}\ }\textbf {\bibinfo {volume} {D90}},\ \bibinfo {pages}
  {023526} (\bibinfo {year} {2014})},\ \Eprint {http://arxiv.org/abs/1402.4090}
  {arXiv:1402.4090 [astro-ph.HE]} \BibitemShut {NoStop}%
\bibitem [{\citenamefont {Daylan}\ \emph {et~al.}(2016)\citenamefont {Daylan},
  \citenamefont {Finkbeiner}, \citenamefont {Hooper}, \citenamefont {Linden},
  \citenamefont {Portillo}, \citenamefont {Rodd},\ and\ \citenamefont
  {Slatyer}}]{Daylan:2014rsa}%
  \BibitemOpen
  \bibfield  {author} {\bibinfo {author} {\bibfnamefont {T.}~\bibnamefont
  {Daylan}}, \bibinfo {author} {\bibfnamefont {D.~P.}\ \bibnamefont
  {Finkbeiner}}, \bibinfo {author} {\bibfnamefont {D.}~\bibnamefont {Hooper}},
  \bibinfo {author} {\bibfnamefont {T.}~\bibnamefont {Linden}}, \bibinfo
  {author} {\bibfnamefont {S.~K.~N.}\ \bibnamefont {Portillo}}, \bibinfo
  {author} {\bibfnamefont {N.~L.}\ \bibnamefont {Rodd}}, \ and\ \bibinfo
  {author} {\bibfnamefont {T.~R.}\ \bibnamefont {Slatyer}},\ }\href {\doibase
  10.1016/j.dark.2015.12.005} {\bibfield  {journal} {\bibinfo  {journal} {Phys.
  Dark Univ.}\ }\textbf {\bibinfo {volume} {12}},\ \bibinfo {pages} {1}
  (\bibinfo {year} {2016})},\ \Eprint {http://arxiv.org/abs/1402.6703}
  {arXiv:1402.6703 [astro-ph.HE]} \BibitemShut {NoStop}%
\bibitem [{\citenamefont {Calore}\ \emph
  {et~al.}(2015{\natexlab{a}})\citenamefont {Calore}, \citenamefont {Cholis},
  \citenamefont {McCabe},\ and\ \citenamefont {Weniger}}]{Calore:2014nla}%
  \BibitemOpen
  \bibfield  {author} {\bibinfo {author} {\bibfnamefont {F.}~\bibnamefont
  {Calore}}, \bibinfo {author} {\bibfnamefont {I.}~\bibnamefont {Cholis}},
  \bibinfo {author} {\bibfnamefont {C.}~\bibnamefont {McCabe}}, \ and\ \bibinfo
  {author} {\bibfnamefont {C.}~\bibnamefont {Weniger}},\ }\href {\doibase
  10.1103/PhysRevD.91.063003} {\bibfield  {journal} {\bibinfo  {journal} {Phys.
  Rev. D}\ }\textbf {\bibinfo {volume} {91}},\ \bibinfo {pages} {063003}
  (\bibinfo {year} {2015}{\natexlab{a}})},\ \Eprint
  {http://arxiv.org/abs/1411.4647} {arXiv:1411.4647 [hep-ph]} \BibitemShut
  {NoStop}%
\bibitem [{\citenamefont {Calore}\ \emph
  {et~al.}(2015{\natexlab{b}})\citenamefont {Calore}, \citenamefont {Cholis},\
  and\ \citenamefont {Weniger}}]{Calore:2014xka}%
  \BibitemOpen
  \bibfield  {author} {\bibinfo {author} {\bibfnamefont {F.}~\bibnamefont
  {Calore}}, \bibinfo {author} {\bibfnamefont {I.}~\bibnamefont {Cholis}}, \
  and\ \bibinfo {author} {\bibfnamefont {C.}~\bibnamefont {Weniger}},\ }\href
  {\doibase 10.1088/1475-7516/2015/03/038} {\bibfield  {journal} {\bibinfo
  {journal} {JCAP}\ }\textbf {\bibinfo {volume} {1503}},\ \bibinfo {pages}
  {038} (\bibinfo {year} {2015}{\natexlab{b}})},\ \Eprint
  {http://arxiv.org/abs/1409.0042} {arXiv:1409.0042 [astro-ph.CO]} \BibitemShut
  {NoStop}%
\bibitem [{\citenamefont {Ajello}\ \emph {et~al.}(2016)\citenamefont {Ajello}
  \emph {et~al.}}]{TheFermi-LAT:2015kwa}%
  \BibitemOpen
  \bibfield  {author} {\bibinfo {author} {\bibfnamefont {M.}~\bibnamefont
  {Ajello}} \emph {et~al.} (\bibinfo {collaboration} {Fermi-LAT}),\ }\href
  {\doibase 10.3847/0004-637X/819/1/44} {\bibfield  {journal} {\bibinfo
  {journal} {Astrophys. J.}\ }\textbf {\bibinfo {volume} {819}},\ \bibinfo
  {pages} {44} (\bibinfo {year} {2016})},\ \Eprint
  {http://arxiv.org/abs/1511.02938} {arXiv:1511.02938 [astro-ph.HE]}
  \BibitemShut {NoStop}%
\bibitem [{\citenamefont {Ackermann}\ \emph {et~al.}(2017)\citenamefont
  {Ackermann} \emph {et~al.}}]{TheFermi-LAT:2017vmf}%
  \BibitemOpen
  \bibfield  {author} {\bibinfo {author} {\bibfnamefont {M.}~\bibnamefont
  {Ackermann}} \emph {et~al.} (\bibinfo {collaboration} {Fermi-LAT}),\ }\href
  {\doibase 10.3847/1538-4357/aa6cab} {\bibfield  {journal} {\bibinfo
  {journal} {Astrophys. J.}\ }\textbf {\bibinfo {volume} {840}},\ \bibinfo
  {pages} {43} (\bibinfo {year} {2017})},\ \Eprint
  {http://arxiv.org/abs/1704.03910} {arXiv:1704.03910 [astro-ph.HE]}
  \BibitemShut {NoStop}%
\bibitem [{\citenamefont {Di~Mauro}\ \emph {et~al.}(2019)\citenamefont
  {Di~Mauro}, \citenamefont {Hou}, \citenamefont {Eckner}, \citenamefont
  {Zaharijas},\ and\ \citenamefont {Charles}}]{DiMauro:2019frs}%
  \BibitemOpen
  \bibfield  {author} {\bibinfo {author} {\bibfnamefont {M.}~\bibnamefont
  {Di~Mauro}}, \bibinfo {author} {\bibfnamefont {X.}~\bibnamefont {Hou}},
  \bibinfo {author} {\bibfnamefont {C.}~\bibnamefont {Eckner}}, \bibinfo
  {author} {\bibfnamefont {G.}~\bibnamefont {Zaharijas}}, \ and\ \bibinfo
  {author} {\bibfnamefont {E.}~\bibnamefont {Charles}},\ }\href {\doibase
  10.1103/PhysRevD.99.123027} {\bibfield  {journal} {\bibinfo  {journal} {Phys.
  Rev. D}\ }\textbf {\bibinfo {volume} {99}},\ \bibinfo {pages} {123027}
  (\bibinfo {year} {2019})},\ \Eprint {http://arxiv.org/abs/1904.10977}
  {arXiv:1904.10977 [astro-ph.HE]} \BibitemShut {NoStop}%
\bibitem [{\citenamefont {Di~Mauro}(2021)}]{DiMauro:2021raz}%
  \BibitemOpen
  \bibfield  {author} {\bibinfo {author} {\bibfnamefont {M.}~\bibnamefont
  {Di~Mauro}},\ }\href {\doibase 10.1103/PhysRevD.103.063029} {\bibfield
  {journal} {\bibinfo  {journal} {Phys. Rev. D}\ }\textbf {\bibinfo {volume}
  {103}},\ \bibinfo {pages} {063029} (\bibinfo {year} {2021})},\ \Eprint
  {http://arxiv.org/abs/2101.04694} {arXiv:2101.04694 [astro-ph.HE]}
  \BibitemShut {NoStop}%
\bibitem [{\citenamefont {Cholis}\ \emph {et~al.}(2022)\citenamefont {Cholis},
  \citenamefont {Zhong}, \citenamefont {McDermott},\ and\ \citenamefont
  {Surdutovich}}]{Cholis:2021rpp}%
  \BibitemOpen
  \bibfield  {author} {\bibinfo {author} {\bibfnamefont {I.}~\bibnamefont
  {Cholis}}, \bibinfo {author} {\bibfnamefont {Y.-M.}\ \bibnamefont {Zhong}},
  \bibinfo {author} {\bibfnamefont {S.~D.}\ \bibnamefont {McDermott}}, \ and\
  \bibinfo {author} {\bibfnamefont {J.~P.}\ \bibnamefont {Surdutovich}},\
  }\href {\doibase 10.1103/PhysRevD.105.103023} {\bibfield  {journal} {\bibinfo
   {journal} {Phys. Rev. D}\ }\textbf {\bibinfo {volume} {105}},\ \bibinfo
  {pages} {103023} (\bibinfo {year} {2022})},\ \Eprint
  {http://arxiv.org/abs/2112.09706} {arXiv:2112.09706 [astro-ph.HE]}
  \BibitemShut {NoStop}%
\bibitem [{\citenamefont {Di~Mauro}\ and\ \citenamefont
  {Winkler}(2021)}]{DiMauro:2021qcf}%
  \BibitemOpen
  \bibfield  {author} {\bibinfo {author} {\bibfnamefont {M.}~\bibnamefont
  {Di~Mauro}}\ and\ \bibinfo {author} {\bibfnamefont {M.~W.}\ \bibnamefont
  {Winkler}},\ }\href {\doibase 10.1103/PhysRevD.103.123005} {\bibfield
  {journal} {\bibinfo  {journal} {Phys. Rev. D}\ }\textbf {\bibinfo {volume}
  {103}},\ \bibinfo {pages} {123005} (\bibinfo {year} {2021})},\ \Eprint
  {http://arxiv.org/abs/2101.11027} {arXiv:2101.11027 [astro-ph.HE]}
  \BibitemShut {NoStop}%
\bibitem [{\citenamefont {Koechler}\ and\ \citenamefont
  {Di~Mauro}(2025)}]{Koechler:2025ryv}%
  \BibitemOpen
  \bibfield  {author} {\bibinfo {author} {\bibfnamefont {J.}~\bibnamefont
  {Koechler}}\ and\ \bibinfo {author} {\bibfnamefont {M.}~\bibnamefont
  {Di~Mauro}},\ }\href@noop {} {\  (\bibinfo {year} {2025})},\ \Eprint
  {http://arxiv.org/abs/2508.02775} {arXiv:2508.02775 [hep-ph]} \BibitemShut
  {NoStop}%
\bibitem [{\citenamefont {Bartels}\ \emph {et~al.}(2016)\citenamefont
  {Bartels}, \citenamefont {Krishnamurthy},\ and\ \citenamefont
  {Weniger}}]{Bartels:2015aea}%
  \BibitemOpen
  \bibfield  {author} {\bibinfo {author} {\bibfnamefont {R.}~\bibnamefont
  {Bartels}}, \bibinfo {author} {\bibfnamefont {S.}~\bibnamefont
  {Krishnamurthy}}, \ and\ \bibinfo {author} {\bibfnamefont {C.}~\bibnamefont
  {Weniger}},\ }\href {\doibase 10.1103/PhysRevLett.116.051102} {\bibfield
  {journal} {\bibinfo  {journal} {Phys. Rev. Lett.}\ }\textbf {\bibinfo
  {volume} {116}},\ \bibinfo {pages} {051102} (\bibinfo {year} {2016})},\
  \Eprint {http://arxiv.org/abs/1506.05104} {arXiv:1506.05104 [astro-ph.HE]}
  \BibitemShut {NoStop}%
\bibitem [{\citenamefont {Lee}\ \emph {et~al.}(2016)\citenamefont {Lee},
  \citenamefont {Lisanti}, \citenamefont {Safdi}, \citenamefont {Slatyer},\
  and\ \citenamefont {Xue}}]{Lee:2015fea}%
  \BibitemOpen
  \bibfield  {author} {\bibinfo {author} {\bibfnamefont {S.~K.}\ \bibnamefont
  {Lee}}, \bibinfo {author} {\bibfnamefont {M.}~\bibnamefont {Lisanti}},
  \bibinfo {author} {\bibfnamefont {B.~R.}\ \bibnamefont {Safdi}}, \bibinfo
  {author} {\bibfnamefont {T.~R.}\ \bibnamefont {Slatyer}}, \ and\ \bibinfo
  {author} {\bibfnamefont {W.}~\bibnamefont {Xue}},\ }\href {\doibase
  10.1103/PhysRevLett.116.051103} {\bibfield  {journal} {\bibinfo  {journal}
  {Phys. Rev. Lett.}\ }\textbf {\bibinfo {volume} {116}},\ \bibinfo {pages}
  {051103} (\bibinfo {year} {2016})},\ \Eprint
  {http://arxiv.org/abs/1506.05124} {arXiv:1506.05124 [astro-ph.HE]}
  \BibitemShut {NoStop}%
\bibitem [{\citenamefont {Macias}\ \emph {et~al.}(2018)\citenamefont {Macias},
  \citenamefont {Gordon}, \citenamefont {Crocker}, \citenamefont {Coleman},
  \citenamefont {Paterson}, \citenamefont {Horiuchi},\ and\ \citenamefont
  {Pohl}}]{Macias:2016nev}%
  \BibitemOpen
  \bibfield  {author} {\bibinfo {author} {\bibfnamefont {O.}~\bibnamefont
  {Macias}}, \bibinfo {author} {\bibfnamefont {C.}~\bibnamefont {Gordon}},
  \bibinfo {author} {\bibfnamefont {R.~M.}\ \bibnamefont {Crocker}}, \bibinfo
  {author} {\bibfnamefont {B.}~\bibnamefont {Coleman}}, \bibinfo {author}
  {\bibfnamefont {D.}~\bibnamefont {Paterson}}, \bibinfo {author}
  {\bibfnamefont {S.}~\bibnamefont {Horiuchi}}, \ and\ \bibinfo {author}
  {\bibfnamefont {M.}~\bibnamefont {Pohl}},\ }\href {\doibase
  10.1038/s41550-018-0414-3} {\bibfield  {journal} {\bibinfo  {journal} {Nat.
  Astron.}\ }\textbf {\bibinfo {volume} {2}},\ \bibinfo {pages} {387} (\bibinfo
  {year} {2018})},\ \Eprint {http://arxiv.org/abs/1611.06644} {arXiv:1611.06644
  [astro-ph.HE]} \BibitemShut {NoStop}%
\bibitem [{\citenamefont {Bartels}\ \emph {et~al.}(2018)\citenamefont
  {Bartels}, \citenamefont {Storm}, \citenamefont {Weniger},\ and\
  \citenamefont {Calore}}]{Bartels:2017vsx}%
  \BibitemOpen
  \bibfield  {author} {\bibinfo {author} {\bibfnamefont {R.}~\bibnamefont
  {Bartels}}, \bibinfo {author} {\bibfnamefont {E.}~\bibnamefont {Storm}},
  \bibinfo {author} {\bibfnamefont {C.}~\bibnamefont {Weniger}}, \ and\
  \bibinfo {author} {\bibfnamefont {F.}~\bibnamefont {Calore}},\ }\href
  {\doibase 10.1038/s41550-018-0531-z} {\bibfield  {journal} {\bibinfo
  {journal} {Nat. Astron.}\ }\textbf {\bibinfo {volume} {2}},\ \bibinfo {pages}
  {819} (\bibinfo {year} {2018})},\ \Eprint {http://arxiv.org/abs/1711.04778}
  {arXiv:1711.04778 [astro-ph.HE]} \BibitemShut {NoStop}%
\bibitem [{\citenamefont {Manconi}\ \emph {et~al.}(2024)\citenamefont
  {Manconi}, \citenamefont {Calore},\ and\ \citenamefont
  {Donato}}]{Manconi:2024tgh}%
  \BibitemOpen
  \bibfield  {author} {\bibinfo {author} {\bibfnamefont {S.}~\bibnamefont
  {Manconi}}, \bibinfo {author} {\bibfnamefont {F.}~\bibnamefont {Calore}}, \
  and\ \bibinfo {author} {\bibfnamefont {F.}~\bibnamefont {Donato}},\ }\href
  {\doibase 10.1103/PhysRevD.109.123042} {\bibfield  {journal} {\bibinfo
  {journal} {Phys. Rev. D}\ }\textbf {\bibinfo {volume} {109}},\ \bibinfo
  {pages} {123042} (\bibinfo {year} {2024})},\ \Eprint
  {http://arxiv.org/abs/2402.04733} {arXiv:2402.04733 [astro-ph.HE]}
  \BibitemShut {NoStop}%
\bibitem [{\citenamefont {Leane}\ and\ \citenamefont
  {Slatyer}(2019)}]{Leane:2019uhc}%
  \BibitemOpen
  \bibfield  {author} {\bibinfo {author} {\bibfnamefont {R.~K.}\ \bibnamefont
  {Leane}}\ and\ \bibinfo {author} {\bibfnamefont {T.~R.}\ \bibnamefont
  {Slatyer}},\ }\href {\doibase 10.1103/PhysRevLett.123.241101} {\bibfield
  {journal} {\bibinfo  {journal} {Phys. Rev. Lett.}\ }\textbf {\bibinfo
  {volume} {123}},\ \bibinfo {pages} {241101} (\bibinfo {year} {2019})},\
  \Eprint {http://arxiv.org/abs/1904.08430} {arXiv:1904.08430 [astro-ph.HE]}
  \BibitemShut {NoStop}%
\bibitem [{\citenamefont {Chang}\ \emph {et~al.}(2019)\citenamefont {Chang},
  \citenamefont {Mishra-Sharma}, \citenamefont {Lisanti}, \citenamefont
  {Buschmann}, \citenamefont {Rodd},\ and\ \citenamefont
  {Safdi}}]{Chang:2019ars}%
  \BibitemOpen
  \bibfield  {author} {\bibinfo {author} {\bibfnamefont {L.~J.}\ \bibnamefont
  {Chang}}, \bibinfo {author} {\bibfnamefont {S.}~\bibnamefont
  {Mishra-Sharma}}, \bibinfo {author} {\bibfnamefont {M.}~\bibnamefont
  {Lisanti}}, \bibinfo {author} {\bibfnamefont {M.}~\bibnamefont {Buschmann}},
  \bibinfo {author} {\bibfnamefont {N.~L.}\ \bibnamefont {Rodd}}, \ and\
  \bibinfo {author} {\bibfnamefont {B.~R.}\ \bibnamefont {Safdi}},\ }\href@noop
  {} {\  (\bibinfo {year} {2019})},\ \Eprint {http://arxiv.org/abs/1908.10874}
  {arXiv:1908.10874 [astro-ph.CO]} \BibitemShut {NoStop}%
\bibitem [{\citenamefont {Zhong}\ \emph {et~al.}(2019)\citenamefont {Zhong},
  \citenamefont {McDermott}, \citenamefont {Cholis},\ and\ \citenamefont
  {Fox}}]{Zhong:2019ycb}%
  \BibitemOpen
  \bibfield  {author} {\bibinfo {author} {\bibfnamefont {Y.-M.}\ \bibnamefont
  {Zhong}}, \bibinfo {author} {\bibfnamefont {S.~D.}\ \bibnamefont
  {McDermott}}, \bibinfo {author} {\bibfnamefont {I.}~\bibnamefont {Cholis}}, \
  and\ \bibinfo {author} {\bibfnamefont {P.~J.}\ \bibnamefont {Fox}},\
  }\href@noop {} {\  (\bibinfo {year} {2019})},\ \Eprint
  {http://arxiv.org/abs/1911.12369} {arXiv:1911.12369 [astro-ph.HE]}
  \BibitemShut {NoStop}%
\bibitem [{\citenamefont {Calore}\ \emph {et~al.}(2021)\citenamefont {Calore},
  \citenamefont {Donato},\ and\ \citenamefont {Manconi}}]{Calore:2021jvg}%
  \BibitemOpen
  \bibfield  {author} {\bibinfo {author} {\bibfnamefont {F.}~\bibnamefont
  {Calore}}, \bibinfo {author} {\bibfnamefont {F.}~\bibnamefont {Donato}}, \
  and\ \bibinfo {author} {\bibfnamefont {S.}~\bibnamefont {Manconi}},\ }\href
  {\doibase 10.1103/PhysRevLett.127.161102} {\bibfield  {journal} {\bibinfo
  {journal} {Phys. Rev. Lett.}\ }\textbf {\bibinfo {volume} {127}},\ \bibinfo
  {pages} {161102} (\bibinfo {year} {2021})},\ \Eprint
  {http://arxiv.org/abs/2102.12497} {arXiv:2102.12497 [astro-ph.HE]}
  \BibitemShut {NoStop}%
\bibitem [{\citenamefont {List}\ \emph {et~al.}(2025)\citenamefont {List},
  \citenamefont {Park}, \citenamefont {Rodd}, \citenamefont {Schoen},\ and\
  \citenamefont {Wolf}}]{List:2025qbx}%
  \BibitemOpen
  \bibfield  {author} {\bibinfo {author} {\bibfnamefont {F.}~\bibnamefont
  {List}}, \bibinfo {author} {\bibfnamefont {Y.}~\bibnamefont {Park}}, \bibinfo
  {author} {\bibfnamefont {N.~L.}\ \bibnamefont {Rodd}}, \bibinfo {author}
  {\bibfnamefont {E.}~\bibnamefont {Schoen}}, \ and\ \bibinfo {author}
  {\bibfnamefont {F.}~\bibnamefont {Wolf}},\ }\href@noop {} {\  (\bibinfo
  {year} {2025})},\ \Eprint {http://arxiv.org/abs/2507.17804} {arXiv:2507.17804
  [astro-ph.HE]} \BibitemShut {NoStop}%
\bibitem [{\citenamefont {Aalbers}\ \emph {et~al.}(2023)\citenamefont {Aalbers}
  \emph {et~al.}}]{LZ:2023}%
  \BibitemOpen
  \bibfield  {author} {\bibinfo {author} {\bibfnamefont {J.}~\bibnamefont
  {Aalbers}} \emph {et~al.},\ }\href {\doibase 10.1103/PhysRevLett.131.041002}
  {\bibfield  {journal} {\bibinfo  {journal} {Phys. Rev. Lett.}\ }\textbf
  {\bibinfo {volume} {131}},\ \bibinfo {pages} {041002} (\bibinfo {year}
  {2023})},\ \Eprint {http://arxiv.org/abs/2207.03764} {arXiv:2207.03764
  [hep-ex]} \BibitemShut {NoStop}%
\bibitem [{\citenamefont {Aprile}\ \emph {et~al.}(2023)\citenamefont {Aprile}
  \emph {et~al.}}]{Aprile:2023XENONnT}%
  \BibitemOpen
  \bibfield  {author} {\bibinfo {author} {\bibfnamefont {E.}~\bibnamefont
  {Aprile}} \emph {et~al.},\ }\href {\doibase 10.1103/PhysRevLett.131.041003}
  {\bibfield  {journal} {\bibinfo  {journal} {Phys. Rev. Lett.}\ }\textbf
  {\bibinfo {volume} {131}},\ \bibinfo {pages} {041003} (\bibinfo {year}
  {2023})}\BibitemShut {NoStop}%
\bibitem [{\citenamefont {Aalbers}\ \emph {et~al.}(2025)\citenamefont {Aalbers}
  \emph {et~al.}}]{LZ:2024zvo}%
  \BibitemOpen
  \bibfield  {author} {\bibinfo {author} {\bibfnamefont {J.}~\bibnamefont
  {Aalbers}} \emph {et~al.} (\bibinfo {collaboration} {LZ}),\ }\href {\doibase
  10.1103/4dyc-z8zf} {\bibfield  {journal} {\bibinfo  {journal} {Phys. Rev.
  Lett.}\ }\textbf {\bibinfo {volume} {135}},\ \bibinfo {pages} {011802}
  (\bibinfo {year} {2025})},\ \Eprint {http://arxiv.org/abs/2410.17036}
  {arXiv:2410.17036 [hep-ex]} \BibitemShut {NoStop}%
\bibitem [{\citenamefont {Arcadi}\ \emph
  {et~al.}(2018{\natexlab{a}})\citenamefont {Arcadi}, \citenamefont {Dutra},
  \citenamefont {Ghosh}, \citenamefont {Lindner}, \citenamefont {Mambrini},
  \citenamefont {Pierre}, \citenamefont {Profumo},\ and\ \citenamefont
  {Queiroz}}]{Arcadi:2017kky}%
  \BibitemOpen
  \bibfield  {author} {\bibinfo {author} {\bibfnamefont {G.}~\bibnamefont
  {Arcadi}}, \bibinfo {author} {\bibfnamefont {M.}~\bibnamefont {Dutra}},
  \bibinfo {author} {\bibfnamefont {P.}~\bibnamefont {Ghosh}}, \bibinfo
  {author} {\bibfnamefont {M.}~\bibnamefont {Lindner}}, \bibinfo {author}
  {\bibfnamefont {Y.}~\bibnamefont {Mambrini}}, \bibinfo {author}
  {\bibfnamefont {M.}~\bibnamefont {Pierre}}, \bibinfo {author} {\bibfnamefont
  {S.}~\bibnamefont {Profumo}}, \ and\ \bibinfo {author} {\bibfnamefont
  {F.~S.}\ \bibnamefont {Queiroz}},\ }\href {\doibase
  10.1140/epjc/s10052-018-5662-y} {\bibfield  {journal} {\bibinfo  {journal}
  {Eur. Phys. J. C}\ }\textbf {\bibinfo {volume} {78}},\ \bibinfo {pages} {203}
  (\bibinfo {year} {2018}{\natexlab{a}})},\ \Eprint
  {http://arxiv.org/abs/1703.07364} {arXiv:1703.07364 [hep-ph]} \BibitemShut
  {NoStop}%
\bibitem [{\citenamefont {Arcadi}\ \emph
  {et~al.}(2020{\natexlab{a}})\citenamefont {Arcadi}, \citenamefont {Djouadi},\
  and\ \citenamefont {Raidal}}]{Arcadi:2019lka}%
  \BibitemOpen
  \bibfield  {author} {\bibinfo {author} {\bibfnamefont {G.}~\bibnamefont
  {Arcadi}}, \bibinfo {author} {\bibfnamefont {A.}~\bibnamefont {Djouadi}}, \
  and\ \bibinfo {author} {\bibfnamefont {M.}~\bibnamefont {Raidal}},\ }\href
  {\doibase 10.1016/j.physrep.2019.11.003} {\bibfield  {journal} {\bibinfo
  {journal} {Phys. Rept.}\ }\textbf {\bibinfo {volume} {842}},\ \bibinfo
  {pages} {1} (\bibinfo {year} {2020}{\natexlab{a}})},\ \Eprint
  {http://arxiv.org/abs/1903.03616} {arXiv:1903.03616 [hep-ph]} \BibitemShut
  {NoStop}%
\bibitem [{\citenamefont {Di~Mauro}\ \emph {et~al.}(2023)\citenamefont
  {Di~Mauro}, \citenamefont {Arina}, \citenamefont {Fornengo}, \citenamefont
  {Heisig},\ and\ \citenamefont {Massaro}}]{DiMauro:2023tho}%
  \BibitemOpen
  \bibfield  {author} {\bibinfo {author} {\bibfnamefont {M.}~\bibnamefont
  {Di~Mauro}}, \bibinfo {author} {\bibfnamefont {C.}~\bibnamefont {Arina}},
  \bibinfo {author} {\bibfnamefont {N.}~\bibnamefont {Fornengo}}, \bibinfo
  {author} {\bibfnamefont {J.}~\bibnamefont {Heisig}}, \ and\ \bibinfo {author}
  {\bibfnamefont {D.}~\bibnamefont {Massaro}},\ }\href {\doibase
  10.1103/PhysRevD.108.095008} {\bibfield  {journal} {\bibinfo  {journal}
  {Phys. Rev. D}\ }\textbf {\bibinfo {volume} {108}},\ \bibinfo {pages}
  {095008} (\bibinfo {year} {2023})},\ \Eprint
  {http://arxiv.org/abs/2305.11937} {arXiv:2305.11937 [hep-ph]} \BibitemShut
  {NoStop}%
\bibitem [{\citenamefont {Arcadi}\ \emph {et~al.}(2024)\citenamefont {Arcadi},
  \citenamefont {Cabo-Almeida}, \citenamefont {Dutra}, \citenamefont {Ghosh},
  \citenamefont {Lindner}, \citenamefont {Mambrini}, \citenamefont {Neto},
  \citenamefont {Pierre}, \citenamefont {Profumo},\ and\ \citenamefont
  {Queiroz}}]{Arcadi:2024ukq}%
  \BibitemOpen
  \bibfield  {author} {\bibinfo {author} {\bibfnamefont {G.}~\bibnamefont
  {Arcadi}}, \bibinfo {author} {\bibfnamefont {D.}~\bibnamefont
  {Cabo-Almeida}}, \bibinfo {author} {\bibfnamefont {M.}~\bibnamefont {Dutra}},
  \bibinfo {author} {\bibfnamefont {P.}~\bibnamefont {Ghosh}}, \bibinfo
  {author} {\bibfnamefont {M.}~\bibnamefont {Lindner}}, \bibinfo {author}
  {\bibfnamefont {Y.}~\bibnamefont {Mambrini}}, \bibinfo {author}
  {\bibfnamefont {J.~P.}\ \bibnamefont {Neto}}, \bibinfo {author}
  {\bibfnamefont {M.}~\bibnamefont {Pierre}}, \bibinfo {author} {\bibfnamefont
  {S.}~\bibnamefont {Profumo}}, \ and\ \bibinfo {author} {\bibfnamefont
  {F.~S.}\ \bibnamefont {Queiroz}},\ }\href@noop {} {\  (\bibinfo {year}
  {2024})},\ \Eprint {http://arxiv.org/abs/2403.15860} {arXiv:2403.15860
  [hep-ph]} \BibitemShut {NoStop}%
\bibitem [{\citenamefont {Di~Mauro}\ and\ \citenamefont
  {Xie}(2025)}]{DiMauro:2025jia}%
  \BibitemOpen
  \bibfield  {author} {\bibinfo {author} {\bibfnamefont {M.}~\bibnamefont
  {Di~Mauro}}\ and\ \bibinfo {author} {\bibfnamefont {B.}~\bibnamefont {Xie}},\
  }\href@noop {} {\  (\bibinfo {year} {2025})},\ \Eprint
  {http://arxiv.org/abs/2510.08677} {arXiv:2510.08677 [hep-ph]} \BibitemShut
  {NoStop}%
\bibitem [{\citenamefont {Pospelov}\ \emph {et~al.}(2008)\citenamefont
  {Pospelov}, \citenamefont {Ritz},\ and\ \citenamefont
  {Voloshin}}]{Pospelov:2007mp}%
  \BibitemOpen
  \bibfield  {author} {\bibinfo {author} {\bibfnamefont {M.}~\bibnamefont
  {Pospelov}}, \bibinfo {author} {\bibfnamefont {A.}~\bibnamefont {Ritz}}, \
  and\ \bibinfo {author} {\bibfnamefont {M.~B.}\ \bibnamefont {Voloshin}},\
  }\href {\doibase 10.1016/j.physletb.2008.02.052} {\bibfield  {journal}
  {\bibinfo  {journal} {Phys. Lett. B}\ }\textbf {\bibinfo {volume} {662}},\
  \bibinfo {pages} {53} (\bibinfo {year} {2008})},\ \Eprint
  {http://arxiv.org/abs/0711.4866} {arXiv:0711.4866 [hep-ph]} \BibitemShut
  {NoStop}%
\bibitem [{\citenamefont {Pospelov}\ and\ \citenamefont
  {Ritz}(2009)}]{Pospelov:2008jd}%
  \BibitemOpen
  \bibfield  {author} {\bibinfo {author} {\bibfnamefont {M.}~\bibnamefont
  {Pospelov}}\ and\ \bibinfo {author} {\bibfnamefont {A.}~\bibnamefont
  {Ritz}},\ }\href {\doibase 10.1016/j.physletb.2008.12.012} {\bibfield
  {journal} {\bibinfo  {journal} {Phys. Lett. B}\ }\textbf {\bibinfo {volume}
  {671}},\ \bibinfo {pages} {391} (\bibinfo {year} {2009})},\ \Eprint
  {http://arxiv.org/abs/0810.1502} {arXiv:0810.1502 [hep-ph]} \BibitemShut
  {NoStop}%
\bibitem [{\citenamefont {Di~Mauro}\ and\ \citenamefont
  {Wang}(2025)}]{DiMauro:2025jsb}%
  \BibitemOpen
  \bibfield  {author} {\bibinfo {author} {\bibfnamefont {M.}~\bibnamefont
  {Di~Mauro}}\ and\ \bibinfo {author} {\bibfnamefont {Y.}~\bibnamefont
  {Wang}},\ }\href@noop {} {\  (\bibinfo {year} {2025})},\ \Eprint
  {http://arxiv.org/abs/2510.23771} {arXiv:2510.23771 [hep-ph]} \BibitemShut
  {NoStop}%
\bibitem [{\citenamefont {Di~Mauro}(2025)}]{DiMauro:2025uxt}%
  \BibitemOpen
  \bibfield  {author} {\bibinfo {author} {\bibfnamefont {M.}~\bibnamefont
  {Di~Mauro}},\ }\href@noop {} {\  (\bibinfo {year} {2025})},\ \Eprint
  {http://arxiv.org/abs/2511.19622} {arXiv:2511.19622 [hep-ph]} \BibitemShut
  {NoStop}%
\bibitem [{\citenamefont {McDaniel}\ \emph {et~al.}(2024)\citenamefont
  {McDaniel}, \citenamefont {Ajello}, \citenamefont {Karwin}, \citenamefont
  {Di~Mauro}, \citenamefont {Drlica-Wagner},\ and\ \citenamefont
  {S\'anchez-Conde}}]{McDaniel:2023bju}%
  \BibitemOpen
  \bibfield  {author} {\bibinfo {author} {\bibfnamefont {A.}~\bibnamefont
  {McDaniel}}, \bibinfo {author} {\bibfnamefont {M.}~\bibnamefont {Ajello}},
  \bibinfo {author} {\bibfnamefont {C.~M.}\ \bibnamefont {Karwin}}, \bibinfo
  {author} {\bibfnamefont {M.}~\bibnamefont {Di~Mauro}}, \bibinfo {author}
  {\bibfnamefont {A.}~\bibnamefont {Drlica-Wagner}}, \ and\ \bibinfo {author}
  {\bibfnamefont {M.~A.}\ \bibnamefont {S\'anchez-Conde}},\ }\href {\doibase
  10.1103/PhysRevD.109.063024} {\bibfield  {journal} {\bibinfo  {journal}
  {Phys. Rev. D}\ }\textbf {\bibinfo {volume} {109}},\ \bibinfo {pages}
  {063024} (\bibinfo {year} {2024})},\ \Eprint
  {http://arxiv.org/abs/2311.04982} {arXiv:2311.04982 [astro-ph.HE]}
  \BibitemShut {NoStop}%
\bibitem [{\citenamefont {Abdallah}\ \emph {et~al.}(2015)\citenamefont
  {Abdallah}, \citenamefont {Araujo}, \citenamefont {Arbey}, \citenamefont
  {Ashkenazi}, \citenamefont {Belyaev}, \citenamefont {Berger}, \citenamefont
  {Boehm}, \citenamefont {Boveia}, \citenamefont {Brennan}, \citenamefont
  {Brooke}, \citenamefont {Buchmueller}, \citenamefont {Buckley}, \citenamefont
  {Busoni}, \citenamefont {Calibbi}, \citenamefont {Chauhan}, \citenamefont
  {Daci}, \citenamefont {Davies}, \citenamefont {{De Bruyn}}, \citenamefont
  {{De Jong}}, \citenamefont {{De Roeck}}, \citenamefont {{de Vries}},
  \citenamefont {{Del Re}}, \citenamefont {{De Simone}}, \citenamefont {{Di
  Simone}}, \citenamefont {Doglioni}, \citenamefont {Dolan}, \citenamefont
  {Dreiner}, \citenamefont {Ellis}, \citenamefont {Eno}, \citenamefont
  {Etzion}, \citenamefont {Fairbairn}, \citenamefont {Feldstein}, \citenamefont
  {Flaecher}, \citenamefont {Feng}, \citenamefont {Fox}, \citenamefont
  {Genest}, \citenamefont {Gouskos}, \citenamefont {Gramling}, \citenamefont
  {Haisch}, \citenamefont {Harnik}, \citenamefont {Hibbs}, \citenamefont {Hoh},
  \citenamefont {Hopkins}, \citenamefont {Ippolito}, \citenamefont {Jacques},
  \citenamefont {Kahlhoefer}, \citenamefont {Khoze}, \citenamefont {Kirk},
  \citenamefont {Korn}, \citenamefont {Kotov}, \citenamefont {Kunori},
  \citenamefont {Landsberg}, \citenamefont {Liem}, \citenamefont {Lin},
  \citenamefont {Lowette}, \citenamefont {Lucas}, \citenamefont {Malgeri},
  \citenamefont {Malik}, \citenamefont {McCabe}, \citenamefont {Mete},
  \citenamefont {Morgante}, \citenamefont {Mrenna}, \citenamefont {Nakahama},
  \citenamefont {Newbold}, \citenamefont {Nordstrom}, \citenamefont {Pani},
  \citenamefont {Papucci}, \citenamefont {Pataraia}, \citenamefont {Penning},
  \citenamefont {Pinna}, \citenamefont {Polesello}, \citenamefont {Racco},
  \citenamefont {Re}, \citenamefont {Riotto}, \citenamefont {Rizzo},
  \citenamefont {Salek}, \citenamefont {Sarkar}, \citenamefont {Schramm},
  \citenamefont {Skubic}, \citenamefont {Slone}, \citenamefont {Smirnov},
  \citenamefont {Soreq}, \citenamefont {Sumner}, \citenamefont {Tait},
  \citenamefont {Thomas}, \citenamefont {Tomalin}, \citenamefont {Tunnell},
  \citenamefont {Vichi}, \citenamefont {Volansky}, \citenamefont {Weiner},
  \citenamefont {West}, \citenamefont {Wielers}, \citenamefont {Worm},
  \citenamefont {Yavin}, \citenamefont {Zaldivar}, \citenamefont {Zhou},\ and\
  \citenamefont {Zurek}}]{ABDALLAH20158}%
  \BibitemOpen
  \bibfield  {author} {\bibinfo {author} {\bibfnamefont {J.}~\bibnamefont
  {Abdallah}}, \bibinfo {author} {\bibfnamefont {H.}~\bibnamefont {Araujo}},
  \bibinfo {author} {\bibfnamefont {A.}~\bibnamefont {Arbey}}, \bibinfo
  {author} {\bibfnamefont {A.}~\bibnamefont {Ashkenazi}}, \bibinfo {author}
  {\bibfnamefont {A.}~\bibnamefont {Belyaev}}, \bibinfo {author} {\bibfnamefont
  {J.}~\bibnamefont {Berger}}, \bibinfo {author} {\bibfnamefont
  {C.}~\bibnamefont {Boehm}}, \bibinfo {author} {\bibfnamefont
  {A.}~\bibnamefont {Boveia}}, \bibinfo {author} {\bibfnamefont
  {A.}~\bibnamefont {Brennan}}, \bibinfo {author} {\bibfnamefont
  {J.}~\bibnamefont {Brooke}}, \bibinfo {author} {\bibfnamefont
  {O.}~\bibnamefont {Buchmueller}}, \bibinfo {author} {\bibfnamefont
  {M.}~\bibnamefont {Buckley}}, \bibinfo {author} {\bibfnamefont
  {G.}~\bibnamefont {Busoni}}, \bibinfo {author} {\bibfnamefont
  {L.}~\bibnamefont {Calibbi}}, \bibinfo {author} {\bibfnamefont
  {S.}~\bibnamefont {Chauhan}}, \bibinfo {author} {\bibfnamefont
  {N.}~\bibnamefont {Daci}}, \bibinfo {author} {\bibfnamefont {G.}~\bibnamefont
  {Davies}}, \bibinfo {author} {\bibfnamefont {I.}~\bibnamefont {{De Bruyn}}},
  \bibinfo {author} {\bibfnamefont {P.}~\bibnamefont {{De Jong}}}, \bibinfo
  {author} {\bibfnamefont {A.}~\bibnamefont {{De Roeck}}}, \bibinfo {author}
  {\bibfnamefont {K.}~\bibnamefont {{de Vries}}}, \bibinfo {author}
  {\bibfnamefont {D.}~\bibnamefont {{Del Re}}}, \bibinfo {author}
  {\bibfnamefont {A.}~\bibnamefont {{De Simone}}}, \bibinfo {author}
  {\bibfnamefont {A.}~\bibnamefont {{Di Simone}}}, \bibinfo {author}
  {\bibfnamefont {C.}~\bibnamefont {Doglioni}}, \bibinfo {author}
  {\bibfnamefont {M.}~\bibnamefont {Dolan}}, \bibinfo {author} {\bibfnamefont
  {H.~K.}\ \bibnamefont {Dreiner}}, \bibinfo {author} {\bibfnamefont
  {J.}~\bibnamefont {Ellis}}, \bibinfo {author} {\bibfnamefont
  {S.}~\bibnamefont {Eno}}, \bibinfo {author} {\bibfnamefont {E.}~\bibnamefont
  {Etzion}}, \bibinfo {author} {\bibfnamefont {M.}~\bibnamefont {Fairbairn}},
  \bibinfo {author} {\bibfnamefont {B.}~\bibnamefont {Feldstein}}, \bibinfo
  {author} {\bibfnamefont {H.}~\bibnamefont {Flaecher}}, \bibinfo {author}
  {\bibfnamefont {E.}~\bibnamefont {Feng}}, \bibinfo {author} {\bibfnamefont
  {P.}~\bibnamefont {Fox}}, \bibinfo {author} {\bibfnamefont {M.-H.}\
  \bibnamefont {Genest}}, \bibinfo {author} {\bibfnamefont {L.}~\bibnamefont
  {Gouskos}}, \bibinfo {author} {\bibfnamefont {J.}~\bibnamefont {Gramling}},
  \bibinfo {author} {\bibfnamefont {U.}~\bibnamefont {Haisch}}, \bibinfo
  {author} {\bibfnamefont {R.}~\bibnamefont {Harnik}}, \bibinfo {author}
  {\bibfnamefont {A.}~\bibnamefont {Hibbs}}, \bibinfo {author} {\bibfnamefont
  {S.}~\bibnamefont {Hoh}}, \bibinfo {author} {\bibfnamefont {W.}~\bibnamefont
  {Hopkins}}, \bibinfo {author} {\bibfnamefont {V.}~\bibnamefont {Ippolito}},
  \bibinfo {author} {\bibfnamefont {T.}~\bibnamefont {Jacques}}, \bibinfo
  {author} {\bibfnamefont {F.}~\bibnamefont {Kahlhoefer}}, \bibinfo {author}
  {\bibfnamefont {V.~V.}\ \bibnamefont {Khoze}}, \bibinfo {author}
  {\bibfnamefont {R.}~\bibnamefont {Kirk}}, \bibinfo {author} {\bibfnamefont
  {A.}~\bibnamefont {Korn}}, \bibinfo {author} {\bibfnamefont {K.}~\bibnamefont
  {Kotov}}, \bibinfo {author} {\bibfnamefont {S.}~\bibnamefont {Kunori}},
  \bibinfo {author} {\bibfnamefont {G.}~\bibnamefont {Landsberg}}, \bibinfo
  {author} {\bibfnamefont {S.}~\bibnamefont {Liem}}, \bibinfo {author}
  {\bibfnamefont {T.}~\bibnamefont {Lin}}, \bibinfo {author} {\bibfnamefont
  {S.}~\bibnamefont {Lowette}}, \bibinfo {author} {\bibfnamefont
  {R.}~\bibnamefont {Lucas}}, \bibinfo {author} {\bibfnamefont
  {L.}~\bibnamefont {Malgeri}}, \bibinfo {author} {\bibfnamefont
  {S.}~\bibnamefont {Malik}}, \bibinfo {author} {\bibfnamefont
  {C.}~\bibnamefont {McCabe}}, \bibinfo {author} {\bibfnamefont {A.~S.}\
  \bibnamefont {Mete}}, \bibinfo {author} {\bibfnamefont {E.}~\bibnamefont
  {Morgante}}, \bibinfo {author} {\bibfnamefont {S.}~\bibnamefont {Mrenna}},
  \bibinfo {author} {\bibfnamefont {Y.}~\bibnamefont {Nakahama}}, \bibinfo
  {author} {\bibfnamefont {D.}~\bibnamefont {Newbold}}, \bibinfo {author}
  {\bibfnamefont {K.}~\bibnamefont {Nordstrom}}, \bibinfo {author}
  {\bibfnamefont {P.}~\bibnamefont {Pani}}, \bibinfo {author} {\bibfnamefont
  {M.}~\bibnamefont {Papucci}}, \bibinfo {author} {\bibfnamefont
  {S.}~\bibnamefont {Pataraia}}, \bibinfo {author} {\bibfnamefont
  {B.}~\bibnamefont {Penning}}, \bibinfo {author} {\bibfnamefont
  {D.}~\bibnamefont {Pinna}}, \bibinfo {author} {\bibfnamefont
  {G.}~\bibnamefont {Polesello}}, \bibinfo {author} {\bibfnamefont
  {D.}~\bibnamefont {Racco}}, \bibinfo {author} {\bibfnamefont
  {E.}~\bibnamefont {Re}}, \bibinfo {author} {\bibfnamefont {A.~W.}\
  \bibnamefont {Riotto}}, \bibinfo {author} {\bibfnamefont {T.}~\bibnamefont
  {Rizzo}}, \bibinfo {author} {\bibfnamefont {D.}~\bibnamefont {Salek}},
  \bibinfo {author} {\bibfnamefont {S.}~\bibnamefont {Sarkar}}, \bibinfo
  {author} {\bibfnamefont {S.}~\bibnamefont {Schramm}}, \bibinfo {author}
  {\bibfnamefont {P.}~\bibnamefont {Skubic}}, \bibinfo {author} {\bibfnamefont
  {O.}~\bibnamefont {Slone}}, \bibinfo {author} {\bibfnamefont
  {J.}~\bibnamefont {Smirnov}}, \bibinfo {author} {\bibfnamefont
  {Y.}~\bibnamefont {Soreq}}, \bibinfo {author} {\bibfnamefont
  {T.}~\bibnamefont {Sumner}}, \bibinfo {author} {\bibfnamefont {T.~M.}\
  \bibnamefont {Tait}}, \bibinfo {author} {\bibfnamefont {M.}~\bibnamefont
  {Thomas}}, \bibinfo {author} {\bibfnamefont {I.}~\bibnamefont {Tomalin}},
  \bibinfo {author} {\bibfnamefont {C.}~\bibnamefont {Tunnell}}, \bibinfo
  {author} {\bibfnamefont {A.}~\bibnamefont {Vichi}}, \bibinfo {author}
  {\bibfnamefont {T.}~\bibnamefont {Volansky}}, \bibinfo {author}
  {\bibfnamefont {N.}~\bibnamefont {Weiner}}, \bibinfo {author} {\bibfnamefont
  {S.~M.}\ \bibnamefont {West}}, \bibinfo {author} {\bibfnamefont
  {M.}~\bibnamefont {Wielers}}, \bibinfo {author} {\bibfnamefont
  {S.}~\bibnamefont {Worm}}, \bibinfo {author} {\bibfnamefont {I.}~\bibnamefont
  {Yavin}}, \bibinfo {author} {\bibfnamefont {B.}~\bibnamefont {Zaldivar}},
  \bibinfo {author} {\bibfnamefont {N.}~\bibnamefont {Zhou}}, \ and\ \bibinfo
  {author} {\bibfnamefont {K.}~\bibnamefont {Zurek}},\ }\href {\doibase
  https://doi.org/10.1016/j.dark.2015.08.001} {\bibfield  {journal} {\bibinfo
  {journal} {Physics of the Dark Universe}\ }\textbf {\bibinfo {volume}
  {9-10}},\ \bibinfo {pages} {8} (\bibinfo {year} {2015})}\BibitemShut
  {NoStop}%
\bibitem [{\citenamefont {Arina}(2018)}]{Arina:2018zcq}%
  \BibitemOpen
  \bibfield  {author} {\bibinfo {author} {\bibfnamefont {C.}~\bibnamefont
  {Arina}},\ }\href {\doibase 10.3389/fspas.2018.00030} {\bibfield  {journal}
  {\bibinfo  {journal} {Front. Astron. Space Sci.}\ }\textbf {\bibinfo {volume}
  {5}},\ \bibinfo {pages} {30} (\bibinfo {year} {2018})},\ \Eprint
  {http://arxiv.org/abs/1805.04290} {arXiv:1805.04290 [hep-ph]} \BibitemShut
  {NoStop}%
\bibitem [{\citenamefont {Chang}\ \emph
  {et~al.}(2023{\natexlab{a}})\citenamefont {Chang}, \citenamefont {Scott},
  \citenamefont {Gonzalo}, \citenamefont {Kahlhoefer}, \citenamefont
  {Kvellestad},\ and\ \citenamefont {White}}]{Chang:2022jgo}%
  \BibitemOpen
  \bibfield  {author} {\bibinfo {author} {\bibfnamefont {C.}~\bibnamefont
  {Chang}}, \bibinfo {author} {\bibfnamefont {P.}~\bibnamefont {Scott}},
  \bibinfo {author} {\bibfnamefont {T.~E.}\ \bibnamefont {Gonzalo}}, \bibinfo
  {author} {\bibfnamefont {F.}~\bibnamefont {Kahlhoefer}}, \bibinfo {author}
  {\bibfnamefont {A.}~\bibnamefont {Kvellestad}}, \ and\ \bibinfo {author}
  {\bibfnamefont {M.}~\bibnamefont {White}},\ }\href {\doibase
  10.1140/epjc/s10052-023-11399-w} {\bibfield  {journal} {\bibinfo  {journal}
  {Eur. Phys. J. C}\ }\textbf {\bibinfo {volume} {83}},\ \bibinfo {pages} {249}
  (\bibinfo {year} {2023}{\natexlab{a}})},\ \Eprint
  {http://arxiv.org/abs/2209.13266} {arXiv:2209.13266 [hep-ph]} \BibitemShut
  {NoStop}%
\bibitem [{\citenamefont {Chang}\ \emph
  {et~al.}(2023{\natexlab{b}})\citenamefont {Chang}, \citenamefont {Scott},
  \citenamefont {Gonzalo}, \citenamefont {Kahlhoefer},\ and\ \citenamefont
  {White}}]{Chang:2023cki}%
  \BibitemOpen
  \bibfield  {author} {\bibinfo {author} {\bibfnamefont {C.}~\bibnamefont
  {Chang}}, \bibinfo {author} {\bibfnamefont {P.}~\bibnamefont {Scott}},
  \bibinfo {author} {\bibfnamefont {T.~E.}\ \bibnamefont {Gonzalo}}, \bibinfo
  {author} {\bibfnamefont {F.}~\bibnamefont {Kahlhoefer}}, \ and\ \bibinfo
  {author} {\bibfnamefont {M.}~\bibnamefont {White}},\ }\href {\doibase
  10.1140/epjc/s10052-023-11859-3} {\bibfield  {journal} {\bibinfo  {journal}
  {Eur. Phys. J. C}\ }\textbf {\bibinfo {volume} {83}},\ \bibinfo {pages} {692}
  (\bibinfo {year} {2023}{\natexlab{b}})},\ \bibinfo {note} {[Erratum:
  Eur.Phys.J.C 83, 768 (2023)]},\ \Eprint {http://arxiv.org/abs/2303.08351}
  {arXiv:2303.08351 [hep-ph]} \BibitemShut {NoStop}%
\bibitem [{\citenamefont {Cline}\ \emph {et~al.}(2013)\citenamefont {Cline},
  \citenamefont {Kainulainen}, \citenamefont {Scott},\ and\ \citenamefont
  {Weniger}}]{Cline:2013gha}%
  \BibitemOpen
  \bibfield  {author} {\bibinfo {author} {\bibfnamefont {J.~M.}\ \bibnamefont
  {Cline}}, \bibinfo {author} {\bibfnamefont {K.}~\bibnamefont {Kainulainen}},
  \bibinfo {author} {\bibfnamefont {P.}~\bibnamefont {Scott}}, \ and\ \bibinfo
  {author} {\bibfnamefont {C.}~\bibnamefont {Weniger}},\ }\href {\doibase
  10.1103/PhysRevD.88.055025} {\bibfield  {journal} {\bibinfo  {journal} {Phys.
  Rev. D}\ }\textbf {\bibinfo {volume} {88}},\ \bibinfo {pages} {055025}
  (\bibinfo {year} {2013})},\ \bibinfo {note} {[Erratum: Phys.Rev.D 92, 039906
  (2015)]},\ \Eprint {http://arxiv.org/abs/1306.4710} {arXiv:1306.4710
  [hep-ph]} \BibitemShut {NoStop}%
\bibitem [{\citenamefont {Beniwal}\ \emph {et~al.}(2016)\citenamefont
  {Beniwal}, \citenamefont {Rajec}, \citenamefont {Savage}, \citenamefont
  {Scott}, \citenamefont {Weniger}, \citenamefont {White},\ and\ \citenamefont
  {Williams}}]{Beniwal:2015sdl}%
  \BibitemOpen
  \bibfield  {author} {\bibinfo {author} {\bibfnamefont {A.}~\bibnamefont
  {Beniwal}}, \bibinfo {author} {\bibfnamefont {F.}~\bibnamefont {Rajec}},
  \bibinfo {author} {\bibfnamefont {C.}~\bibnamefont {Savage}}, \bibinfo
  {author} {\bibfnamefont {P.}~\bibnamefont {Scott}}, \bibinfo {author}
  {\bibfnamefont {C.}~\bibnamefont {Weniger}}, \bibinfo {author} {\bibfnamefont
  {M.}~\bibnamefont {White}}, \ and\ \bibinfo {author} {\bibfnamefont {A.~G.}\
  \bibnamefont {Williams}},\ }\href {\doibase 10.1103/PhysRevD.93.115016}
  {\bibfield  {journal} {\bibinfo  {journal} {Phys. Rev. D}\ }\textbf {\bibinfo
  {volume} {93}},\ \bibinfo {pages} {115016} (\bibinfo {year} {2016})},\
  \Eprint {http://arxiv.org/abs/1512.06458} {arXiv:1512.06458 [hep-ph]}
  \BibitemShut {NoStop}%
\bibitem [{\citenamefont {Farzan}\ and\ \citenamefont
  {Akbarieh}(2012)}]{Farzan:2012hh}%
  \BibitemOpen
  \bibfield  {author} {\bibinfo {author} {\bibfnamefont {Y.}~\bibnamefont
  {Farzan}}\ and\ \bibinfo {author} {\bibfnamefont {A.~R.}\ \bibnamefont
  {Akbarieh}},\ }\href {\doibase 10.1088/1475-7516/2012/10/026} {\bibfield
  {journal} {\bibinfo  {journal} {JCAP}\ }\textbf {\bibinfo {volume} {10}},\
  \bibinfo {pages} {026} (\bibinfo {year} {2012})},\ \Eprint
  {http://arxiv.org/abs/1207.4272} {arXiv:1207.4272 [hep-ph]} \BibitemShut
  {NoStop}%
\bibitem [{\citenamefont {Ko}\ \emph {et~al.}(2014)\citenamefont {Ko},
  \citenamefont {Park},\ and\ \citenamefont {Tang}}]{Ko:2014gha}%
  \BibitemOpen
  \bibfield  {author} {\bibinfo {author} {\bibfnamefont {P.}~\bibnamefont
  {Ko}}, \bibinfo {author} {\bibfnamefont {W.-I.}\ \bibnamefont {Park}}, \ and\
  \bibinfo {author} {\bibfnamefont {Y.}~\bibnamefont {Tang}},\ }\href {\doibase
  10.1088/1475-7516/2014/09/013} {\bibfield  {journal} {\bibinfo  {journal}
  {JCAP}\ }\textbf {\bibinfo {volume} {09}},\ \bibinfo {pages} {013} (\bibinfo
  {year} {2014})},\ \Eprint {http://arxiv.org/abs/1404.5257} {arXiv:1404.5257
  [hep-ph]} \BibitemShut {NoStop}%
\bibitem [{\citenamefont {Baek}\ \emph {et~al.}(2014)\citenamefont {Baek},
  \citenamefont {Ko}, \citenamefont {Park},\ and\ \citenamefont
  {Tang}}]{Baek:2014goa}%
  \BibitemOpen
  \bibfield  {author} {\bibinfo {author} {\bibfnamefont {S.}~\bibnamefont
  {Baek}}, \bibinfo {author} {\bibfnamefont {P.}~\bibnamefont {Ko}}, \bibinfo
  {author} {\bibfnamefont {W.-I.}\ \bibnamefont {Park}}, \ and\ \bibinfo
  {author} {\bibfnamefont {Y.}~\bibnamefont {Tang}},\ }\href {\doibase
  10.1088/1475-7516/2014/06/046} {\bibfield  {journal} {\bibinfo  {journal}
  {JCAP}\ }\textbf {\bibinfo {volume} {06}},\ \bibinfo {pages} {046} (\bibinfo
  {year} {2014})},\ \Eprint {http://arxiv.org/abs/1402.2115} {arXiv:1402.2115
  [hep-ph]} \BibitemShut {NoStop}%
\bibitem [{\citenamefont {Duch}\ \emph {et~al.}(2015)\citenamefont {Duch},
  \citenamefont {Grzadkowski},\ and\ \citenamefont {McGarrie}}]{Duch:2015jta}%
  \BibitemOpen
  \bibfield  {author} {\bibinfo {author} {\bibfnamefont {M.}~\bibnamefont
  {Duch}}, \bibinfo {author} {\bibfnamefont {B.}~\bibnamefont {Grzadkowski}}, \
  and\ \bibinfo {author} {\bibfnamefont {M.}~\bibnamefont {McGarrie}},\ }\href
  {\doibase 10.1007/JHEP09(2015)162} {\bibfield  {journal} {\bibinfo  {journal}
  {JHEP}\ }\textbf {\bibinfo {volume} {09}},\ \bibinfo {pages} {162} (\bibinfo
  {year} {2015})},\ \Eprint {http://arxiv.org/abs/1506.08805} {arXiv:1506.08805
  [hep-ph]} \BibitemShut {NoStop}%
\bibitem [{\citenamefont {Arcadi}\ \emph
  {et~al.}(2020{\natexlab{b}})\citenamefont {Arcadi}, \citenamefont {Djouadi},\
  and\ \citenamefont {Kado}}]{Arcadi:2020jqf}%
  \BibitemOpen
  \bibfield  {author} {\bibinfo {author} {\bibfnamefont {G.}~\bibnamefont
  {Arcadi}}, \bibinfo {author} {\bibfnamefont {A.}~\bibnamefont {Djouadi}}, \
  and\ \bibinfo {author} {\bibfnamefont {M.}~\bibnamefont {Kado}},\ }\href
  {\doibase 10.1016/j.physletb.2020.135427} {\bibfield  {journal} {\bibinfo
  {journal} {Phys. Lett. B}\ }\textbf {\bibinfo {volume} {805}},\ \bibinfo
  {pages} {135427} (\bibinfo {year} {2020}{\natexlab{b}})},\ \Eprint
  {http://arxiv.org/abs/2001.10750} {arXiv:2001.10750 [hep-ph]} \BibitemShut
  {NoStop}%
\bibitem [{\citenamefont {Arcadi}\ \emph {et~al.}(2015)\citenamefont {Arcadi},
  \citenamefont {Mambrini},\ and\ \citenamefont {Richard}}]{Arcadi:2014lta}%
  \BibitemOpen
  \bibfield  {author} {\bibinfo {author} {\bibfnamefont {G.}~\bibnamefont
  {Arcadi}}, \bibinfo {author} {\bibfnamefont {Y.}~\bibnamefont {Mambrini}}, \
  and\ \bibinfo {author} {\bibfnamefont {F.}~\bibnamefont {Richard}},\ }\href
  {\doibase 10.1088/1475-7516/2015/03/018} {\bibfield  {journal} {\bibinfo
  {journal} {JCAP}\ }\textbf {\bibinfo {volume} {03}},\ \bibinfo {pages} {018}
  (\bibinfo {year} {2015})},\ \Eprint {http://arxiv.org/abs/1411.2985}
  {arXiv:1411.2985 [hep-ph]} \BibitemShut {NoStop}%
\bibitem [{\citenamefont {Foot}(1991)}]{Foot:1990mn}%
  \BibitemOpen
  \bibfield  {author} {\bibinfo {author} {\bibfnamefont {R.}~\bibnamefont
  {Foot}},\ }\href {\doibase 10.1142/S0217732391000543} {\bibfield  {journal}
  {\bibinfo  {journal} {Mod. Phys. Lett. A}\ }\textbf {\bibinfo {volume} {6}},\
  \bibinfo {pages} {527} (\bibinfo {year} {1991})}\BibitemShut {NoStop}%
\bibitem [{\citenamefont {{He}}\ \emph {et~al.}(1991)\citenamefont {{He}},
  \citenamefont {{Joshi}}, \citenamefont {{Lew}},\ and\ \citenamefont
  {{Volkas}}}]{1991PhRvD..43...22H}%
  \BibitemOpen
  \bibfield  {author} {\bibinfo {author} {\bibfnamefont {X.~G.}\ \bibnamefont
  {{He}}}, \bibinfo {author} {\bibfnamefont {G.~C.}\ \bibnamefont {{Joshi}}},
  \bibinfo {author} {\bibfnamefont {H.}~\bibnamefont {{Lew}}}, \ and\ \bibinfo
  {author} {\bibfnamefont {R.~R.}\ \bibnamefont {{Volkas}}},\ }\href {\doibase
  10.1103/PhysRevD.43.R22} {\bibfield  {journal} {\bibinfo  {journal} {\prd}\
  }\textbf {\bibinfo {volume} {43}},\ \bibinfo {pages} {R22} (\bibinfo {year}
  {1991})}\BibitemShut {NoStop}%
\bibitem [{\citenamefont {Heeck}\ and\ \citenamefont
  {Rodejohann}(2011)}]{Heeck:2011wj}%
  \BibitemOpen
  \bibfield  {author} {\bibinfo {author} {\bibfnamefont {J.}~\bibnamefont
  {Heeck}}\ and\ \bibinfo {author} {\bibfnamefont {W.}~\bibnamefont
  {Rodejohann}},\ }\href {\doibase 10.1103/PhysRevD.84.075007} {\bibfield
  {journal} {\bibinfo  {journal} {Phys. Rev. D}\ }\textbf {\bibinfo {volume}
  {84}},\ \bibinfo {pages} {075007} (\bibinfo {year} {2011})},\ \Eprint
  {http://arxiv.org/abs/1107.5238} {arXiv:1107.5238 [hep-ph]} \BibitemShut
  {NoStop}%
\bibitem [{\citenamefont {Bauer}\ \emph {et~al.}(2018)\citenamefont {Bauer},
  \citenamefont {Foldenauer},\ and\ \citenamefont {Jaeckel}}]{Bauer:2018onh}%
  \BibitemOpen
  \bibfield  {author} {\bibinfo {author} {\bibfnamefont {M.}~\bibnamefont
  {Bauer}}, \bibinfo {author} {\bibfnamefont {P.}~\bibnamefont {Foldenauer}}, \
  and\ \bibinfo {author} {\bibfnamefont {J.}~\bibnamefont {Jaeckel}},\ }\href
  {\doibase 10.1007/JHEP07(2018)094} {\bibfield  {journal} {\bibinfo  {journal}
  {JHEP}\ }\textbf {\bibinfo {volume} {07}},\ \bibinfo {pages} {094} (\bibinfo
  {year} {2018})},\ \Eprint {http://arxiv.org/abs/1803.05466} {arXiv:1803.05466
  [hep-ph]} \BibitemShut {NoStop}%
\bibitem [{\citenamefont {Alloul}\ \emph {et~al.}(2014)\citenamefont {Alloul},
  \citenamefont {Christensen}, \citenamefont {Degrande}, \citenamefont {Duhr},\
  and\ \citenamefont {Fuks}}]{Alloul:2013bka}%
  \BibitemOpen
  \bibfield  {author} {\bibinfo {author} {\bibfnamefont {A.}~\bibnamefont
  {Alloul}}, \bibinfo {author} {\bibfnamefont {N.~D.}\ \bibnamefont
  {Christensen}}, \bibinfo {author} {\bibfnamefont {C.}~\bibnamefont
  {Degrande}}, \bibinfo {author} {\bibfnamefont {C.}~\bibnamefont {Duhr}}, \
  and\ \bibinfo {author} {\bibfnamefont {B.}~\bibnamefont {Fuks}},\ }\href
  {\doibase 10.1016/j.cpc.2014.04.012} {\bibfield  {journal} {\bibinfo
  {journal} {Comput. Phys. Commun.}\ }\textbf {\bibinfo {volume} {185}},\
  \bibinfo {pages} {2250} (\bibinfo {year} {2014})},\ \Eprint
  {http://arxiv.org/abs/1310.1921} {arXiv:1310.1921 [hep-ph]} \BibitemShut
  {NoStop}%
\bibitem [{\citenamefont {Degrande}\ \emph {et~al.}(2012)\citenamefont
  {Degrande}, \citenamefont {Duhr}, \citenamefont {Fuks}, \citenamefont
  {Grellscheid}, \citenamefont {Mattelaer},\ and\ \citenamefont
  {Reiter}}]{Degrande:2011ua}%
  \BibitemOpen
  \bibfield  {author} {\bibinfo {author} {\bibfnamefont {C.}~\bibnamefont
  {Degrande}}, \bibinfo {author} {\bibfnamefont {C.}~\bibnamefont {Duhr}},
  \bibinfo {author} {\bibfnamefont {B.}~\bibnamefont {Fuks}}, \bibinfo {author}
  {\bibfnamefont {D.}~\bibnamefont {Grellscheid}}, \bibinfo {author}
  {\bibfnamefont {O.}~\bibnamefont {Mattelaer}}, \ and\ \bibinfo {author}
  {\bibfnamefont {T.}~\bibnamefont {Reiter}},\ }\href {\doibase
  10.1016/j.cpc.2012.01.022} {\bibfield  {journal} {\bibinfo  {journal}
  {Comput. Phys. Commun.}\ }\textbf {\bibinfo {volume} {183}},\ \bibinfo
  {pages} {1201} (\bibinfo {year} {2012})},\ \Eprint
  {http://arxiv.org/abs/1108.2040} {arXiv:1108.2040 [hep-ph]} \BibitemShut
  {NoStop}%
\bibitem [{\citenamefont {Belyaev}\ \emph {et~al.}(2013)\citenamefont
  {Belyaev}, \citenamefont {Christensen},\ and\ \citenamefont
  {Pukhov}}]{Belyaev:2012qa}%
  \BibitemOpen
  \bibfield  {author} {\bibinfo {author} {\bibfnamefont {A.}~\bibnamefont
  {Belyaev}}, \bibinfo {author} {\bibfnamefont {N.~D.}\ \bibnamefont
  {Christensen}}, \ and\ \bibinfo {author} {\bibfnamefont {A.}~\bibnamefont
  {Pukhov}},\ }\href {\doibase 10.1016/j.cpc.2013.01.014} {\bibfield  {journal}
  {\bibinfo  {journal} {Comput. Phys. Commun.}\ }\textbf {\bibinfo {volume}
  {184}},\ \bibinfo {pages} {1729} (\bibinfo {year} {2013})},\ \Eprint
  {http://arxiv.org/abs/1207.6082} {arXiv:1207.6082 [hep-ph]} \BibitemShut
  {NoStop}%
\bibitem [{\citenamefont {Backovic}\ \emph {et~al.}(2014)\citenamefont
  {Backovic}, \citenamefont {Kong},\ and\ \citenamefont
  {McCaskey}}]{Backovic:2013dpa}%
  \BibitemOpen
  \bibfield  {author} {\bibinfo {author} {\bibfnamefont {M.}~\bibnamefont
  {Backovic}}, \bibinfo {author} {\bibfnamefont {K.}~\bibnamefont {Kong}}, \
  and\ \bibinfo {author} {\bibfnamefont {M.}~\bibnamefont {McCaskey}},\ }\href
  {\doibase 10.1016/j.dark.2014.04.001} {\bibfield  {journal} {\bibinfo
  {journal} {Physics of the Dark Universe}\ }\textbf {\bibinfo {volume}
  {5-6}},\ \bibinfo {pages} {18} (\bibinfo {year} {2014})},\ \Eprint
  {http://arxiv.org/abs/1308.4955} {arXiv:1308.4955 [hep-ph]} \BibitemShut
  {NoStop}%
\bibitem [{\citenamefont {Ambrogi}\ \emph {et~al.}(2019)\citenamefont
  {Ambrogi}, \citenamefont {Arina}, \citenamefont {Backovic}, \citenamefont
  {Heisig}, \citenamefont {Maltoni}, \citenamefont {Mantani}, \citenamefont
  {Mattelaer},\ and\ \citenamefont {Mohlabeng}}]{Ambrogi:2018jqj}%
  \BibitemOpen
  \bibfield  {author} {\bibinfo {author} {\bibfnamefont {F.}~\bibnamefont
  {Ambrogi}}, \bibinfo {author} {\bibfnamefont {C.}~\bibnamefont {Arina}},
  \bibinfo {author} {\bibfnamefont {M.}~\bibnamefont {Backovic}}, \bibinfo
  {author} {\bibfnamefont {J.}~\bibnamefont {Heisig}}, \bibinfo {author}
  {\bibfnamefont {F.}~\bibnamefont {Maltoni}}, \bibinfo {author} {\bibfnamefont
  {L.}~\bibnamefont {Mantani}}, \bibinfo {author} {\bibfnamefont
  {O.}~\bibnamefont {Mattelaer}}, \ and\ \bibinfo {author} {\bibfnamefont
  {G.}~\bibnamefont {Mohlabeng}},\ }\href {\doibase 10.1016/j.dark.2018.11.009}
  {\bibfield  {journal} {\bibinfo  {journal} {Phys. Dark Univ.}\ }\textbf
  {\bibinfo {volume} {24}},\ \bibinfo {pages} {100249} (\bibinfo {year}
  {2019})},\ \Eprint {http://arxiv.org/abs/1804.00044} {arXiv:1804.00044
  [hep-ph]} \BibitemShut {NoStop}%
\bibitem [{\citenamefont {Arina}\ \emph {et~al.}(2023)\citenamefont {Arina},
  \citenamefont {Heisig}, \citenamefont {Maltoni}, \citenamefont {Massaro},\
  and\ \citenamefont {Mattelaer}}]{Arina:2021gfn}%
  \BibitemOpen
  \bibfield  {author} {\bibinfo {author} {\bibfnamefont {C.}~\bibnamefont
  {Arina}}, \bibinfo {author} {\bibfnamefont {J.}~\bibnamefont {Heisig}},
  \bibinfo {author} {\bibfnamefont {F.}~\bibnamefont {Maltoni}}, \bibinfo
  {author} {\bibfnamefont {D.}~\bibnamefont {Massaro}}, \ and\ \bibinfo
  {author} {\bibfnamefont {O.}~\bibnamefont {Mattelaer}},\ }\href {\doibase
  10.1140/epjc/s10052-023-11377-2} {\bibfield  {journal} {\bibinfo  {journal}
  {Eur. Phys. J. C}\ }\textbf {\bibinfo {volume} {83}},\ \bibinfo {pages} {241}
  (\bibinfo {year} {2023})},\ \Eprint {http://arxiv.org/abs/2107.04598}
  {arXiv:2107.04598 [hep-ph]} \BibitemShut {NoStop}%
\bibitem [{\citenamefont {Belanger}\ \emph {et~al.}(2007)\citenamefont
  {Belanger}, \citenamefont {Boudjema}, \citenamefont {Pukhov},\ and\
  \citenamefont {Semenov}}]{Belanger:2006is}%
  \BibitemOpen
  \bibfield  {author} {\bibinfo {author} {\bibfnamefont {G.}~\bibnamefont
  {Belanger}}, \bibinfo {author} {\bibfnamefont {F.}~\bibnamefont {Boudjema}},
  \bibinfo {author} {\bibfnamefont {A.}~\bibnamefont {Pukhov}}, \ and\ \bibinfo
  {author} {\bibfnamefont {A.}~\bibnamefont {Semenov}},\ }\href {\doibase
  10.1016/j.cpc.2006.11.008} {\bibfield  {journal} {\bibinfo  {journal}
  {Comput. Phys. Commun.}\ }\textbf {\bibinfo {volume} {176}},\ \bibinfo
  {pages} {367} (\bibinfo {year} {2007})},\ \Eprint
  {http://arxiv.org/abs/hep-ph/0607059} {arXiv:hep-ph/0607059} \BibitemShut
  {NoStop}%
\bibitem [{\citenamefont {Belanger}\ \emph {et~al.}(2014)\citenamefont
  {Belanger}, \citenamefont {Boudjema}, \citenamefont {Pukhov},\ and\
  \citenamefont {Semenov}}]{Belanger:2013oya}%
  \BibitemOpen
  \bibfield  {author} {\bibinfo {author} {\bibfnamefont {G.}~\bibnamefont
  {Belanger}}, \bibinfo {author} {\bibfnamefont {F.}~\bibnamefont {Boudjema}},
  \bibinfo {author} {\bibfnamefont {A.}~\bibnamefont {Pukhov}}, \ and\ \bibinfo
  {author} {\bibfnamefont {A.}~\bibnamefont {Semenov}},\ }\href {\doibase
  10.1016/j.cpc.2013.10.016} {\bibfield  {journal} {\bibinfo  {journal}
  {Comput. Phys. Commun.}\ }\textbf {\bibinfo {volume} {185}},\ \bibinfo
  {pages} {960} (\bibinfo {year} {2014})},\ \Eprint
  {http://arxiv.org/abs/1305.0237} {arXiv:1305.0237 [hep-ph]} \BibitemShut
  {NoStop}%
\bibitem [{\citenamefont {B\'elanger}\ \emph {et~al.}(2018)\citenamefont
  {B\'elanger}, \citenamefont {Boudjema}, \citenamefont {Goudelis},
  \citenamefont {Pukhov},\ and\ \citenamefont {Zaldivar}}]{Belanger:2018ccd}%
  \BibitemOpen
  \bibfield  {author} {\bibinfo {author} {\bibfnamefont {G.}~\bibnamefont
  {B\'elanger}}, \bibinfo {author} {\bibfnamefont {F.}~\bibnamefont
  {Boudjema}}, \bibinfo {author} {\bibfnamefont {A.}~\bibnamefont {Goudelis}},
  \bibinfo {author} {\bibfnamefont {A.}~\bibnamefont {Pukhov}}, \ and\ \bibinfo
  {author} {\bibfnamefont {B.}~\bibnamefont {Zaldivar}},\ }\href {\doibase
  10.1016/j.cpc.2018.04.027} {\bibfield  {journal} {\bibinfo  {journal}
  {Comput. Phys. Commun.}\ }\textbf {\bibinfo {volume} {231}},\ \bibinfo
  {pages} {173} (\bibinfo {year} {2018})},\ \Eprint
  {http://arxiv.org/abs/1801.03509} {arXiv:1801.03509 [hep-ph]} \BibitemShut
  {NoStop}%
\bibitem [{\citenamefont {Alguero}\ \emph {et~al.}(2024)\citenamefont
  {Alguero}, \citenamefont {Belanger}, \citenamefont {Boudjema}, \citenamefont
  {Chakraborti}, \citenamefont {Goudelis}, \citenamefont {Kraml}, \citenamefont
  {Mjallal},\ and\ \citenamefont {Pukhov}}]{Alguero:2023zol}%
  \BibitemOpen
  \bibfield  {author} {\bibinfo {author} {\bibfnamefont {G.}~\bibnamefont
  {Alguero}}, \bibinfo {author} {\bibfnamefont {G.}~\bibnamefont {Belanger}},
  \bibinfo {author} {\bibfnamefont {F.}~\bibnamefont {Boudjema}}, \bibinfo
  {author} {\bibfnamefont {S.}~\bibnamefont {Chakraborti}}, \bibinfo {author}
  {\bibfnamefont {A.}~\bibnamefont {Goudelis}}, \bibinfo {author}
  {\bibfnamefont {S.}~\bibnamefont {Kraml}}, \bibinfo {author} {\bibfnamefont
  {A.}~\bibnamefont {Mjallal}}, \ and\ \bibinfo {author} {\bibfnamefont
  {A.}~\bibnamefont {Pukhov}},\ }\href {\doibase 10.1016/j.cpc.2024.109133}
  {\bibfield  {journal} {\bibinfo  {journal} {Comput. Phys. Commun.}\ }\textbf
  {\bibinfo {volume} {299}},\ \bibinfo {pages} {109133} (\bibinfo {year}
  {2024})},\ \Eprint {http://arxiv.org/abs/2312.14894} {arXiv:2312.14894
  [hep-ph]} \BibitemShut {NoStop}%
\bibitem [{\citenamefont {Fitzpatrick}\ \emph {et~al.}(2013)\citenamefont
  {Fitzpatrick}, \citenamefont {Haxton}, \citenamefont {Katz}, \citenamefont
  {Lubbers},\ and\ \citenamefont {Xu}}]{Fitzpatrick:2012ix}%
  \BibitemOpen
  \bibfield  {author} {\bibinfo {author} {\bibfnamefont {A.~L.}\ \bibnamefont
  {Fitzpatrick}}, \bibinfo {author} {\bibfnamefont {W.}~\bibnamefont {Haxton}},
  \bibinfo {author} {\bibfnamefont {E.}~\bibnamefont {Katz}}, \bibinfo {author}
  {\bibfnamefont {N.}~\bibnamefont {Lubbers}}, \ and\ \bibinfo {author}
  {\bibfnamefont {Y.}~\bibnamefont {Xu}},\ }\href {\doibase
  10.1088/1475-7516/2013/02/004} {\bibfield  {journal} {\bibinfo  {journal}
  {JCAP}\ }\textbf {\bibinfo {volume} {02}},\ \bibinfo {pages} {004} (\bibinfo
  {year} {2013})},\ \Eprint {http://arxiv.org/abs/1203.3542} {arXiv:1203.3542
  [hep-ph]} \BibitemShut {NoStop}%
\bibitem [{\citenamefont {Arina}\ \emph {et~al.}(2015)\citenamefont {Arina},
  \citenamefont {Del~Nobile},\ and\ \citenamefont {Panci}}]{Arina:2014yna}%
  \BibitemOpen
  \bibfield  {author} {\bibinfo {author} {\bibfnamefont {C.}~\bibnamefont
  {Arina}}, \bibinfo {author} {\bibfnamefont {E.}~\bibnamefont {Del~Nobile}}, \
  and\ \bibinfo {author} {\bibfnamefont {P.}~\bibnamefont {Panci}},\ }\href
  {\doibase 10.1103/PhysRevLett.114.011301} {\bibfield  {journal} {\bibinfo
  {journal} {Phys. Rev. Lett.}\ }\textbf {\bibinfo {volume} {114}},\ \bibinfo
  {pages} {011301} (\bibinfo {year} {2015})},\ \Eprint
  {http://arxiv.org/abs/1406.5542} {arXiv:1406.5542 [hep-ph]} \BibitemShut
  {NoStop}%
\bibitem [{\citenamefont {Dolan}\ \emph {et~al.}(2015)\citenamefont {Dolan},
  \citenamefont {Kahlhoefer}, \citenamefont {McCabe},\ and\ \citenamefont
  {Schmidt-Hoberg}}]{Dolan:2014ska}%
  \BibitemOpen
  \bibfield  {author} {\bibinfo {author} {\bibfnamefont {M.~J.}\ \bibnamefont
  {Dolan}}, \bibinfo {author} {\bibfnamefont {F.}~\bibnamefont {Kahlhoefer}},
  \bibinfo {author} {\bibfnamefont {C.}~\bibnamefont {McCabe}}, \ and\ \bibinfo
  {author} {\bibfnamefont {K.}~\bibnamefont {Schmidt-Hoberg}},\ }\href
  {\doibase 10.1007/JHEP03(2015)171} {\bibfield  {journal} {\bibinfo  {journal}
  {JHEP}\ }\textbf {\bibinfo {volume} {03}},\ \bibinfo {pages} {171} (\bibinfo
  {year} {2015})},\ \bibinfo {note} {[Erratum: JHEP 07, 103 (2015)]},\ \Eprint
  {http://arxiv.org/abs/1412.5174} {arXiv:1412.5174 [hep-ph]} \BibitemShut
  {NoStop}%
\bibitem [{\citenamefont {Abe}\ \emph {et~al.}(2019)\citenamefont {Abe},
  \citenamefont {Fujiwara},\ and\ \citenamefont {Hisano}}]{Abe:2018emu}%
  \BibitemOpen
  \bibfield  {author} {\bibinfo {author} {\bibfnamefont {T.}~\bibnamefont
  {Abe}}, \bibinfo {author} {\bibfnamefont {M.}~\bibnamefont {Fujiwara}}, \
  and\ \bibinfo {author} {\bibfnamefont {J.}~\bibnamefont {Hisano}},\ }\href
  {\doibase 10.1007/JHEP02(2019)028} {\bibfield  {journal} {\bibinfo  {journal}
  {JHEP}\ }\textbf {\bibinfo {volume} {02}},\ \bibinfo {pages} {028} (\bibinfo
  {year} {2019})},\ \Eprint {http://arxiv.org/abs/1810.01039} {arXiv:1810.01039
  [hep-ph]} \BibitemShut {NoStop}%
\bibitem [{\citenamefont {Ertas}\ and\ \citenamefont
  {Kahlhoefer}(2019)}]{Ertas:2019dew}%
  \BibitemOpen
  \bibfield  {author} {\bibinfo {author} {\bibfnamefont {F.}~\bibnamefont
  {Ertas}}\ and\ \bibinfo {author} {\bibfnamefont {F.}~\bibnamefont
  {Kahlhoefer}},\ }\href {\doibase 10.1007/JHEP06(2019)052} {\bibfield
  {journal} {\bibinfo  {journal} {JHEP}\ }\textbf {\bibinfo {volume} {06}},\
  \bibinfo {pages} {052} (\bibinfo {year} {2019})},\ \Eprint
  {http://arxiv.org/abs/1902.11070} {arXiv:1902.11070 [hep-ph]} \BibitemShut
  {NoStop}%
\bibitem [{\citenamefont {Silveira}\ and\ \citenamefont
  {Zee}(1985)}]{Silveira1985136}%
  \BibitemOpen
  \bibfield  {author} {\bibinfo {author} {\bibfnamefont {V.}~\bibnamefont
  {Silveira}}\ and\ \bibinfo {author} {\bibfnamefont {A.}~\bibnamefont {Zee}},\
  }\href {\doibase https://doi.org/10.1016/0370-2693(85)90624-0} {\bibfield
  {journal} {\bibinfo  {journal} {Physics Letters B}\ }\textbf {\bibinfo
  {volume} {161}},\ \bibinfo {pages} {136} (\bibinfo {year}
  {1985})}\BibitemShut {NoStop}%
\bibitem [{\citenamefont {{ATLAS collaboration}}(2020)}]{ATLAS-CONF-2020-008}%
  \BibitemOpen
  \bibfield  {author} {\bibinfo {author} {\bibnamefont {{ATLAS
  collaboration}}},\ }\href {https://cds.cern.ch/record/2715447} {\emph
  {\bibinfo {title} {{Search for invisible Higgs boson decays with vector boson
  fusion signatures with the ATLAS detector using an integrated luminosity of
  139 fb$^{-1}$}}}},\ \bibinfo {type} {Tech. Rep.}\ (\bibinfo  {institution}
  {CERN},\ \bibinfo {address} {Geneva},\ \bibinfo {year} {2020})\ \bibinfo
  {note} {all figures including auxiliary figures are available at
  https://atlas.web.cern.ch/Atlas/GROUPS/PHYSICS/
  CONFNOTES/ATLAS-CONF-2020-008}\BibitemShut {NoStop}%
\bibitem [{\citenamefont {Sirunyan}\ \emph {et~al.}(2019)\citenamefont
  {Sirunyan} \emph {et~al.}}]{CMS:2018yfx}%
  \BibitemOpen
  \bibfield  {author} {\bibinfo {author} {\bibfnamefont {A.~M.}\ \bibnamefont
  {Sirunyan}} \emph {et~al.} (\bibinfo {collaboration} {CMS}),\ }\href
  {\doibase 10.1016/j.physletb.2019.04.025} {\bibfield  {journal} {\bibinfo
  {journal} {Phys. Lett. B}\ }\textbf {\bibinfo {volume} {793}},\ \bibinfo
  {pages} {520} (\bibinfo {year} {2019})},\ \Eprint
  {http://arxiv.org/abs/1809.05937} {arXiv:1809.05937 [hep-ex]} \BibitemShut
  {NoStop}%
\bibitem [{\citenamefont {Burgess}\ \emph {et~al.}(2008)\citenamefont
  {Burgess}, \citenamefont {de~Rham}, \citenamefont {Hoover},\ and\
  \citenamefont {Tolley}}]{Burgess:2008ri}%
  \BibitemOpen
  \bibfield  {author} {\bibinfo {author} {\bibfnamefont {C.}~\bibnamefont
  {Burgess}}, \bibinfo {author} {\bibfnamefont {C.}~\bibnamefont {de~Rham}},
  \bibinfo {author} {\bibfnamefont {D.}~\bibnamefont {Hoover}}, \ and\ \bibinfo
  {author} {\bibfnamefont {A.~J.}\ \bibnamefont {Tolley}},\ }\href {\doibase
  10.1088/1126-6708/2008/09/033} {\bibfield  {journal} {\bibinfo  {journal}
  {JHEP}\ }\textbf {\bibinfo {volume} {09}},\ \bibinfo {pages} {033} (\bibinfo
  {year} {2008})},\ \Eprint {http://arxiv.org/abs/0804.1075} {arXiv:0804.1075
  [hep-th]} \BibitemShut {NoStop}%
\bibitem [{\citenamefont {Evans}\ \emph {et~al.}(2018)\citenamefont {Evans},
  \citenamefont {Gori},\ and\ \citenamefont {Shelton}}]{Evans:2017kti}%
  \BibitemOpen
  \bibfield  {author} {\bibinfo {author} {\bibfnamefont {J.~A.}\ \bibnamefont
  {Evans}}, \bibinfo {author} {\bibfnamefont {S.}~\bibnamefont {Gori}}, \ and\
  \bibinfo {author} {\bibfnamefont {J.}~\bibnamefont {Shelton}},\ }\href
  {\doibase 10.1007/JHEP02(2018)100} {\bibfield  {journal} {\bibinfo  {journal}
  {JHEP}\ }\textbf {\bibinfo {volume} {02}},\ \bibinfo {pages} {100} (\bibinfo
  {year} {2018})},\ \Eprint {http://arxiv.org/abs/1712.03974} {arXiv:1712.03974
  [hep-ph]} \BibitemShut {NoStop}%
\bibitem [{\citenamefont {Arcadi}\ \emph
  {et~al.}(2018{\natexlab{b}})\citenamefont {Arcadi}, \citenamefont {Hugle},\
  and\ \citenamefont {Queiroz}}]{Arcadi:2018tly}%
  \BibitemOpen
  \bibfield  {author} {\bibinfo {author} {\bibfnamefont {G.}~\bibnamefont
  {Arcadi}}, \bibinfo {author} {\bibfnamefont {T.}~\bibnamefont {Hugle}}, \
  and\ \bibinfo {author} {\bibfnamefont {F.~S.}\ \bibnamefont {Queiroz}},\
  }\href {\doibase 10.1016/j.physletb.2018.07.028} {\bibfield  {journal}
  {\bibinfo  {journal} {Phys. Lett. B}\ }\textbf {\bibinfo {volume} {784}},\
  \bibinfo {pages} {151} (\bibinfo {year} {2018}{\natexlab{b}})},\ \Eprint
  {http://arxiv.org/abs/1803.05723} {arXiv:1803.05723 [hep-ph]} \BibitemShut
  {NoStop}%
\bibitem [{\citenamefont {Helm}(1956)}]{Helm:1956zz}%
  \BibitemOpen
  \bibfield  {author} {\bibinfo {author} {\bibfnamefont {R.~H.}\ \bibnamefont
  {Helm}},\ }\href {\doibase 10.1103/PhysRev.104.1466} {\bibfield  {journal}
  {\bibinfo  {journal} {Phys. Rev.}\ }\textbf {\bibinfo {volume} {104}},\
  \bibinfo {pages} {1466} (\bibinfo {year} {1956})}\BibitemShut {NoStop}%
\bibitem [{\citenamefont {Duda}\ \emph {et~al.}(2007)\citenamefont {Duda},
  \citenamefont {Kemper},\ and\ \citenamefont {Gondolo}}]{Duda:2006uk}%
  \BibitemOpen
  \bibfield  {author} {\bibinfo {author} {\bibfnamefont {G.}~\bibnamefont
  {Duda}}, \bibinfo {author} {\bibfnamefont {A.}~\bibnamefont {Kemper}}, \ and\
  \bibinfo {author} {\bibfnamefont {P.}~\bibnamefont {Gondolo}},\ }\href
  {\doibase 10.1088/1475-7516/2007/04/012} {\bibfield  {journal} {\bibinfo
  {journal} {JCAP}\ }\textbf {\bibinfo {volume} {04}},\ \bibinfo {pages} {012}
  (\bibinfo {year} {2007})},\ \Eprint {http://arxiv.org/abs/hep-ph/0608035}
  {arXiv:hep-ph/0608035} \BibitemShut {NoStop}%
\bibitem [{\citenamefont {Aghanim}\ \emph {et~al.}(2020)\citenamefont {Aghanim}
  \emph {et~al.}}]{Planck:2018vyg}%
  \BibitemOpen
  \bibfield  {author} {\bibinfo {author} {\bibfnamefont {N.}~\bibnamefont
  {Aghanim}} \emph {et~al.} (\bibinfo {collaboration} {Planck}),\ }\href
  {\doibase 10.1051/0004-6361/201833910} {\bibfield  {journal} {\bibinfo
  {journal} {Astron. Astrophys.}\ }\textbf {\bibinfo {volume} {641}},\ \bibinfo
  {pages} {A6} (\bibinfo {year} {2020})},\ \bibinfo {note} {[Erratum:
  Astron.Astrophys. 652, C4 (2021)]},\ \Eprint
  {http://arxiv.org/abs/1807.06209} {arXiv:1807.06209 [astro-ph.CO]}
  \BibitemShut {NoStop}%
\bibitem [{\citenamefont {Baudis}(2024)}]{DARWIN}%
  \BibitemOpen
  \bibfield  {author} {\bibinfo {author} {\bibfnamefont {L.}~\bibnamefont
  {Baudis}},\ }\href {\doibase https://doi.org/10.1016/j.nuclphysb.2024.116473}
  {\bibfield  {journal} {\bibinfo  {journal} {Nuclear Physics B}\ }\textbf
  {\bibinfo {volume} {1003}},\ \bibinfo {pages} {116473} (\bibinfo {year}
  {2024})},\ \bibinfo {note} {special Issue of Nobel Symposium 182 on Dark
  Matter}\BibitemShut {NoStop}%
\bibitem [{\citenamefont {Aad}\ \emph {et~al.}(2024)\citenamefont {Aad} \emph
  {et~al.}}]{ATLAS:2024kpy}%
  \BibitemOpen
  \bibfield  {author} {\bibinfo {author} {\bibfnamefont {G.}~\bibnamefont
  {Aad}} \emph {et~al.} (\bibinfo {collaboration} {ATLAS}),\ }\href {\doibase
  10.1140/epjc/s10052-024-13215-5} {\bibfield  {journal} {\bibinfo  {journal}
  {Eur. Phys. J. C}\ }\textbf {\bibinfo {volume} {84}},\ \bibinfo {pages}
  {1102} (\bibinfo {year} {2024})},\ \Eprint {http://arxiv.org/abs/2404.15930}
  {arXiv:2404.15930 [hep-ex]} \BibitemShut {NoStop}%
\end{thebibliography}%

\end{document}